\documentclass[11pt]{article}
\newcommand{\Comment}[1]{{}}
\usepackage{amsfonts,amsthm,amsmath,amssymb,slashed}
\usepackage[textwidth = 430 pt, textheight = 630 pt]{geometry}
\usepackage{color} 
\usepackage{hyperref}
\usepackage[compat=1.1.0]{tikz-feynman}
\usepackage{comment}
\usepackage{subfigure}
\usepackage{placeins}

\Comment{\usepackage{color}
\definecolor{MyDarkBlue}{rgb}{0.15,0.15,0.45}
\usepackage[linktocpage=true]{hyperref}
\hypersetup{
colorlinks=true,
citecolor=MyDarkBlue,
linkcolor=MyDarkBlue,
urlcolor=MyDarkBlue,
pdfauthor={Matheus Cravo and Horatiu Nastase },
pdftitle={3-point function of $SO(3)$ global symmetry current in a three-dimensional toy model for holographic cosmology.},
pdfsubject={hep-th}
}

\usepackage[numbers,sort&compress]{natbib}
\usepackage{hypernat}}
\usepackage{graphicx}
\usepackage{cite}

\newcommand\ignore[1]{}
\def\one{{\,\hbox{1\kern-.8mm l}}}

\def\n{\nu}

\def\d{\partial}

\newcommand{\Cset}{{\,\,{{{^{_{\pmb{\mid}}}}\kern-.45em{\mathrm C}}}}}

\newcommand{\be}{\begin{equation}}
\newcommand{\bea}{\begin{eqnarray}}

\newcommand{\ee}{\end{equation}}
\newcommand{\eea}{\end{eqnarray}}

\begin{document}

\renewcommand{\thefootnote}{\fnsymbol{footnote}}

\makeatletter
\@addtoreset{equation}{section}
\makeatother
\renewcommand{\theequation}{\thesection.\arabic{equation}}

\rightline{}
\rightline{}
%   \vspace{1.8truecm}

%\begin{flushright}
% preprint nrs.
%\end{flushright}

%\vspace{10pt}

%%%%%%%%%%%%%%%%%

\begin{center}
{\LARGE \bf{\sc 3-point function of currents for holographic cosmology and monopole non-Gaussianities}}
\end{center} 
 \vspace{1truecm}
\thispagestyle{empty} \centerline{
{\large \bf {\sc Matheus Cravo${}^{a}$}}\footnote{E-mail address: \Comment{\href{mailto:matheus.cravo@unesp.br}}{\tt matheus.cravo@unesp.br}}
{\bf{\sc and}}
{\large \bf {\sc Horatiu Nastase${}^{a}$}}\footnote{E-mail address: \Comment{\href{mailto:horatiu.nastase@unesp.br}}
{\tt horatiu.nastase@unesp.br}}
                                                        }

\vspace{.5cm}

%\vspace{.3cm}

\centerline{{\it ${}^a$Instituto de F\'{i}sica Te\'{o}rica, UNESP-Universidade Estadual Paulista}} 
\centerline{{\it R. Dr. Bento T. Ferraz 271, Bl. II, Sao Paulo 01140-070, SP, Brazil}}
\vspace{.3cm}

%\vspace{.3cm}
%\centerline{{\it ${}^c$ Simons Center for Geometry and Physics,}} 
%\centerline{{\it SUNY, Stony Brook, NY 11794, USA }} 
\vspace{1truecm}

%%%%%%%%%%%%%%%%%
\thispagestyle{empty}

\centerline{\sc Abstract}

\vspace{.4truecm}

\begin{center}
\begin{minipage}[c]{380pt}
{\noindent

In this paper we present the calculation of the three-point function in momentum space of currents
for a $SO(3)$ global symmetry, in a three-dimensional toy model within phenomenological holographic
cosmology. Since the two-point function gives, via electric-magnetic duality, the resolution of the cosmological monopole 
problem, the three-point function is related to the monopole non-Gaussianities. 
%We first review the phenomenological holographic approach to cosmology and the solution to the monopole problem, presenting the toy model we will use in the rest of the paper. Then, motivated by the fact that the result of the 3-point function for the energy-momentum tensor --- related to non-Gaussianities in the CMBR --- already exists in the literature, we further explore the current operator by calculating its 3-point function. 
We check that the final result is UV and IR finite and satisfies the transverse Ward identities and consider the $k_1\ll k_2,k_3$
case, relevant for cosmology. We also show that the 
two-loop result for the 3-point function is completely independent of the explicit form of the potential, meaning that, like the 
solution to the monopole problem, also the non-Gaussianities are universal within the phenomenological holographic cosmology.
%by considering the Feynman diagrams with insertions of the quartic interaction vertex.
}
\end{minipage}
\end{center}

\vspace{.5cm}

\setcounter{page}{0}
\setcounter{tocdepth}{2}

\newpage

\tableofcontents
\renewcommand{\thefootnote}{\arabic{footnote}}
\setcounter{footnote}{0}

\linespread{1.1}
\parskip 4pt

%{}~
%{}~

%---------------------------------------------------------
%%%%%%%%%%%%%%%%%%%%%%%%%%%%%%%%%%%%%%%%%%%%%
%%%%%%%%%%%%%%%%%%%%%%%%%%%%%%%%%%%%%%%%%%%%%
\section{Introduction}
\label{sec:intro}
%%%%%%%%%%%%%%%%%%%%%%%%%%%%%%%%%%%%%%%%%%%%%
%%%%%%%%%%%%%%%%%%%%%%%%%%%%%%%%%%%%%%%%%%%%%

The standard model of cosmology ($\Lambda$CDM plus inflation) is currently one of the most accurate descriptions 
of the physics of the early Universe 
\cite{Brout:1977ix,Starobinsky1979ty,Starobinsky1980te,Sato:1980yn, Guth:1980zm,Linde:1981mu,Hawking1982cz,Guth1982ec,Albrecht:1982wi}. 
In particular, inflation was proposed to solve the problems with the original big bang cosmology, such as the flatness, 
horizon, entropy, and relic (monopole) problems (see \cite{Nastase:2019mhe} for a review). Besides that, 
it also provides a quantum dynamical explanation for the fluctuations in the temperature of the cosmic 
microwave background radiation (CMBR) and fits the current experimental data \cite{Planck:2018nkj}. 

Despite its successes, there are reasons to think that inflation alone may not be sufficient to explain all the phenomena 
in the early Universe and its evolution. For instance, inflation is a semiclassical theory, meaning that although the 
inflaton field is quantized, the background spacetime remains classical. On the other hand, quantum effects in 
gravity become important at small scales, of the order of the Planck length.  This suggests that we should 
view inflation as an effective low-energy description of some more complete theory of quantum gravity. 
Furthermore, there are theoretical aspects of inflation that remain unsatisfactory, such as the trans-Planckian problem 
\cite{Brandenberger:2012aj, Starobinsky:2001kn} (see also \cite{Turok:2002yq,Brandenberger:1999sw,Dvali:2020cgt} 
for a critical review of inflation and its problems; for the problems with the existence of de Sitter in quantum gravity see 
the conjecture \cite{Obied:2018sgi}, and for trans-Planckian issues related to the conjectures see \cite{Bedroya_2020,Bedroya:2019tba,Brahma:2019vpl,Bernardo:2019bbi}).

On the other hand, there are several attempts to apply the holographic dictionary of AdS/CFT, or more generally 
gauge/gravity dualities, to cosmology. The idea that quantum gravity is holographic was initially developed in 
\cite{tHooft:1993dmi, Susskind:1994vu}, but became concrete in the AdS/CFT correspondence \cite{Maldacena:1997re}
(see the books  \cite{Nastase:2015wjb,Ammon:2015wua} for more information). The applications of holography to cosmology 
were started by \cite{Witten:2001kn, Strominger:2001pn, Strominger:2001gp, Maldacena:2002vr}, and the fact that the 
usual weakly coupled inflation can be described through holography via a strongly coupled QFT was developed, for 
instance, in \cite{Maldacena:2011nz, Hartle:2012qb, Hartle:2012tv,Schalm:2012pi, Bzowski:2012ih, Mata:2012bx, Garriga:2013rpa, McFadden:2013ria, Ghosh:2014kba, Garriga:2014ema, Kundu:2014gxa, Garriga:2014fda, McFadden:2014nta, Arkani-Hamed:2015bza, Kundu:2015xta, Hertog:2015nia,Garriga:2015tea,  Garriga:2016poh,Hawking:2017wrd,Arkani-Hamed:2018kmz}.
Holography in cosmology, however, sometimes meant different things, for instance in 
\cite{Strominger:2001pn, Strominger:2001gp, Das:2006wg,Larsen:2002et, Anninos:2011ui}. 
In the ones that hope to replace inflation by a strongly coupled gravity phase, 
the objective is to describe cosmological observables in terms of quantum field theory variables 
in one dimension less, without gravity. But our Universe is not AdS or even dS, and we do 
not have a formal top-down construction of a "dS/CFT" correspondence from string theory or other theories of quantum gravity.

The alternative is to use phenomenological models, notably the one proposed by McFadden and Skenderis 
\cite{McFadden:2009fg,McFadden:2010na}, which we will refer to just as  \emph{holographic cosmology}. 
These models and their implications were developed in \cite{McFadden:2009fg, McFadden:2010na, McFadden:2010vh, McFadden:2011kk, Bzowski:2011ab, Coriano:2012hd, Kawai:2014vxa,McFadden:2010jw}, using methods from 
\cite{Skenderis:2002wp,Papadimitriou:2004ap, Papadimitriou:2004rz, Maldacena:2002vr}.
One replaces inflation with a phase of strongly-coupled gravity, dual to a weakly-coupled (1+2)-dimensional field theory 
with a generalized conformal structure. The phenomenological action for this model includes a gauge field, scalars, and 
fermions, all transforming in the adjoint representation of $SU(N)$, in the large $N$ limit.
Holographic cosmology can solve all the pre-inflationary problems with Big Bang cosmology 
\cite{Nastase:2019rsn,Nastase:2020uon}, 
provides an accurate fit to the CMBR data using the same 
number of free parameters as inflation 
\cite{Easther:2011wh,Afshordi:2016dvb,Afshordi:2017ihr}, and it 
is equivalent to the 'Wavefunction of the Universe' approach proposed by Maldacena to describe de-Sitter inflation 
\cite{Maldacena:2002vr}.

The solution to the monopole (and to all pre-inflationary problems) in holographic cosmology was first presented in 
\cite{Nastase:2019rsn}, and  a detailed discussion can be found in \cite{Nastase:2020uon,Nastase:2020lvn}. 
The general strategy involves selecting a particular toy model, within the general class of holographic cosmology models,
with global $SO(3)$ symmetry to match the $SO(3)$ gauge symmetry responsible for the production of monopoles in cosmology 
through the Kibble mechanism. Then one calculates the 2-point function $\langle j_\mu(p) j_\nu(-p) \rangle$ of the $SO(3)$ 
Noether current to extract its anomalous dimension and finds that the electric current is a marginally irrelevant operator. Due to the 
electric-magnetic duality, this implies that the magnetic current (which couples to the monopole field $\tilde{A}_\mu$ in gauge/gravity duality) is a marginally relevant operator. In holographic cosmology, the cosmological time evolution is mapped to the 
inverse of the RG flow in the dual field theory (from the IR to the UV). This explains the dilution of monopoles at late times
(the monopole problem), 
similarly to the solution to the flatness problem in holographic cosmology, where one shows that the energy-momentum tensor 
$T_{\mu \nu}$ (which couples with the deviation from the flat metric $h_{\mu \nu}$) is a marginally relevant operator.

In this paper, we further explore the properties of the current operator in holographic cosmology by computing its
3-point function, $\langle j_{\mu}^{a}(p_{1})j_{\nu}^{b}(p_{2})j_{\rho}^{c}(p_{3})\rangle$, in momentum space, at one-loop. 
The 3-point function for the energy-momentum tensor $\langle T_{\mu \nu}T_{\rho \sigma}T_{\lambda \xi} \rangle $ in 
momentum space already exists in the literature and gives information about non-Gaussianities in 
holographic cosmology  \cite{Bzowski:2011ab}. Similarly, the 3-point function of currents will give non-Gaussianities 
for the monopole distribution via the same holographic cosmology, though one can also consider it as a new calculation 
in 3-dimensional quantum field theory. 

In addition to its application to holographic cosmology, the techniques we use to solve tensorial Feynman integrals are 
interesting on their own. We check that the final result satisfies the transverse Ward identity at 1-loop and it is IR and UV finite. 
Two particular cases are also considered in this respect: with one and two external light-like momenta. 
We regularize the IR divergences in these cases using mass regularization, and check the Ward identities. 
For the application to cosmology, we consider the relevant case of general (non-lightlike) momenta, which can be Wick 
rotated to the Euclidean case, and specialize for the $k_1\ll k_2,k_3$ case.
In the end, we also comment on the two-loop result, to be calculated explicitly in the future, but we already find that it is 
independent on the form of the quartic scalar potential, so is universal within the phenomenological 
models, for such theories that admit vortex solutions. 

This paper is organized as follows. We begin by reviewing the holographic cosmology approach in section 2, with 
particular attention to the solution to the monopole problem considering the same three-dimensional toy model used in 
the rest of this paper. In section 3, we compute the 3-point function for the current operator first for two particular cases: 
with one and two light-like external momenta, with details of the calculation given in the appendices. Then we compute 
the final expression for the 3-point function, without any assumption about the external momenta, as is most relevant for 
cosmology. We discuss the (in)dependence of the two-loop result on the form of the potential in section 4. 
Finally, we conclude in section 5 and provide prospects for future work.

\section{Review: holographic cosmology, toy models, and the solution to the monopole problem} 
\label{section_review_holographic_cosmology}

We review the main results of holographic cosmology in this section --- one can find the details in 
\cite{McFadden:2009fg,McFadden:2010na,Easther:2011wh,Afshordi:2016dvb, Nastase:2019rsn, Nastase:2020uon, Nastase:2020lvn, Bzowski:2011ab}. 

The starting point is the one-to-one correspondence between $d=1+3$ FLRW cosmologies and $d=4$ 
Euclidean domain-wall spacetime, called domain-wall/cosmology correspondence. We can write the metric for both systems as
\begin{equation}
    ds^2  = \eta dz^2 + a(z) d\Vec{x}^2,
\end{equation}
where, for $\eta = +1$ and $\eta = -1$ we have the domain-wall and FLRW solutions, respectively. 
For the first, $z$ represents the holographic radial coordinate, while for the latter, $z$ is a time parameter in cosmology. 
The connection between these two solutions at metric level is the analytical continuation
\begin{equation}
    t \rightarrow it.
\end{equation}

For a single scalar field $\phi(z)$ minimally coupled to gravity, we write the action
\begin{equation}
    S = \frac{\eta}{2 \kappa^2} \int d^4x \sqrt{-g} \left[ -R + \partial_\mu \phi \partial^\mu \phi  + 2\kappa^2 V(\phi) \right],
\end{equation}
where $\kappa$ is the Newton's constant, $R$ is the Ricci scalar, and $V(\phi)$ is the scalar potential. 
The equations of motion reveal that for every FLRW solution with potential $+V(\phi)$, we have a 
domain-wall solution with potential $-V(\phi)$. This is true also at the perturbative level, once we 
perform the analitical continuation in momentum space,
\begin{equation}
    \bar q = -iq.
\end{equation}

From now on, the quantities with a bar are the domain-wall variables. The choice of $\eta=\pm 1$ is equivalent to 
the analytical continuation of Newton's constant, $\bar \kappa^2 = -\kappa ^2$. In the field theory side of the 
holographic duality, this corresponds to $\bar N^2 = -N^2$, with $\bar N$ ($N$) being the rank of the gauge 
group of the field theory dual to the domain-wall (cosmology), since $\bar \kappa^2 \sim (\bar N)^{-2}$. 
Since $z$ is a holographic coordinate, under gauge/gravity duality, time evolution in cosmology is mapped 
to the (inverse) RG flow of the field theory dual to the domain-wall.

We can consider the holographic description of domain-walls for two cases: asymptotically AdS domain-wall and 
power-law domain-walls, corresponding to $a(z) \sim e^z$ and $a(z)\sim z^n$, for $z\rightarrow \infty$, respectively. 
The goal is to compute observables when gravity is strongly coupled and cannot be geometrically described by 
general relativity. This corresponds to study the field theory dual to the domain-wall in the weakly coupled regime, 
making the computations, and then going back to Lorentzian signature to obtain the results dual to cosmology. 
In particular, one can calculate the scalar and tensor power spectrum of primordial fluctuations that leave a 
measurable imprint in the CMBR. Under gauge/gravity duality, the metric in the bulk (gravity side) couples 
to the energy-momentum tensor in the boundary theory (QFT side). Explicitly, this allows us to write the 
scalar and tensor power spectrum in terms of the 2-point function of the energy-momentum tensor of the dual field theory,
\begin{equation}
 \Delta_S^2(q) = -\frac{q^3}{16 \pi^2 \operatorname{Im} B(-i q)}, \hspace*{2cm}
 \Delta_T^2(q)= -\frac{2 q^3}{\pi^2 \operatorname{Im} A(-i q)}, \label{scalar_tensor_powerspectrum_phenomenological_approach}
\end{equation}
where $A(-iq)=A(\bar{q})$ and $B(-iq)=B(\bar q)$ are the coefficients of the decomposition of the energy-momentum tensor into Lorentz structures
    \begin{equation}
        \left\langle T_{i j}(\bar{q}) T_{k l}(-\bar{q})\right\rangle=A(\bar{q}) \Pi_{i j k l}+B(\bar{q}) \pi_{i j} \pi_{k l},
    \end{equation}
with 
\begin{equation}
    \Pi_{i j k l}=\pi_{i(k} \pi_{l) j}-\frac{1}{2} \pi_{i j} \pi_{k l}, \quad \pi_{i j}=\delta_{i j}-\frac{\bar{q}_i \bar{q}_j}{\bar{q}^2}.
\end{equation}

The last step is to write the action for the dual field theory. Since we do not have a top-down construction providing 
the action for the field theory dual to cosmology as for the case of AdS/CFT, we adopt the phenomenological approach, 
writing the most general three-dimensional field theory with generalized conformal structure and gauge group 
$SU(\bar N)$, in the large $\bar N$ limit,
\begin{eqnarray}
    S_{QFT} &=& \frac{1}{g_{Y M}^2} \int d^3 x \operatorname{Tr}\left[\frac{1}{2} F_{i j} F^{i j}
   +\delta_{M_1 M_2} D_i \Phi^{M_1} D^i \Phi^{M_2}+2 \delta_{L_1 L_2} \bar{\psi}^{L_1} \gamma^i D_i \psi^{L_2}\right. \nonumber \\
    && \left.+\sqrt{2} \mu_{M L_1 L_2} \Phi^M \bar{\psi}^{L_1} \psi^{L_2}+\frac{1}{6} \lambda_{M_1 \ldots M_4} \Phi^{M_1} 
    \ldots \Phi^{M_4}\right]. \label{phenomenological_models}
\end{eqnarray}

Pertubatively, the coefficients $A$ and $B$ are given by 
\begin{eqnarray}
  A(q) &=& q^{3} N^{2} f_{T 0}\left[1-f_{T 1} g_{\text {eff }}^{2} \log \left(g_{\text {eff }}^{2}\right)
  +f_{T 2} g_{\text {eff }}^{2}+\mathcal{O}\left(g_{\text {eff }}^{4}\right)\right], \\
  B(q) &= & \frac{1}{4}q^{3} N^{2} f_{0}\left[1-f_{1} g_{\text {eff }}^{2} \log \left(g_{\text {eff }}^{2}\right)
  +f_{2} g_{\text {eff }}^{2}+\mathcal{O}\left(g_{\text {eff }}^{4}\right)\right],
\end{eqnarray}
where the one-loop contribution from  $\langle T_{ij}T_{kl} \rangle$ gives $f_0$ and $f_{T0}$, 
and the two-loop contribution gives $f_1$, $f_2$, $f_{T1}$, and $f_{T2}$. The effective coupling is given by
\begin{equation}
g_{eff}^2 = \frac{g^2 N}{q}.
\end{equation}
 
By defining new variables 
\begin{equation}
    gq_\star = f_1 g_{\text{YM}}^2 N, \quad \quad \ln \frac{1}{\beta} = \frac{f_2}{f_1} + \ln |f_1|,
\end{equation}
we obtain the following parameterization for the scalar power spectrum in cosmology:
\begin{equation}
  \Delta_{S}^{2}(q)=\frac{\left(\Delta_{S}^{(0)}\right)^{2}}{1+\left(g\frac{q_{\star}}{q}\right)\ln\left|\frac{q}{\beta gq_{*}}\right|+
  \mathcal{O}\left(g\frac{q_{*}}{q}\right)^{2}}, \hspace{2cm} (\Delta_S^{(0)})^2 
  = \frac{1}{4\pi ^2 N^2 f_0},\label{power_spectrum_holographic_cosmology}
\end{equation}
where $q_\star$ is a pivot scale identified with the renormalization scale of the dual quantum field theory. 
This result has a different functional form than the power-law form obtained in inflation, though the experimental 
data is consistent with both, by expanding the power law $q^n$ for $n\ll 1$ as $1+n\log q$. Then, as discussed in 
\cite{Afshordi:2016dvb}, holographic cosmology is found to be as good as inflation in terms of the
fit to the CMBR data.

\subsection{Solution to the monopole problem in holographic cosmology}

We still have to discuss how holographic cosmology also solves the pre-inflationary problems. A complete discussion is 
given in \cite{Nastase:2019rsn,Nastase:2020uon}; in this section we will just review the solution to the monopole problem. 

Ideally, we should consider 't Hooft–Polyakov monopoles in cosmology. These are produced in GUT phase transitions 
in the early Universe, when a $SO(3)$ gauge symmetry inside the full gauge group
is spontaneously broken, and the vaccum state is invariant 
under  $U(1)$ symmetry \cite{tHooft:1974kcl,Kibble:1976sj}. Under gauge/gravity duality, the $SO(3)$ gauge symmetry on the 
gravity side is mapped to a $SO(3)$ global symmetry on the QFT side.  The monopole configuration $ A_\mu$ 
is dual to a magnetic current $\tilde j^\mu$ in the QFT side, which is a vortex current in three dimensions. 
The dilution of monopoles at late times occurs if the current $\tilde j^\mu$ is a relevant operator, meaning 
the corresponding monopole field configuration $A_\mu$ in cosmology goes to zero in the UV (late-times in cosmology). 
A summary of this dictionary is presented in  \autoref{holographic_dictionary}.
\begin{table}[h!]
    \centering
    \begin{tabular}{c c c}
  \hline 
  Bulk (cosmology) &  & Boundary (QFT)\tabularnewline
  \hline 
   $SO(3)$ gauge symmetry & $\longleftrightarrow$  & $SO(3)$ global symmetry \tabularnewline

  Monopole field $A^{\mu}$ & $\longleftrightarrow$ & Global current $\widetilde{j}_{\mu}$\tabularnewline

  Magnetic monopole & $\longleftrightarrow$ & Vortex solution \tabularnewline

  Dilution of monopoles  & $\longleftrightarrow$ & $\widetilde{j}_{\mu}$ is a relevant operator\tabularnewline
  \hline

  \end{tabular} 
    \caption{A summary of the holographic dictionary to solve the monopole problem.}
    \label{holographic_dictionary}
\end{table}

To check whether $\tilde j^\mu$ is a relevant operator we have to compute the 2-point function 
$\langle \tilde j_\mu(p) \tilde j_\nu(-p) \rangle$ and extract the anomalous dimension of the magnetic current operator. 
Instead of working with the vortex current, however, we can calculate the 2-point function of the $SO(3)$ 
Noether current $j_\mu$, and by the electric-magnetic duality \cite{Witten:2003ya,Murugan:2014sfa}, 
relate the anomalous dimension of both operators:
\begin{equation}
    \delta(j) = -\delta(\tilde j). \label{electric-magnetic_duality}
\end{equation}

More precisely, the two-point function in a conformal field theory (and therefore also in a theory with generalized 
conformal invariance, since the functional form only depends on this invariance) was found to be, by 
Witten in the Abelian case \cite{Witten:2003ya} and in the non-Abelian case a generalization of that in \cite{Herzog:2007ij},
\be
\langle j_\mu^a(p)j_\nu^b(-p)\rangle =\frac{t}{2\pi}p\delta^{ab}\left(\delta_{\mu\nu}-\frac{p_\mu p_\nu}{p^2}\right)
+\frac{w}{2\pi}\delta^{ab}\epsilon_{\mu\nu\rho}p^\rho\;,
\ee
where $t$ and $w$ are scalar functions of the couplings and parameters. 

Then the action of the $Sl(2,\mathbb{Z})$ duality element $S$ (S-duality) on the 2-point function is the usual one,
acting only on $t$ and $w$ as
\be
t\rightarrow \frac{t}{t^2+w^2}\;,\;\;\;
w\rightarrow \frac{w}{t^2+w^2}\;,
\ee
so that in the parity-invariant case with $w=0$, the dual 2-point function of magnetic currents is found by just 
inverting $t\rightarrow 1/t$, as
\be
\langle \tilde j_\mu^a(p)\tilde j_\nu^b(-p)\rangle =\frac{1}{2\pi t}p\delta^{ab}\left(\delta_{\mu\nu}-\frac{p_\mu p_\nu}{p^2}\right).
\ee

Then, if at two-loop we have $t\simeq t_0 p^{\delta}\simeq t_0(1+\delta \ln p)$, going to the magnetic currents 
amounts to just $\delta\rightarrow \tilde \delta=-\delta$. If the Noether current is irrelevant, 
then the vortex current is relevant, diluting monopoles as time goes forward 
in cosmology (inverse RG flow in field theory), and thus solving the monopole 
problem in holographic cosmology. 

The phenomenological field theory action for holographic cosmology doesn't have generically any global symmetry, 
so we must consider special cases to do calculations. 
The toy model proposed in \cite{Nastase:2019rsn,Nastase:2020uon} to perform the calculation 
of the 2-point function $\langle j_\mu(p)j_\nu(-p)  \rangle$ has action
\begin{equation}
  S = \int d^{3}x\text{Tr}\left[-\frac{1}{2}F_{\mu\nu}F^{\mu\nu}-2\sum_{i=1,2}|D_{\mu}\vec{\Phi}_i|^{2}-
  4\lambda\left|\vec{\Phi}_{1}\times\vec{\Phi}_{2}\right|^{2}\right] ,\label{toy_model_horatiu}
\end{equation}
where $\vec{\Phi}_i$ is an $SO(3)$ vector, in the following written explicitly as $\phi_i^a$, $a\in SO(3)$. 
The potential allows the existence of vortex solutions of the type
\begin{eqnarray}
  \phi_{1}^{a}&=&\phi_{1}(r)f^{a}e^{i\alpha}, \label{vortex_anzatz1}\\
  \phi_{2}^{a} &=&\phi_{2}(r)f^{a}e^{i\alpha},  \label{vortex_anzatz2}
\end{eqnarray}
where $f^a$ is a vector in the internal space of $SO(3)$. The Noether current is given by 
\begin{equation}
    j_{\mu}^{a} = i\epsilon^{abc}\sum_{i=1,2}\left[\left(\partial_{\mu}\phi_{i}^{*b}\right)\phi_{i}^{c}
    +\left(\partial_{\mu}\phi_{i}^{b}\right)\phi_{i}^{*c}+igA_{\mu}\left(\phi_{i}^{*b}\phi_{i}^{c}-\phi_{i}^{b}\phi_{i}^{*c}\right)\right],
\end{equation}
from which we derive the following Feynman rules:
\begin{itemize}
    \item[-] for the external current insertion:
\end{itemize}
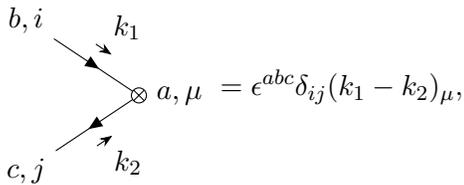
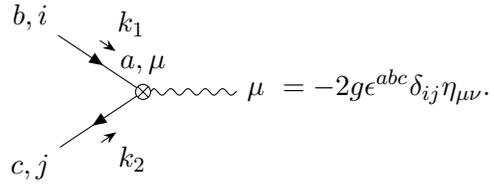
\begin{equation*}
  \begin{tikzpicture}[baseline = (m.base),arrowlabel/.style={
      /tikzfeynman/momentum/.cd, % means that the following keys are read from the /tikzfeynman/momentum family
      arrow shorten=#1,arrow distance=2.5mm
    },
    arrowlabel/.default=0.4]
    \def\leglength{1.5}
    \begin{feynman} [inline =(m.base) ]
      \vertex[crossed dot, label=right:{\(a,\mu \)}] (m) at (0, 0) {};
      \vertex (n) at (-\leglength,1cm) {\(b,i\)};
      \vertex (p) at ( -\leglength,-1cm) {\(c,j\)};
      \diagram*{
        (n) -- [fermion,momentum={[arrowlabel]$k_1$}] (m) -- [fermion,reversed momentum={[arrowlabel]$k_2$}] (p),
        };  
    \end{feynman}
  \end{tikzpicture} = \epsilon^{abc}\delta_{ij} (k_1 - k_2)_\mu, \hspace{1cm}  \begin{tikzpicture}[baseline = (c.base), arrowlabel/.style={
      /tikzfeynman/momentum/.cd, % means that the following keys are read from the /tikzfeynman/momentum family
      arrow shorten=#1,arrow distance=2.5mm
    },
    arrowlabel/.default=0.4]
    \def\leglength{1.5}
    \begin{feynman} [inline =(c.base) ]
      \vertex[crossed dot, label=above:{\(a,\mu\)}] (m) at (0, 0) {};
      \vertex (a) at (-\leglength,1cm) {\(b,i\)};
      \vertex (b) at (-\leglength,-1cm) {\(c,j\)};
      \vertex (c) at (\leglength,0) {\(\mu\)};
  
  \diagram*{
    (a) -- [fermion,momentum={[arrowlabel]$k_1$}] (m) -- [fermion,reversed momentum={[arrowlabel]$k_2$}] (b),
    (c) -- [photon] (m),
  };
  \end{feynman}
  \end{tikzpicture} = -2 g\epsilon^{abc} \delta_{ij} \eta_{\mu\nu} .
  \end{equation*}
\begin{itemize}
    \item[-] for the interactions:
\end{itemize}
\begin{equation*}
  \begin{tikzpicture}[baseline = (a.base),arrowlabel/.style={
      /tikzfeynman/momentum/.cd, % means that the following keys are read from the /tikzfeynman/momentum family
      arrow shorten=#1,arrow distance=2.5mm
    },
    arrowlabel/.default=0.4]
      \def\leglength{2}
      \begin{feynman} [inline =(a.base) ]
          \vertex[dot] (m) at (0, 0) {};
          \vertex (a) at (-\leglength,0) {\(b,i\)};
          \vertex (b) at ( \leglength,0) {\(c,j\)};
          \vertex (c) at (0,- \leglength) {\(\mu\)};
  
  \diagram*{
      (a) -- [fermion,momentum={[arrowlabel]$k_1$}] (m) -- [fermion,reversed momentum={[arrowlabel]$k_2$}] (b),
      (c) -- [photon,momentum={[arrowlabel]$k_3$}] (m),
  };
      \end{feynman}
  \end{tikzpicture} = g\delta^{ab}\delta_{ij}\left(k_{1}-k_{2}\right)_{\mu}\left(2\pi\right)^{3}\delta^{3}(k_{1}+k_{2}+k_{3}),
\end{equation*}
\begin{equation*}
  \begin{tikzpicture}[baseline = (m.base),arrowlabel/.style={
      /tikzfeynman/momentum/.cd, % means that the following keys are read from the /tikzfeynman/momentum family
      arrow shorten=#1,arrow distance=2.5mm
    },
    arrowlabel/.default=0.4]
      \def\leglength{1.5}
      \begin{feynman} [inline =(a.base) ]
          \vertex[dot] (m) at (0, 0) {};
          \vertex (a) at (-\leglength,\leglength) {\(b,i\)};
          \vertex (b) at ( \leglength,\leglength) {\(c,j\)};
          \vertex (c) at (-\leglength,- \leglength) {\(\mu\)};
          \vertex (d) at (\leglength,- \leglength) {\(\nu\)};
  
  \diagram*{
      (a) -- [fermion,momentum'={[arrowlabel]$k_1$}] (m) -- [fermion,reversed momentum={[arrowlabel]$k_2$}] (b),
      (c) -- [photon,momentum'={[arrowlabel]$k_3$}] (m) -- [photon,reversed momentum={[arrowlabel]$k_4$}] (d),
  };
      \end{feynman}
  \end{tikzpicture} = -2g^{2}\delta_{ij}\delta^{ab}\eta_{\mu\nu}\left(2\pi\right)^{3}\delta^{3}(k_{1}+k_{2}+k_{3}+k_{4}),
\end{equation*}
\begin{equation*}
    \hspace{-0.8cm}\begin{tikzpicture}[baseline = (m.base),arrowlabel/.style={
        /tikzfeynman/momentum/.cd, % means that the following keys are read from the /tikzfeynman/momentum family
        arrow shorten=#1,arrow distance=2.5mm
      },
      arrowlabel/.default=0.4]
        \def\leglength{1.4}
        \begin{feynman} [inline =(a.base) ]
            \vertex[dot] (m) at (0, 0) {};
            \vertex (a) at (-\leglength,\leglength) {\(a,1\)};
            \vertex (b) at ( \leglength,\leglength) {\(c,1\)};
            \vertex (c) at (-\leglength,- \leglength) {\(b,2\)};
            \vertex (d) at (\leglength,- \leglength) {\(d,2\)};
    
    \diagram*{
        (a) -- [fermion,momentum'={[arrowlabel]$k_1$}] (m) -- [fermion,reversed momentum={[arrowlabel]$k_3$}] (b),
        (c) -- [fermion,momentum'={[arrowlabel]$k_2$}] (m) -- [fermion,reversed momentum={[arrowlabel]$k_4$}] (d),
    };
        \end{feynman}
    \end{tikzpicture} =  -\lambda\epsilon^{ecd}\epsilon^{eab}(2\pi)^{3}\delta^{3}(k_{1}+k_{2}+k_{3}+k_{4}). \
\end{equation*}
\begin{itemize}
    \item[-] for the propagators:
\end{itemize}
\begin{equation*}
    \begin{tikzpicture}[baseline = (m.base),arrowlabel/.style={
        /tikzfeynman/momentum/.cd, % means that the following keys are read from the /tikzfeynman/momentum family
        arrow shorten=#1,arrow distance=2.5mm
      },
      arrowlabel/.default=0.4]
        \def\leglength{2}
        \begin{feynman} [inline =(a.base) ]
            \vertex (a) at (-\leglength,0) {\(a,i\)};
            \vertex (b) at ( \leglength,0) {\(b,j\)};
    \diagram*{
        (a) -- [fermion,momentum={[arrowlabel]$k$}] (b),
    };
        \end{feynman}
    \end{tikzpicture} = \frac{\delta_{ij}\delta^{ab}}{k^2},
\end{equation*}
\begin{equation*}
    \begin{tikzpicture}[baseline = (m.base),arrowlabel/.style={
        /tikzfeynman/momentum/.cd, % means that the following keys are read from the /tikzfeynman/momentum family
        arrow shorten=#1,arrow distance=2.5mm
      },
      arrowlabel/.default=0.4]
        \def\leglength{2}
        \begin{feynman} [inline =(a.base) ]
            \vertex (a) at (-\leglength,0) {\(\mu\)};
            \vertex (b) at ( \leglength,0) {\(\nu\)};
    \diagram*{
        (a) -- [photon,momentum={[arrowlabel]$k$}] (b),
    };
        \end{feynman}
    \end{tikzpicture} = \frac{\eta_{\mu\nu}}{k^2}.
\end{equation*}

After writing all possible diagrams contributing to the 2-point function at two-loops and performing the calculations, we find 
\begin{equation}
  \langle j^a_\mu(p)  j^a_\nu(-p) \rangle = N^{2}\frac{p}{4}\delta^{ab}\pi_{\mu\nu}\left[1-64\delta^{ab}\frac{g^{2}N}{p}
  \left(2J_{0}-\frac{34}{J_{0}}+\frac{B_{0}^{2}p^{2}}{2}\right)\right],
\end{equation}
where $J_0$ has divergences. Using dimensional regularization with $d=3+\epsilon$, we find
\begin{equation}
  J_{0}\simeq-\frac{2\pi p^{2\epsilon}}{(4\pi)^{3}}\left(\frac{1}{\epsilon}+\text{finite}\right).
\end{equation}

Dropping terms of the type $1/\epsilon$, the renormalized result after expanding around small $\epsilon$ is 
\begin{equation}
\langle j^a_\mu(p)  j^a_\nu(-p) \rangle 
=  N^{2}\frac{p}{4}\delta^{ab}\pi_{\mu\nu}\left(1+\delta^{ab}\frac{16}{\pi^{2}}\frac{g^{2}N}{p}\ln p+\text{finite}\right),
\end{equation}
 which gives the anomalous dimension of the $SO(3)$ Noether current, such that the (two-loop) correction changes $
 p\rightarrow p^{1+\delta}$ in the one-loop result, as
\begin{equation}
  \delta = \frac{8}{\pi^2} \frac{g^2 N}{p}.
\end{equation}

Since $\delta >0$, we conclude that $j_\mu^a$ is an irrelevant operator and, as a consequence of equation 
\eqref{electric-magnetic_duality}, $\tilde j_\mu^a$ is a relevant operator, thus solving the monopole problem. 
As highlighted in \cite{Nastase:2020lvn}, this result is independent of the explicit form of the potential since all 
Feynman diagrams containing the quartic vertex are zero in dimensional regularization. Therefore, the 
solution to the monopole problem within this toy model can be extended to the whole phenomenological 
class of models with only bosons and a potential that allows the existence of vortex solutions.

\section{Calculation of the  3-point function for the current operator in momentum space}

Now we want to go even further and compute the 3-point function of the Noether current, using the same toy model 
\eqref{toy_model_horatiu}. This will be relevant to the non-Gaussianities in the cosmological monopole distribution, 
just as the scalar 3-point function gives non-Gaussianities in the CMBR. 

There is only one diagram contributing to  $\langle j^a_\mu(p_1) j^b_\nu(-p_2) j^c_\rho(-p_3) 
\rangle$ at one loop,
\begin{equation}
    \langle j^a_\mu(p_1) j^b_\nu(-p_2) j^c_\rho(-p_3) \rangle = \begin{tikzpicture}[baseline = (m.base),arrowlabel/.style={
        /tikzfeynman/momentum/.cd, % means that the following keys are read from the /tikzfeynman/momentum family
        arrow shorten=#1,arrow distance=2.5mm
      },
      arrowlabel/.default=0.4]
        \def\leglength{1}
        \begin{feynman} [inline =(a.base) ]
            \vertex[crossed dot, label = left:{$a,\mu$}] (a) at (-\leglength,0) {};
            \vertex[crossed dot, label = right:{$b,\nu$}]  (b) at ( \leglength,\leglength) {};
            \vertex[crossed dot, label = right:{$c,\rho$}]  (c) at (\leglength,- \leglength) {};
    
    \diagram*{
        (a) -- [fermion,momentum ={[arrowlabel]$q+p_2$}] (b) -- [fermion, momentum={[arrowlabel]$q$}] 
        (c) -- [fermion, momentum={[arrowlabel]$q- p_3$}] (a),
        (l1) [momentum ={[arrowlabel]$q+p_2$}] (a),
    };
        \end{feynman}
    \end{tikzpicture},
\end{equation}
where the momentum $p_1$ is flowing in the diagram from the left and the momentum $p_2+p_3$ is flowing out on the 
right-hand side. Conservation of the total momentum implies that
\begin{equation}
    p_{1\mu}= p_{2\mu} + p_{3\mu}.
\end{equation}

Using the Feynman rules, we find the integral expression
\begin{multline}
    \langle j^a_\mu(p_1) j^b_\nu(-p_2) j^c_\rho(-p_3) \rangle = 2N^2\epsilon^{abc} \left\{ 8I_{\mu\nu\rho}
    +4\left[\left(p_{2\mu}-p_{3\mu}\right)I_{\nu\rho}+p_{2\nu}I_{\mu\rho}-p_{3\rho}I_{\mu\nu}\right]\right.  \\
     \left. \left.+2\left[\left(p_{2\mu}-p_{3\mu}\right)\left(p_{2\nu}I_{\rho}-p_{3\rho}I_{\nu}\right)-p_{2\nu}p_{3\rho}I_{\mu}\right]
     + \left( p_{3\mu}p_{2\nu}p_{3\rho}-p_{2\mu}p_{2\nu}p_{3\rho} \right) I_0 \right. \right\},
\end{multline}
where
\begin{align}
     I_0 &=  \int \frac{d^d q}{(2\pi)^d} \frac{1}{q^2 (q+p_2)^2 (q-p_3)^2 }, &  I_\mu &
     =  \int \frac{d^d q}{(2\pi)^d} \frac{q_\mu}{q^2 (q+p_2)^2 (q-p_3)^2 }, \label{set_of_integrals_1}\\
     I_{\mu \nu} &= \int \frac{d^d q}{(2\pi)^d} \frac{q_\mu q_\nu}{q^2 (q+p_2)^2 (q-p_3)^2 }, & I_{\mu \nu \rho} &
     = \int \frac{d^d q}{(2\pi)^d} \frac{q_\mu q_\nu q_\rho}{q^2 (q+p_2)^2 (q-p_3)^2 }.  \label{set_of_integrals_2}
\end{align}

These tensorial integrals can reduced into scalar integrals by using the Feynman parametrization
\begin{equation}
    \frac{1}{ABC} = 2 \int_{0}^{1}dx \int_{0}^{1-x}dy \frac{1}{\left[ Ax+By+C(1-x-y) \right]^3}.
\end{equation}

Choosing $A=q^{2}$, $B=(q+p_{2})^{2}$, and $C=(q-p_{3})^{2}$, we have 
\begin{eqnarray}
    I_{0} &=& \int\frac{d^{d}q}{(2\pi)^{d}}\frac{1}{q^{2}(q+p_{2})^{2}(q-p_{3})^{2}} \nonumber \\
    &=& 2\int_{0}^{1}dx\int_{0}^{1-x}dy\int\frac{d^{d}q}{(2\pi)^{d}}\frac{1}{\left[q^{2}+2q\cdot\left[2yp_{2}
    -(1-x-y)p_{3}\right]+yp_{2}^{2}+(1-x-y)p_{3}^{2}\right]^{3}}, \nonumber \\
\end{eqnarray}
and changing the integration variable as
\begin{equation}
    q_\mu \rightarrow q_\mu  -yp_{2\mu}+(1-x-y)p_{3\mu},
\end{equation}
we get 
\begin{equation}
    I_{0}=2\int_{0}^{1}dx\int_{0}^{1-x}dy\int\frac{d^{d}{q}}{(2\pi)^{d}}\frac{1}{({q}^{2}+\Delta^{2})^{3}},
\end{equation}
where 
\begin{equation}
    \Delta^2 = xyp_2^2 + (1-x-y)(xp_3^2 + yp_1^2). \label{denominator_calculations}
\end{equation}

We can use the general result in dimension regularization for integrals of this type,
\begin{equation}
    \int\frac{d^{d}q}{(2\pi)^{d}}\frac{1}{(q^{2}+m^{2})^{n}}=\frac{\Gamma(n-d/2)}{(4\pi)^{d/2}\Gamma(n)}(\Delta^{2})^{d/2-n},
\end{equation}
such that, after defining the notation 
\begin{equation}
    I^{(a,b,c)}=\int_{0}^{1}dx\int_{0}^{1-x}dy\frac{x^{a}y^{b}}{\left(\Delta^{2}\right)^{c}}, \label{notation_integrals_feynman_parameters}
\end{equation}
and setting $d=3$ (there are no divergences for $d=3$ in the Gamma functions), we find
\begin{equation}
    I_0 = \frac{1}{16\pi} I^{(0,0,3/2)}.
\end{equation}

We can do the same for the remaining integrals, after the decomposition in terms of Lorentz tensors (see 
\autoref{appendix_integrals_in_momentum_space} for details)
\begin{align}
    I_\mu &= \frac{1}{16\pi}f_{\mu}, & I_{\mu \nu} &= \frac{1}{16\pi}\left(f_{\mu\nu}+\delta_{\mu\nu} I^{(0,0,1/2)}\right),
\end{align}
\begin{eqnarray}
    I_{\mu \nu \rho} &=& \frac{1}{16\pi} \left( \delta_{\mu\nu} f_{\rho}^{\prime} + \delta_{\mu\rho} f_{\nu}^{\prime} + \delta_{\nu\rho} f_{\mu}^{\prime} + f_{\mu\nu\rho} \right),\nonumber \\
\end{eqnarray}
where
\begin{align}
f_{\mu}(p_{1},p_{2},p_{3}) &= p_{3\mu}\left( I^{(0,0,3/2)} - I^{(1,0,3/2)}  \right)-p_{1\mu} I^{(0,1,3/2)}, \nonumber \\
f_{\mu\nu}(p_{1},p_{2},p_{3})& = \left(I^{(0,0,3/2)} - 2 I^{(1,0,3/2)} +  I^{(2,0,3/2)} \right)p_{3\mu}p_{3\nu} \nonumber \\
& \hspace{-0.5cm}  +\left(I^{(1,1,3/2)}-I^{(0,1,3/2)}\right)	 \left(p_{3\mu}p_{1\nu}+p_{1\mu}p_{3\nu}\right) 
+I^{(0,2,3/2)}p_{1\mu}p_{1\nu},\nonumber \\
 f_{\mu}^{\prime}(p_{1},p_{2},p_{3}) &=  p_{3\mu}\left( I^{(0,0,1/2)} - I^{(1,0,1/2)}  \right) - p_{1\mu} I^{(0,1,1/2)}, \nonumber \\
 f_{\mu\nu\rho} (p_{1},p_{2},p_{3}) &= \left(I^{(0,0,3/2)}-3I^{(1,0,3/2)}+3I^{(2,0,3/2)}-I^{(3,0,3/2)}\right)
 p_{3\mu}p_{3\nu}p_{3\rho} \nonumber \\
&  \hspace{-0.5cm}-\left(I^{(0,1,3/2)}+I^{(2,1,3/2)}-2I^{(1,1,3/2)}\right)(p_{3\mu}p_{3\nu}p_{1\rho}
+p_{3\mu}p_{1\nu}p_{3\rho}+p_{1\mu}p_{3\nu}p_{3\rho}) \nonumber \\
&  \hspace{-0.5cm} +\left(I^{(0,2,3/2)}-I^{(1,2,3/2)}\right)(p_{3\mu}p_{1\nu}p_{1\rho}+p_{1\mu}p_{3\nu}p_{1\rho}
+p_{1\mu}p_{1\nu}p_{3\rho}) \nonumber  \\
&  \hspace{-0.5cm} - I^{(0,3,3/2)}p_{1\mu}p_{1\nu}p_{1\rho}.     \label{intermetiade_integrals}
\end{align}

\begin{comment}
\begin{align}
    f_{\mu}(p_{1},p_{2},p_{3}) = p_{3\mu}\left( I^{(0,0,3/2)} - I^{(1,0,3/2)}  \right)-p_{1\mu} I^{(0,1,3/2)}, 
\end{align}
\begin{align}
    & f_{\mu\nu}(p_{1},p_{2},p_{3}) = \left(f_{0} - 2 I^{(1,0,3/2)} +  I^{(2,0,3/2)} \right)p_{3\mu}p_{3\nu} \nonumber \\
    & \quad +\left(I^{(1,1,3/2)}-I^{(0,1,3/2)}\right)\left(p_{3\mu}p_{1\nu}+p_{1\mu}p_{3\nu}\right) +I^{(0,2,3/2)}p_{1\mu}p_{1\nu},
\end{align}
\begin{align}
    & f_{\mu}^{\prime}(p_{1},p_{2},p_{3}) =  p_{3\mu}\left( I^{(0,0,1/2)} - I^{(1,0,1/2)}  \right) - p_{1\mu} I^{(0,1,1/2)},
\end{align}
\begin{align}
    & f_{\mu\nu\rho} (p_{1},p_{2},p_{3}) = \left(f_{0}-3I^{(1,0,3/2)}+3I^{(2,0,3/2)}
    -I^{(3,0,3/2)}\right)p_{3\mu}p_{3\nu}p_{3\rho} \nonumber \\
    & -\left(I^{(0,1,3/2)}+I^{(2,1,3/2)}-2I^{(1,1,3/2)}\right)(p_{3\mu}p_{3\nu}p_{1\rho}
    +p_{3\mu}p_{1\nu}p_{3\rho}+p_{1\mu}p_{3\nu}p_{3\rho}) \nonumber \\
    & \quad +\left(I^{(0,2,3/2)}-I^{(1,2,3/2)}\right)(p_{3\mu}p_{1\nu}p_{1\rho}
    +p_{1\mu}p_{3\nu}p_{1\rho}+p_{1\mu}p_{1\nu}p_{3\rho}) \nonumber  \\
    & \quad \quad - I^{(0,3,3/2)}p_{1\mu}p_{1\nu}p_{1\rho}. 
\end{align}
\end{comment}

It is convenient to write these integrals in terms of the absolute value of the external four-momenta, $p_1, p_2, p_3$. 
In the end, we will 
use momentum conservation to eliminate the dependence on $p_{1\nu}$ and $p_{1\rho}$, keeping explicitly only 
$p_{1\mu}$, only in order to check the Ward identities (otherwise we could eliminate it in terms of $p_{2\mu}, p_{3,\mu}$
as well, to get an unambiguous result. The problem of computing the $\langle j^a_\mu(p_1) j^b_\nu(-p_2) j^c_\rho(-p_3) 
\rangle$ correlation function reduces to the computation of thirteen scalar integrals in the remaining Feynman parameters. 
\begin{comment}
\begin{align}
    I^{(0,0,3/2)} &= \int_0^1 dx \int_0^{1-x} dy \frac{1}{(\Delta^2)^{3/2}}, & I^{(1,0,3/2)} &
    = \int_0^1 dx \int_0^{1-x} dy \frac{x}{(\Delta^2)^{3/2}}, \nonumber\\
    I^{(2,0,3/2)} &= \int_0^1 dx \int_0^{1-x} dy \frac{x^2}{(\Delta^2)^{3/2}}, & I^{(3,0,3/2)} &
   = \int_0^1 dx \int_0^{1-x} dy \frac{x^3}{(\Delta^2)^{3/2}},\nonumber\\
    I^{(1,1,3/2)} &= \int_0^1 dx \int_0^{1-x} dy \frac{xy}{(\Delta^2)^{3/2}} & I^{(2,1,3/2)} &
    = \int_0^1 dx \int_0^{1-x} dy \frac{x^2y}{(\Delta^2)^{3/2}} \nonumber \\
    I^{(0,0,1/2)} &= \int_0^1 dx \int_0^{1-x} dy \frac{1}{(\Delta^2)^{1/2}},&  I^{(1,0,1/2)} &
    = \int_0^1 dx \int_0^{1-x} dy \frac{x}{(\Delta^2)^{1/2}},\nonumber\\
    \label{all_integrals_feynman_parameters}
\end{align}
%Fullset of integrals in the comment below%
\end{comment}
\begin{align}
    I^{(0,0,3/2)} &= \int_0^1 dx \int_0^{1-x} dy \frac{1}{(\Delta^2)^{3/2}}, & I^{(1,0,3/2)} &
    = \int_0^1 dx \int_0^{1-x} dy \frac{x}{(\Delta^2)^{3/2}}, \nonumber\\
    I^{(2,0,3/2)} &= \int_0^1 dx \int_0^{1-x} dy \frac{x^2}{(\Delta^2)^{3/2}}, & I^{(3,0,3/2)} &
    = \int_0^1 dx \int_0^{1-x} dy \frac{x^3}{(\Delta^2)^{3/2}},\nonumber\\
    I^{(0,1,3/2)} &= \int_0^1 dx \int_0^{1-x} dy \frac{y}{(\Delta^2)^{3/2}}, & I^{(0,2,3/2)} &
    = \int_0^1 dx \int_0^{1-x} dy \frac{y^2}{(\Delta^2)^{3/2}},\nonumber\\
    I^{(0,3,3/2)} &= \int_0^1 dx \int_0^{1-x} dy \frac{y^3}{(\Delta^2)^{3/2}},&
    I^{(1,1,3/2)} &= \int_0^1 dx \int_0^{1-x} dy \frac{xy}{(\Delta^2)^{3/2}}\nonumber \\
    I^{(1,2,3/2)} &= \int_0^1 dx \int_0^{1-x} dy \frac{xy^2}{(\Delta^2)^{3/2}}, & 
    I^{(2,1,3/2)} &= \int_0^1 dx \int_0^{1-x} dy \frac{x^2y}{(\Delta^2)^{3/2}}\nonumber \\
    I^{(0,0,1/2)} &= \int_0^1 dx \int_0^{1-x} dy \frac{1}{(\Delta^2)^{1/2}},&  I^{(1,0,1/2)} &
    = \int_0^1 dx \int_0^{1-x} dy \frac{x}{(\Delta^2)^{1/2}},\nonumber\\
    I^{(0,1,1/2)} &= \int_0^1 dx \int_0^{1-x} dy \frac{y}{(\Delta^2)^{1/2}}. 
    \label{all_integrals_feynman_parameters}
\end{align}

This set of integrals is not completely independent, since 
\begin{equation}
    I^{(a,b,c)}(p_1,p_2,p_3)= I^{(b,a,c)}(p_3,p_2,p_1)
\end{equation}
as demonstrated in \autoref{appendix_integral_Feynman_parameters}.

As we will see, the 1-loop result is finite in the IR and UV, meaning that the integrals \eqref{all_integrals_feynman_parameters} 
are finite in the domain $0\leq x \leq 1$ and $0\leq y \leq 1-x$. 

We will first solve these integrals  for two particular cases: 
(i) $p_1^2 = p_3^2 = 0$ and (ii) $p_1^2 = 0$. This choice of external light-like momenta will induce infrared divergences, 
which we can regularize by adding a mass term in the Lagrangian and then taking the massless limit at the end of the 
calculation. This amounts to adding a mass in the scalar propagator,
\begin{equation}
    \frac{1}{k^2 } \rightarrow \frac{1}{k^2 + m^2}.
\end{equation}

Of course, while the UV divergences belong to the theory, the IR divergences are not physical, and arise due to not 
calculating something experimentally measurable; once we do so, by summing up with other $n$-point function(s), 
the measurable result should always be IR divergence free. But the IR finiteness of the $n$-point functions in the 
3-dimensional QFTs relevant for holographic cosmology was only conjectured until recently, when it was proven in 
\cite{Lee:2022mlh,Rocha:2022sfj}.

Note that these special cases with lightlike external momenta are only relevant in the Minkowski signature, 
for the case of quantum field theory applications, whereas for the 
application to holographic cosmology we need to Wick rotate to Euclidean space, since the 3-dimensional momenta become spatial 
momenta in cosmology. In that case, $p_i^2>0$ for the external momenta (in practice, what is measured is the correlation 
function in 3-dimensional $x$-space, and one can do a Fourier transform to $p$ space). But we will use $p_i^2=0$ as 
a testing ground for our formulas, as the formulas are simpler in these special cases.

As a check, we will also show that the full 3-point function satisfies the transverse Ward identities even in these cases, with a small 
caveat for the first case, which will be explained. The details are provided in  \autoref{appendix_integral_Feynman_parameters}.

After we deal with these two cases, we present the full computation of the 3-point function, without any assumption 
about the external momenta, and we show that the result is finite and satisfies the Ward identities as well.

Finally, we will apply our full result to holographic cosmology.

\subsection{Two external light-like momenta ("on-shell", $p_1^2 = p_3^2 = 0$)}

In the present case, the denominator \eqref{denominator_calculations} (with a mass term) becomes
\begin{equation}
    \Delta^{2}_m =  xyp_{2}^{2} + m^2,
\end{equation}
and the integrals will depend only on $p_2^2$.
After taking the massless limit, we find the following results for the integrals \eqref{all_integrals_feynman_parameters}:
\begin{align}
    I^{(0,0,3/2)}(p_2,m) &= \frac{2}{p_{2}^{2}}\lim_{m\rightarrow0}\frac{1}{m}\log\frac{p_{2}^{2}}{4m^{2}},  \\
    I^{(1,0,3/2)}(p_2,m) &= I^{(0,1,3/2)}(p_2,m) = \frac{2}{p_2 ^2} \lim_{m\rightarrow 0} \frac{1}{m} -\frac{2\pi}{p_2^3},   \\
    I^{(2,0,3/2)}(p_2,m) &=I^{(0,2,3/2)}(p_2,m)= \frac{1}{p_2 ^2} \lim_{m\rightarrow 0} \frac{1}{m} - \frac{\pi}{p_2^3},  \\
    I^{(3,0,3/2)}(p_2,m) &= I^{(0,3,3/2)}(p_2,m) = \frac{2}{3p_{2}^{2}}\lim_{m\rightarrow0}\frac{1}{m} -\frac{3\pi}{4p_{2}^{3}},  \\
    I^{(1,1,3/2)}(p_2,m) &= \frac{\pi}{p_2^3},  \\
    I^{(2,1,3/2)}(p_2,m) &= I^{(1,2,3/2)}(p_2,m) =  \frac{\pi}{4p_2^3},  \\
    I^{(0,0,1/2)}(p_2,m) &= \frac{\pi}{p_2},  \\
    I^{(1,0,1/2)}(p_2,m) &= I^{(0,1,1/2)}(p_2,m) = \frac{\pi}{4p_2},  
\end{align}
\begin{comment}
    \begin{align}
    I^{(0,0,3/2)} &= \frac{2}{p_{2}^{2}}\lim_{m\rightarrow0}\frac{1}{m}\log\frac{p_{2}^{2}}{4m^{2}}, & I^{(1,0,3/2)} &
    = \frac{2}{p_2 ^2} \lim_{m\rightarrow 0} \frac{1}{m} -\frac{2\pi}{p_2^3}, \\ 
    I^{(2,0,3/2)} &= \frac{1}{p_2 ^2} \lim_{m\rightarrow 0} \frac{1}{m} - \frac{\pi}{p_2^3}, & I^{(3,0,3/2)} &
    = \frac{2}{3p_{2}^{2}}\lim_{m\rightarrow0}\frac{1}{m} -\frac{3\pi}{4p_{2}^{3}}, \\
    I^{(0,1,3/2)} &= \frac{2}{p_2 ^2} \lim_{m\rightarrow 0} \frac{1}{m} -\frac{2\pi}{p_2^3}, &  I^{(0,2,3/2)} &
    = \frac{1}{p_2 ^2} \lim_{m\rightarrow 0} \frac{1}{m} - \frac{\pi}{p_2^3}, \\
    I^{(0,3,3/2)} &= \frac{2}{3p_{2}^{2}}\lim_{m\rightarrow0}\frac{1}{m} -\frac{3\pi}{4p_{2}^{3}}, & I^{(1,1,3/2)} &
    = \frac{\pi}{p_2^3}, \\
    I^{(1,2,3/2)} &= \frac{\pi}{4p_2^3}, & I^{(2,1,3/2)} &= \frac{\pi}{4p_2^3}, \\
    I^{(0,0,1/2)} &= \frac{\pi}{p_2}, & I^{(1,0,1/2)} &= \frac{\pi}{4p_2}, \\
    I^{(0,1,1/2)} &= \frac{\pi}{4p_2},
\end{align}
\end{comment}
which gives for the finite part of the 3-point function
\begin{multline}
     \langle j^a_\mu(p_1) j^b_\nu(-p_2) j^c_\rho(-p_3) \rangle_{\text{finite}} = N^2\frac{\epsilon^{abc}}{4p_2^3}\left\{ 
     2p_{3\mu}\left(p_{2\nu}p_{2\rho}+2p_{3\nu}p_{2\rho}+5p_{2\nu}p_{3\rho}+10p_{3\nu}p_{3\rho}\right)\right.  \\
     \left. +p_{2\mu}\left(p_{2\nu}p_{2\rho}+2p_{3\nu}p_{2\rho}+8p_{2\nu}p_{3\rho}+16p_{3\nu}p_{3\rho}\right)+
     p_2^2\left[p_{2\mu}\delta_{\nu\rho}+\left(p_{2\nu}+2p_{3\nu}\right)\delta_{\mu\rho}-p_{2\rho}\delta_{\mu\nu}\right]\right\} ,
\end{multline}
where we have written the result only in terms of the independent variables $p_2^\mu, p_3^\mu$, in particular using 
\begin{equation}
    p_1^\mu = p_2^\mu + p_3^\mu, \quad \quad p_1^2 = p_3^2 = 0 \quad \Rightarrow \quad p_2 \cdot p_3 
    = - \frac{p_2^2}{2}.\label{momcons}
\end{equation}

This form is therefore unambiguous. 
But we can also write it in an (ambiguous) form with some $p_1^\mu$ terms emphasized, that will be of use later on, as
\begin{align}
    &\langle j^a_\mu(p_1) j^b_\nu(-p_2) j^c_\rho(-p_3) \rangle_{\text{finite}} = N^2\frac{\epsilon^{abc}}{4p_{2}^{3}}
    \left\{ p_{1\mu}\left[p_{2\nu}p_{2\rho}+4p_{3\nu}p_{3\rho}+2\left(p_{2\nu}p_{3\rho}-p_{3\nu}p_{2\rho}\right)\right]\right. 
    \nonumber \\
    & +2p_{2\mu}\left[2p_{3\nu}p_{2\rho}+3(p_{2\nu}+2p_{3\nu})p_{3\rho}\right] \nonumber  
    +p_{3\mu}\left[(p_{2\nu}+6p_{3\nu})p_{2\rho}+8(p_{2\nu}+2p_{3\nu})p_{3\rho}\right] \nonumber \\
    & \left.+p_{2}^{2}\left[p_{2\mu}\delta_{\nu\rho}-p_{2\rho}\delta_{\mu\nu}+\left(p_{2\nu}
    +2p_{3\nu}\right)\delta_{\mu\rho}\right]\right\}. \label{JJJ_finite_first_case}
\end{align}
\begin{comment}
\begin{multline}
     \langle j^a_\mu(p_1) j^b_\nu(-p_2) j^c_\rho(-p_3) \rangle_{\text{finite}} = \frac{\epsilon^{abc}}{4}\left\{ \frac{2p_{3\mu}}
     {p_{2}^{3}}\left(p_{2\nu}p_{2\rho}+2p_{3\nu}p_{2\rho}+5p_{2\nu}p_{3\rho}+10p_{3\nu}p_{3\rho}\right)\right.  \\
     \left. +p_{2\mu}\left(p_{2\nu}p_{2\rho}+2p_{3\nu}p_{2\rho}+8p_{2\nu}p_{3\rho}+16p_{3\nu}p_{3\rho}\right)
     +\frac{1}{p_{2}}\left[p_{2\mu}\delta_{\nu\rho}+\left(p_{2\nu}+2p_{3\nu}\right)\delta_{\mu\rho}
     -p_{2\rho}\delta_{\mu\nu}\right]\right\} ,
\end{multline}
\end{comment}

The divergent part, after taking $m\rightarrow 0$, is given by 
\begin{multline}
    \langle j_{\mu}^{a}(p_{1})j_{\nu}^{b}(-p_{2})j_{\rho}^{c}(-p_{3})\rangle_{\text{divergent}} 
    = N^2\frac{\epsilon^{abc}}{12\pi mp_{2}^{2}}\left[-2p_{2\rho}\left(p_{2\mu}+p_{3\mu}\right)\left(p_{2\nu}+p_{3\nu}\right) \right.  \\
    -14\left(p_{2\mu}+p_{3\mu}\right)p_{2\nu}p_{3\rho} +2\left(13p_{2\mu}+14p_{3\mu}\right)p_{3\nu}p_{3\rho} 
    +\left.3\left(p_{2\mu}+p_{3\mu}\right)\left(p_{2\nu}+2p_{3\nu}\right)p_{3\rho}\log\left(\frac{p_{2}^{2}}{4m^{2}}\right)\right]. 
\end{multline}
\begin{comment}
{\em HN: Is the -14 on the second line a +14 maybe? This would make it nicer under the Ward identity with $p_2^\nu$ (no 
factors of 14 anywhere, only factors of 1), though perhaps it is what it is.}
\end{comment}
Again, we can write it in an (ambiguous) form with some $p_1^\mu$ terms emphasized as
\begin{align}
    \langle j^a_\mu(p_1) j^b_\nu(-p_2) j^c_\rho(-p_3) \rangle_{\text{divergent}} &= -N^2\frac{\epsilon^{abc}}{6\pi mp_{2}^{2}}
    \left\{ p_{1\mu}\left[p_{1\nu}p_{1\rho}+6p_{1\nu}p_{3\rho}+6p_{3\nu}p_{3\rho}\right. \right. \nonumber \\
    & \left.\left.+\frac{3}{2}(p_{1\nu}+p_{3\nu})p_{3\rho}\log\left(\frac{p_{2}^{2}}{4m²}\right) \right] +p_{3\mu}p_{3\nu}p_{3\rho} \right\}.
\end{align}
\begin{comment}
    \begin{multline}
    \langle j_{\mu}^{a}(p_{1})j_{\nu}^{b}(-p_{2})j_{\rho}^{c}(-p_{3})\rangle_{\text{divergent}} 
    = \frac{\epsilon^{abc}}{12\pi mp_{2}^{2}}\left[-2p_{2\rho}\left(p_{2\mu}+p_{3\mu}\right)\left(p_{2\nu}+p_{3\nu}\right) \right.  \\
    -14\left(p_{2\mu}+p_{3\mu}\right)p_{2\nu}p_{3\rho} +2\left(13p_{2\mu}+14p_{3\mu}\right)p_{3\nu}p_{3\rho} 
    +\left.3\left(p_{2\mu}+p_{3\mu}\right)\left(p_{2\nu}+2p_{3\nu}\right)p_{3\rho}\log\left(\frac{p_{2}^{2}}{4m^{2}}\right)\right]. 
\end{multline}
\end{comment}

The full 3-point function is the sum of both finite and divergent parts,
\begin{equation}
    \langle j^a_\mu(p_1) j^b_\nu(-p_2) j^c_\rho(-p_3) \rangle = \langle j^a_\mu(p_1) j^b_\nu(-p_2) j^c_\rho(-p_3) 
    \rangle_{\text{finite}} + \langle j^a_\mu(p_1) j^b_\nu(-p_2) j^c_\rho(-p_3) \rangle_{\text{divergent}}. 
    \label{3_point_function_finite_plus_divergent}
\end{equation}

The IR divergence regularized by the divergent term $ \sim 1/m$ is a consequence of the choice $p_3^2 = 0$. 
To make this statement clear, we will compute the Ward identity. Contracting \eqref{3_point_function_finite_plus_divergent} 
with $p_1^\mu
=p_2^\mu+p_3^\mu$ and using (\ref{momcons}),
%\begin{equation}
   % p_1^\mu = p_2^\mu + p_3^\mu, \quad \quad p_1^2 = p_3^2 = 0 \quad \Rightarrow \quad p_2 \cdot p_3 = - \frac{p_2^2}{2},
%\end{equation}
%
we find for the finite part
\begin{eqnarray}
     (p_2+p_3)^\mu \langle j_{\mu}^{a}(p_1)j_{\nu}^{b}(-p_{2})j_{\rho}^{c}(-p_{3})\rangle_{\text{finite}} 
     &=&   N^2\frac{\epsilon^{abc}}{8}p_{2}\left(\delta_{\nu\rho}-\frac{p_{2\nu}p_{2\rho}}{p_{2}^{2}}\right)\nonumber \\
     &=& N^2\frac{\epsilon^{adc}}{8}p_{2}\left(\delta_{\nu\rho}-\frac{p_{2\nu}p_{2\rho}}{p_{2}^{2}}\right)\delta^{db}\nonumber \\
     &=& \frac{1}{2}\epsilon^{adc}\langle j_{\nu}^{d}(p_{2})j_{\rho}^{b}(-p_{2})\rangle\;,
\end{eqnarray}
while the divergent part gives
\begin{align}
    (p_2+p_3)^\mu \langle j_{\mu}^{a}(p_1)j_{\nu}^{b}(-p_{2})j_{\rho}^{c}(-p_{3})\rangle_{\text{divergent}} 
    = N^2\frac{\epsilon^{abc}}{12\pi} \lim_{m\rightarrow 0} \frac{p_{3\nu} p_{3\rho}}{m} .
\end{align}

Therefore, the Ward identity for the full correlation function is
\begin{equation}
    (p_2+p_3)^\mu \langle j_{\mu}^{a}(p_1)j_{\nu}^{b}(-p_{2})j_{\rho}^{c}(-p_{3})\rangle
      = \frac{1}{2}\epsilon^{adc}\langle j_{\nu}^{d}(p_{2})j_{\rho}^{b}(-p_{2})\rangle +  
      N^2\frac{\epsilon^{adc}}{12\pi} \lim_{m\rightarrow 0} \frac{p_{3\nu} p_{3\rho}}{m} \delta^{db}.
\end{equation}

Note that the Ward identity is not satisfied in this case. The second term, however, is just a consequence of taking the 
on-shell $p_3^2 \rightarrow 0$ limit in the term proportional to $\langle j(p_3) j(-p_3)\rangle$ in the Ward identity,
\begin{equation}
    \langle j_{\nu}^{d}(p_{3})j_{\rho}^{b}(-p_{3})\rangle \sim \frac{p_{3\nu} p_{3\rho}}{p_3},
\end{equation}
with $m$ being the regulator. In the next section we will show that the Ward identity is satisfied for  
$p_3^2 \neq 0$ without divergences.

We can also check the other Ward identities. Contracting \eqref{3_point_function_finite_plus_divergent}
with $p_3^\rho$ gives for the finite part 
\bea
p_3^\rho\langle j_\mu^a(p_1)j_\nu^b(-p_2)j_\rho^c(-p_3)\rangle_{\rm finite}&=&N^2\frac{\epsilon^{abc}}{8}p_2\left(\delta_{\mu\nu}
-\frac{p_{2\mu}p_{2\nu}}{p_2^2}\right)\cr
&=&\frac{1}{2}\epsilon^{adc}p_2\langle j_\nu^d(p_2)j_\mu^b(-p_2)\rangle\;,
\eea
while for the divergent part, we get 
\be
p_3^\rho\langle j_\mu^a(p_1)j_\nu^b(-p_2)j_\rho^c(-p_3)\rangle_{\rm divergent}= - N^2
\frac{\epsilon^{abc}}{12\pi}\lim_{m\rightarrow 0}
\frac{p_{1\mu}p_{1\nu}}{m}\;,
\ee
so in total 
\be
p_3^\rho\langle j_\mu^a(p_1)j_\nu^b(-p_2)j_\rho^c(-p_3)\rangle=\frac{1}{2}\epsilon^{adc}\langle j_\nu^d(p_2)j_\mu^b(-p_2)\rangle
- N^2\frac{\epsilon^{adc}}{12\pi}\lim_{m\rightarrow 0}\frac{p_{1\mu}p_{1\nu}}{m}.
\ee

As expected, this gives the same result as if exchanging $p_{1\mu}\leftrightarrow p_{3\rho}$ and $p_2\leftrightarrow -p_2$
in the first Ward identity, 
which is indeed a symmetry of the 3-point function. 

The last Ward identity is obtained by contracting \eqref{3_point_function_finite_plus_divergent} with $p_2^\nu$, which 
gives for the finite part 
\be
p_2^\nu\langle j_\mu^a(p_1)j_\nu^b(-p_2)j_\rho^c(-p_3)\rangle_{\rm finite}=0\;,
\ee
while for the divergent part it gives
\bea
p_2^\nu\langle j_\mu^a(p_1)j_\nu^b(-p_2)j_\rho^c(-p_3)\rangle_{\rm divergent} 
= -N^2\frac{\epsilon^{abc}}{12\pi}\left(\frac{p_{1\mu}p_{1\rho}}{m}-\frac{p_{3\mu}p_{3\rho}}{m}\right)
\eea
which again, as expected from symmetry, is invariant under the exchange $p_{1\mu}\leftrightarrow p_{3\rho}$ and 
$p_2\leftrightarrow -p_2$. This time, this would be understood from the Ward identity only if one would take {\em all the momenta}
to be off-shell ($p_1^2\neq0$, $p_2^2\neq 0$, $p_3^2\neq 0$), otherwise there would be vanishing numerators 
$1/p_1^2, 1/p_2^2, 1/p_3^2$. The fact that the finite part of this Ward identity vanishes is simply due to the fact that 
now there is no $p_{2\nu}$ (the index was contracted) available to make the 2-point function with momentum $p_2$.

\subsection{One external light-like momenta ("on-shell", $p_1^2=0$)}

For $p_1^2 = 0$, the denominator with the mass regulator is
\begin{equation}
    \Delta^2_m =  x\left(1-x-y\right)p_{3}^{2}+xyp_{2}^{2} + m^2.
\end{equation}

The result of the integrals \eqref{all_integrals_feynman_parameters}, in this case, is
\begin{align}
    I^{(0,0,3/2)} &= \frac{2}{(p_2^2 - p_3^2)}\log \left(\frac{p_2 ^2}{p_3 ^2}  \right) \lim_{m\rightarrow 0} \frac{1}{m}, 
    &  I^{(1,0,3/2)} &= \frac{2\pi}{p_2 p_3 (p_2 + p_3)}, \\
    I^{(2,0,3/2)} &= \frac{\pi}{p_2 p_3 (p_2 + p_3)}, & I^{(3,0,3/2)} &= \frac{3\pi}{4p_2 p_3 (p_2 + p_3)},
\end{align}
\begin{align}
    I^{(0,1,3/2)} &= -\frac{2\pi}{p_{2}\left(p_{2}+p_{3}\right)^{2}}+\frac{2}{\left(p_{2}^{2}-p_{3}^{2}\right)}
    \left[1-\frac{p_{3}^{2}}{\left(p_{2}^{2}-p_{3}^{2}\right)}\log\left(\frac{p_{2}^{2}}{p_{3}^{2}}\right)\right]
    \lim_{m\rightarrow0}\frac{1}{m},  \\
    I^{(0,2,3/2)} &= -\frac{p_{2}+3p_{3}}{p_{2}\left(p_{2}+p_{3}\right)^{3}}\pi \nonumber \\
    & \quad \quad \quad \quad +\frac{1}{\left(p_{2}^{2}-p_{3}^{2}\right)^{3}}\left[p_{2}^{4}-4p_{2}^{2}p_{3}^{2}
    +3p_{3}^{4}+2p_{3}^{4}\log\left(\frac{p_{2}^{2}}{p_{2}^{3}}\right)\right]\lim_{m\rightarrow0}\frac{1}{m},  \\
    I^{(0,3,3/2)} &= -\frac{3\pi\left(p_{2}^{2}+4p_{2}p_{3}+5p_{3}^{2}\right)}{4p_{2}\left(p_{2}+p_{3}\right)^{4}} \nonumber \\
    & \quad \quad \quad \quad + \frac{2p_{2}^{6}-9p_{2}^{4}p_{3}^{2}+18p_{2}^{2}p_{3}^{4}-11p_{3}^{6}
    -6p_{3}^{2}\log\left(p_{2}^{2}/p_{3}^{2}\right)}{3\left(p_{2}^{2}-p_{3}^{2}\right)^{4}}\lim_{m\rightarrow0}\frac{1}{m} ,
\end{align}
\begin{align}
    I^{(1,1,3/2)} &= \frac{\pi}{p_{2}\left(p_{2}+p_{3}\right)^{2}}, & I^{(1,2,3/2)} &
    = \frac{p_{2}+3p_{3}}{4p_{2}\left(p_{2}+p_{3}\right)^{3}}\pi ,\\
    I^{(2,1,3/2)} &= \frac{\pi}{4p_{2}\left(p_{2}+p_{3}\right)^{2}}, &  I^{(0,0,1/2)} &= \frac{\pi}{p_{2}+p_{3}}, \\ 
    I^{(1,0,1/2)} &=  \frac{\pi}{4\left(p_{2}+p_{3}\right)}, & I^{(0,1,1/2)} &= \frac{p_{2}+2p_{3}}{4\left(p_{2}+p_{3}\right)^{2}} \pi.
\end{align}

Now, the result for the 3-point function is too extensive to write in a single expression. 
Instead of that, we decompose the 3-point function as
\begin{align}
    \langle j_{\mu}^a(p_{1})j_{\nu}^b(-p_{2})j_{\rho}^c(-p_{3})\rangle &= N^2\epsilon^{abc} c_{0} \left[p_{1\mu}\left(c_{1}
    p_{2\nu}p_{2\rho}+c_{2}p_{3\nu}p_{3\rho}+c_{3}p_{3\nu}p_{2\rho}+c_{4}p_{3\rho}p_{2\nu}\right)\right. \nonumber \\
    &\quad\quad\quad  +p_{2\mu}\left(c_{5}p_{2\nu}p_{2\rho}+c_{6}p_{3\nu}p_{3\rho}
    +c_{7}p_{3\nu}p_{2\rho}+c_{8}p_{3\rho}p_{2\nu}\right) \nonumber \\
    &\quad\quad\quad  +p_{3\mu}\left(c_{9}p_{2\nu}p_{2\rho}+c_{10}p_{3\nu}p_{3\rho}
    +c_{11}p_{3\nu}p_{2\rho}+c_{12}p_{3\rho}p_{2\nu}\right) \nonumber \\
    &\quad\quad\quad  +\delta_{\nu\rho}\left(c_{13}p_{1\mu}+c_{14}p_{2\mu}+c_{15}p_{3\mu}\right) \nonumber \\
    & \left. \quad\quad\quad  +\delta_{\mu\rho}\left(c_{16}p_{2\nu}+c_{17}p_{3\nu}\right) 
    +\delta_{\mu\nu}\left(c_{18}p_{2\rho}+c_{19}p_{3\rho}\right)\right], \label{three_point_function_coefficients}
\end{align}
and after splitting the coefficients into a finite piece and a divergent piece
\begin{equation}
    c_i = c_{i}^{\text{finite}} + c_{i}^{\text{divergent}},
\end{equation}
we find, for the finite part of the 3-point function,
\begin{align}
    c_{0}^{\text{finite}} &=\frac{1}{4p_{2}p_{3}\left(p_{2}+p_{3}\right)^{4}}, \nonumber \\
    c_{1}^{\text{finite}} & = p_{3}\left(p_{2}^{2}+4p_{3}p_{2}+9p_{3}^{2}\right), & c_{2}^{\text{finite}} 
    & = 2p_{2}\left(2p_{2}-p_{3}\right)p_{3}, \nonumber \\
    c_{3}^{\text{finite}} & =-2p_{2}p_{3}\left(p_{2}+4p_{3}\right), & c_{4}^{\text{finite}} 
    & = 2p_{3}\left(p_{2}^{2}+2p_{3}^{2}\right), \nonumber \\
    c_{5}^{\text{finite}} & = -4p_{3}^{2}\left(p_{2}+p_{3}\right), & c_{6}^{\text{finite}} 
    & = -4p_{2}^{2}\left(p_{2}+p_{3}\right), \nonumber \\
    c_{7}^{\text{finite}} & = 4p_{2}p_{3}\left(p_{2}+p_{3}\right), & c_{8}^{\text{finite}} 
    & = -2\left(p_{2}+p_{3}\right)\left(p_{2}^{2}+p_{3}^{2}\right), \nonumber \\
    c_{9}^{\text{finite}} & = p_{3}\left(p_{2}-5p_{3}\right)\left(p_{2}+p_{3}\right), 
    & c_{10}^{\text{finite}} & = -p_{2}\left(5p_{2}-p_{3}\right)\left(p_{2}+p_{3}\right), \nonumber \\
    c_{11}^{\text{finite}} & = 6p_{2}p_{3}\left(p_{2}+p_{3}\right), & c_{12}^{\text{finite}} 
    & = -2\left(p_{2}^{3}+p_{3}^{3}\right), \nonumber \\
    c_{13}^{\text{finite}} & = -p_{2}p_{3}\left(p_{2}+2p_{3}\right)\left(p_{2}+p_{3}\right)^{2}, 
    & c_{14}^{\text{finite}} & = 2p_{2}p_{3}\left(p_{2}+p_{3}\right)^{3}, \nonumber \\
    c_{15}^{\text{finite}} & = p_{2}p_{3}\left(p_{2}+p_{3}\right)^{3}, & c_{16}^{\text{finite}} 
    & = p_{2}^{2}p_{3}\left(p_{2}+p_{3}\right)^{2}, \nonumber \\
    c_{17}^{\text{finite}} & = p_{2}p_{3}\left(2p_{2}+p_{3}\right)\left(p_{2}+p_{3}\right)^{2}, 
    & c_{18}^{\text{finite}} & = -p_{2}p_{3}\left(p_{2}+2p_{3}\right)\left(p_{2}+p_{3}\right)^{2}, \nonumber \\
    c_{19}^{\text{finite}} & = -p_{2}p_{3}^{2}\left(p_{2}+p_{3}\right)^{2} \label{finite_coefficients_three_point_function_second_case}
\end{align}
while for the divergent part, we find
\begin{align}
    c_{0}^{\text{div}} &= \frac{1}{12\pi\left(p_{2}^{2}-p_{3}^{2}\right){}^{4}}\lim_{m\rightarrow0}\frac{1}{m}, \nonumber \\ 
    c_{1}^{\text{div}} &= -2p_{2}^{6}+6p_{3}^{2}p_{2}^{4}\left[\log\left(\frac{p_{2}^{2}}{p_{3}^{2}}\right)-1\right]
    +6p_{3}^{4}p_{2}^{2}\left[2\log\left(\frac{p_{2}^{2}}{p_{3}^{2}}\right)-1\right]
    +2p_{3}^{6}\left[3\log\left(\frac{p_{2}^{2}}{p_{3}^{2}}\right)+7\right], \nonumber \\ 
    c_{2}^{\text{div}} &= 2\left\{ p_{3}^{6}+p_{2}^{6}\left[3\log\left(\frac{p_{2}^{2}}{p_{3}^{2}}\right)-7\right]
    +3p_{3}^{2}p_{2}^{4}\left[2\log\left(\frac{p_{2}^{2}}{p_{3}^{2}}\right)+1\right]
    +3p_{3}^{4}p_{2}^{2}\left[\log\left(\frac{p_{2}^{2}}{p_{3}^{2}}\right)+1\right]\right\}, \nonumber \\ 
    c_{3}^{\text{div}} &= 2\left\{ -p_{2}^{6}+p_{3}^{6}+3p_{3}^{2}p_{2}^{4}\left[2\log\left(\frac{p_{2}^{2}}{p_{3}^{2}}\right)
    -3\right]+3p_{3}^{4}p_{2}^{2}\left[2\log\left(\frac{p_{2}^{2}}{p_{3}^{2}}\right)+3\right]\right\}, \nonumber \\ 
    c_{4}^{\text{div}} &= p_{2}^{6}\left[3\log\left(\frac{p_{2}^{2}}{p_{3}^{2}}\right)-8\right]
    +9p_{3}^{2}p_{2}^{4}\log\left(\frac{p_{2}^{2}}{p_{3}^{2}}\right)+9p_{3}^{4}p_{2}^{2}
    \log\left(\frac{p_{2}^{2}}{p_{3}^{2}}\right)+p_{3}^{6}\left[3\log\left(\frac{p_{2}^{2}}{p_{3}^{2}}\right)+8\right]. \nonumber \\
\end{align}
and
\begin{equation}
    c_i^{\text{div}} = 0, \quad \text{for } i=5,\ldots 19.
\end{equation}

\subsubsection{Ward Identity}

Let us check that the finite part of $\langle j_{\mu}^{a}(p_1)j_{\nu}^{b}(-p_{2})j_{\rho}^{c}(-p_{3})\rangle$ satisfies the 
transverse Ward identity. Contracting \eqref{three_point_function_coefficients} with $p_1^{\mu}$ and using the 
conservation of momentum $p_{1\mu} = p_{2\mu}+p_{3\mu}$ to write, for $p_1^2=0$,
\begin{equation}
    p_{2} \cdot p_3 = -\frac{1}{2} (p_2^2+p_3^2)\;,
\end{equation}
we find
\begin{align}
    p_{1}^{\mu}\langle j_{\mu}^{a}(p_1)j_{\nu}^{b}(-p_{2})j_{\rho}^{c}(-p_{3})\rangle & = \frac{N^2}{2}c_{0}\epsilon^{abc}
    \left\{ p_{3\nu}p_{3\rho}\left[2(c_{17}+c_{19})-\left(p_{2}^{2}-p_{3}^{2}\right)(c_{10}-c_{6})\right]\right. \nonumber \\
    & +p_{2\nu}p_{2\rho}\left[2(c_{16}+c_{18})+\left(p_{2}^{2}-p_{3}^{2}\right)(c_{5}-c_{9})\right] \nonumber \\
    & +p_{2\nu}p_{3\rho}\left[2(c_{16}+c_{19})-\left(p_{2}^{2}-p_{3}^{2}\right)(c_{12}-c_{8})\right] \nonumber \\
    & +p_{3\nu}p_{2\rho}\left[2(c_{17}+c_{18})-\left(p_{2}^{2}-p_{3}^{2}\right)(c_{11}-c_{7})\right] \nonumber \\
    & \left.+\delta_{\nu\rho}(c_{14}-c_{15})\left(p_{2}^{2}-p_{3}^{2}\right)\right\}, 
\end{align}
which after substituting the coefficients \eqref{finite_coefficients_three_point_function_second_case}, gives
\begin{align}
    p_1^\mu\langle j_{\mu}^{a}(p_1)j_{\nu}^{b}(-p_{2})j_{\rho}^{c}(-p_{3})\rangle_{\text{finite}}
    & = N^2\frac{\epsilon^{abc}}{8}\left[\left(\delta_{\nu\rho}-\frac{p_{2\nu}p_{2\rho}}{p_{2}^{2}}\right)p_{2}
    -\left(\delta_{\nu\rho}-\frac{p_{3\nu}p_{3\rho}}{p_{3}^{2}}\right)p_{3}\right] \nonumber \\
    &  = N^2\frac{\epsilon^{adc}}{8}\left(\delta_{\nu\rho}-\frac{p_{2\nu}p_{2\rho}}{p_{2}^{2}}\right)p_{2}\delta^{db}
    -\frac{\epsilon^{abd}}{8}\left(\delta_{\nu\rho}-\frac{p_{3\nu}p_{3\rho}}{p_{3}^{2}}\right)p_{3}\delta^{dc} \nonumber \\
    &= \frac{1}{2}\epsilon^{adc}\langle j_{\nu}^{d}(p_{2})j_{\rho}^{b}(-p_{2})\rangle
    -\frac{1}{2}\epsilon^{abd}\langle j_{\nu}^{d}(p_{3})j_{\rho}^{c}(-p_{3})\rangle.
\end{align}

For the divergent part, however, we note that all coefficients besides the ones multiplying 
$p_{1\mu}$ are zero, and since $p_1^2=0$, the divergent part gives
\begin{equation}
    p_1^\mu\langle j_{\mu}^{a}(p_1)j_{\nu}^{b}(-p_{2})j_{\rho}^{c}(-p_{3})\rangle_{\text{divergent}}=0\;,
\end{equation}
as we anticipate. All in all, the full 3-point function satisfies
\begin{equation}
    p_1^\mu\langle j_{\mu}^{a}(p_1)j_{\nu}^{b}(-p_{2})j_{\rho}^{c}(-p_{3})\rangle = \frac{1}{2}\epsilon^{adc}
    \langle j_{\nu}^{d}(p_{2})j_{\rho}^{b}(-p_{2})\rangle-\frac{1}{2}\epsilon^{abd}\langle j_{\nu}^{d}(p_{3})j_{\rho}^{c}(-p_{3})\rangle.
\end{equation}
which is the transverse Ward identity.

\subsubsection{Application to cosmology}

In holographic cosmology, as we mentioned, one interesting case (see for instance Maldacena's non-Gaussianity 
paper \cite{Maldacena:2002vr})
is when $k_1\ll k_2,k_3$, where $k_i=|\vec{k}_i|$ 
are 3-dimensional 
Euclidean momenta, corresponding in cosmology to spatial momenta. So with respect to our calculation, we need to 
make the Wick rotation to Euclidean space, and then take the above limit.

Since we have chosen $p_1^\mu=p_2^\mu+p_3^\mu$, we will work with 
\be
p_1^2\ll p_2^2\simeq p_3^2\equiv p^2\;,\;\;\;
p^\mu\equiv p_2^\mu\simeq -p_3^\mu\;,\;\;\; p_1^\mu\simeq 0.\label{cosmospecial}
\ee

Under this approximation, we see that we can already consider the results in this subsection, and apply directly the 
approximation above. This gives
\bea
\langle j_{\mu}^{a}(p_1)j_{\nu}^{b}(-p_{2})j_{\rho}^{c}(-p_{3})\rangle&\simeq 
& N^2\epsilon^{abc}c_0\left[p_\mu p_\nu p_\rho\left( (c_5+c_6-c_7-c_8)-
(c_9+c_{10}-c_{11}-c_{12})\right)\right.\cr
&&\left.+\delta_{\nu\rho}p_\mu\left(c_{14}-c_{15}\right)+\delta_{\mu\rho}p_\nu\left(c_{16}-c_{17}\right)
+\delta_{\mu\nu}p_\rho\left(c_{18}-c_{19}\right)\right].\cr
&&
\eea

We find for the finite parts
\bea
c_0^{\rm finite}&=& \frac{1}{64p^6}+{\cal O}(p_1)\cr
(c_5+c_6-c_7-c_8)^{\rm finite}&=&-16p^3+{\cal O}(p_1)\;,\cr
(c_9+c_{10}-c_{11}-c_{12})^{\rm finite}&=&-24p^3+{\cal O}(p_1)\cr
(c_{14}-c_{15})^{\rm finite}&=& 8p^5+{\cal O}(p_1)\;,\cr
(c_{16}-c_{17})^{\rm finite}&=& -8p^5+{\cal O}(p_1)\;,\cr
(c_{18}-c_{19})^{\rm finite}&=&-8p^5+{\cal O}(p_1)\;,
\eea
so that
\be
\langle j_\mu(p_1)j_\nu(-p_2)j_\rho(-p_3)\rangle\simeq  N^2\epsilon^{abc}\frac{1}{8p}\left[-\left(\delta_{\mu\nu}-\frac{p_\mu
p_\nu}{p^2}\right)+\left(\delta_{\nu\rho}-\frac{p_\nu p_\rho}{p^2}\right)-\left(\delta_{\mu\rho}-\frac{p_\mu p_\rho}{p^2}\right)\right]\;,
\ee
neglecting the divergent terms. The divergent terms are now
\be
c_i^{\rm div}\simeq 0+{\cal O}(p_1)\;,\;\; i=1,...,4\;,
\ee
but 
\be
c_0^{\rm div}=\frac{1}{12\pi(2p_1\cdot p_3)^4}\lim_{m\rightarrow 0}\frac{1}{m}\sim \frac{1}{m p_1^4}\;,
\ee
so one would actually need to expand to a high order in $p_1$ the $c_i^{\rm div}$'s to say something decisive. 
Instead, we will get a better handle on this case from the next subsection, when all $p_i^2\neq 0$.

\subsection{General case}

Finally, we perform the full calculation without any assumption about the external momenta. 
All integrals in the Feynman parameters are finite, both in the UV and in the IR, 
hence there is no need to add a mass parameter in the propagators. 
We again provide the details in \autoref{appendix_integral_Feynman_parameters}, and obtain
\begin{align}
    I^{(0,0,3/2)}(p_1,p_2,p_3) &= \frac{2 \pi }{p_1 p_2 p_3}   \\
    I^{(1,0,3/2)}(p_1,p_2,p_3) &= \frac{2 \pi }{p_2 p_3 \left(p_1+p_2+p_3\right)}  \\
    I^{(2,0,3/2)}(p_1,p_2,p_3) &= \frac{2 p_1+p_2+p_3}{p_2 p_3 \left(p_1+p_2+p_3\right)^2} \pi   \\
    I^{(3,0,3/2)}(p_1,p_2,p_3) &= \frac{ 8 p_1^2+9 \left(p_2+p_3\right) p_1
    +3 \left(p_2+p_3\right)^2}{4 p_2 p_3 \left(p_1+p_2+p_3\right)^3}\pi    \\
    I^{(0,1,3/2)}(p_1,p_2,p_3) &=  I^{(1,0,3/2)}(p_3,p_2,p_1) = \frac{2 \pi }{p_1 p_2 \left(p_1+p_2+p_3\right)} \\
    I^{(0,2,3/2)}(p_1,p_2,p_3) &=  I^{(2,0,3/2)}(p_3,p_2,p_1) = \frac{  p_1+p_2+2 p_3}{p_1 p_2 \left(p_1+p_2+p_3\right)^2}\pi \\
    I^{(0,3,3/2)}(p_1,p_2,p_3) &=  I^{(3,0,3/2)}(p_3,p_2,p_1) 
    = \frac{ 8 p_3^2+9 p_3 \left(p_1+p_2\right) + 3 \left(p_1+p_2\right)^2}{4 p_1 p_2 \left(p_1+p_2+p_3\right)^3}\pi \\
    I^{(1,1,3/2)}(p_1,p_2,p_3) &= \frac{\pi }{p_2 \left(p_1+p_2+p_3\right)^2}   \\
    I^{(2,1,3/2)}(p_1,p_2,p_3) &= \frac{  3 p_1+p_2+p_3}{4 p_2 \left(p_1+p_2+p_3\right)^3}\pi   \\
    I^{(1,2,3/2)}(p_1,p_2,p_3) &=  I^{(2,1,3/2)}(p_3,p_2,p_1) = \frac{ p_1+p_2+3 p_3}{4 p_2 \left(p_1+p_2+p_3\right)^3}\pi \\
    I^{(0,0,1/2)}(p_1,p_2,p_3) &=  \frac{\pi }{p_1+p_2+p_3} \\
    I^{(1,0,1/2)}(p_1,p_2,p_3) &= \frac{  2 p_1+p_2+p_3}{4 \left(p_1+p_2+p_3\right)^2} \pi\\
    I^{(0,1,1/2)}(p_1,p_2,p_3) & =  I^{(1,0,1/2)}(p_3,p_2,p_1) = \frac{p_1+p_2+2 p_3}{4 \left(p_1+p_2+p_3\right)^2}\pi 
\end{align}

We again decompose the 3-point function as in \eqref{three_point_function_coefficients}. but now the coefficients are given by 
\begin{align}
    c_{0} & = \frac{1}{4p_{1}p_{2}p_{3}\left(p_{1}+p_{2}+p_{3}\right)^{3}} \nonumber \\ 
    c_{1} & = -p_{3}\left[4p_{3}^{2}+\left(p_{1}+p_{2}\right){}^{2}+3p_{3}\left(p_{1}+p_{2}\right)\right], & 
    c_{2} & = p_{2}p_{3}\left(p_{1}-p_{2}+p_{3}\right), \nonumber \\
    c_{3} & = p_{2}p_{3}\left(p_{1}+p_{2}+3p_{3}\right), &
    c_{4} & = -p_{3}\left[2p_{3}^{2}+\left(2p_{1}+p_{2}\right)p_{3}+p_{2}\left(p_{1}+p_{2}\right)\right] \nonumber \\ 
    c_{5} & = 2p_{3}^{2}\left(p_{1}+p_{2}+p_{3}\right),& 
    c_{6} & = 2p_{2}^{2}\left(p_{1}+p_{2}+p_{3}\right) \nonumber \\ 
    c_{7} & = -2p_{2}p_{3}\left(p_{1}+p_{2}+p_{3}\right). & 
    c_{8} & = -\left(p_{1}+p_{2}+p_{3}\right)\left(p_{1}^{2}-p_{2}^{2}-p_{3}^{2}\right) \nonumber \\ 
    c_{9} & = p_{3}\left[p_{1}^{2}+\left(p_{2}+p_{3}\right)p_{1}+2p_{3}\left(p_{2}+p_{3}\right)\right], & 
    c_{10} & = p_{2}\left[p_{1}^{2}+\left(p_{2}+p_{3}\right)p_{1}+2p_{2}\left(p_{2}+p_{3}\right)\right] \nonumber \\ 
    c_{11} & = -2p_{2}p_{3}\left(p_{2}+p_{3}\right),& 
    c_{12} & = p_{1}^{3}+\left(p_{2}+p_{3}\right)\left[p_{1}^{2}+p_{2}^{2}+p_{3}^{2}+\left(p_{2}+p_{3}\right)p_{1}\right] \nonumber \\ 
    c_{13} & = -p_{1}p_{2}p_{3}\left(p_{1}+p_{2}+p_{3}\right)\left(p_{1}+p_{2}+2p_{3}\right),& 
    c_{14} & = 2p_{1}p_{2}p_{3}\left(p_{1}+p_{2}+p_{3}\right){}^{2} \nonumber \\ 
    c_{15} & = p_{1}p_{2}p_{3}\left(p_{2}+p_{3}\right)\left(p_{1}+p_{2}+p_{3}\right),& 
    c_{16} & = p_{1}p_{2}\left(p_{1}+p_{2}\right)p_{3}\left(p_{1}+p_{2}+p_{3}\right) \nonumber \\ 
    c_{17} & = p_{1}p_{2}p_{3}\left(p_{1}+p_{2}+p_{3}\right)\left(p_{1}+2p_{2}+p_{3}\right),& 
    c_{18} & = -p_{1}p_{2}p_{3}\left(p_{1}+p_{2}+p_{3}\right)\left(p_{1}+p_{2}+2p_{3}\right) \nonumber \\ 
    c_{19} & = -p_{1}p_{2}p_{3}\left(p_{1}+p_{3}\right)\left(p_{1}+p_{2}+p_{3}\right) \label{coefficients_final_case}
\end{align}

Thus the 3-point function is completely IR finite, which translates in holographic cosmology to the fact that 
there is no cosmological singularity (corresponding to the IR of the quantum field theory) for the 3-point function.

\subsubsection{Ward identity}

To check if this result satisfies the transverse Ward identity, we multiply the 3-point function by $p^\mu_1$. 
Conservation of momentum implies the kinematic constraint
\begin{equation}
    p_{2} \cdot p_3 = \frac{1}{2} (p_1^2 - p_2^2 - p_3^2),
\end{equation}

After all simplifications, we find the following expression in terms of the coefficients $c_i$:
\begin{align}
    p_{1}^{\mu}\langle j_{\mu}(p_{1})j_{\nu}(p_{2})j_{\rho}(p_{3})\rangle & =\frac{N^2}{2}c_{0}\epsilon^{abc}\left\{ p_{2\nu}p_{2\rho}
    \left[2(c_{16}+c_{18})+p_{1}^{2}+(2c_{1}+c_{5}+c_{9})+\left(p_{2}^{2}-p_{3}^{2}\right)(c_{5}-c_{9})\right]\right. \nonumber \\
    & +p_{3\nu}p_{3\rho}\left[2(c_{16}+c_{18})+p_{1}^{2}(2c_{1}+c_{5}+c_{9})+\left(p_{2}^{2}-p_{3}^{2}\right)(c_{5}-c_{9})\right] 
    \nonumber \\
    & +p_{2\nu}p_{3\rho}\left[2(c_{16}+c_{19})+p_{1}^{2}(c_{12}+2c_{4}+c_{8})-\left(p_{2}^{2}-p_{3}^{2}\right)(c_{12}-c_{8})\right] 
    \nonumber \\
    & +p_{3\nu}p_{2\rho}\left[2(c_{17}+c_{18})+p_{1}^{2}(c_{11}+2c_{3}+c_{7})-\left(p_{2}^{2}-p_{3}^{2}\right)(c_{11}-c_{7})\right] 
    \nonumber \\
    & \left.+\delta_{\nu\rho}\left[p_{1}^{2}(2c_{13}+c_{14}+c_{15})+\left(p_{2}^{2}-p_{3}^{2}\right)(c_{14}-c_{15})\right]\right\}.
    \nonumber \\ \label{transverse_ward_identity_final_case_coefficients}
\end{align}

Substituting \eqref{coefficients_final_case} into  \eqref{transverse_ward_identity_final_case_coefficients}, we are left with
\begin{equation}
    p_1^\mu\langle j_{\mu}(p_{1})j_{\nu}(p_{2})j_{\rho}(p_{3})\rangle = \frac{1}{2}\epsilon^{adc}
    \langle j_{\nu}^{d}(p_{2})j_{\rho}^{b}(-p_{2})\rangle-\frac{1}{2}\epsilon^{abd}\langle j_{\nu}^{d}(p_{3})j_{\rho}^{c}(-p_{3})\rangle\;,
\end{equation}
which again shows that our result satisfies the transverse Ward identity.

At this point, we note that the general result for the 3-point function in (\ref{three_point_function_coefficients}) and 
(\ref{coefficients_final_case}) is consistent with previous results about the possible tensor structures of the 3-point function
allowed by Bose symmetry and the transverse Ward identity found 
in \cite{Bzowski:2013sza} and \cite{Bzowski:2017poo}.\footnote{We thank the anonymous referee for pointing out these 
references, of which we were previously unaware, and this consistency.} Then instead of the 19 coefficients $c_1-c_{19}$
we have only 2 scalar coefficients that are independent.
The check of the match, for which one needs to take into account that \cite{Bzowski:2017poo} use the different sign convention 
for momenta, with $p_1=-p_2-p_3$, instead of our $p_1=p_2+p_3$, is done in Appendix \ref{previous}.

\subsection{Special case applied for cosmology}

Now we consider the case of the approximation (\ref{cosmospecial}), relevant to cosmology non-Gaussianities for 
holographic cosmology. Applying it directly to the general case gives more relevant information (and no divergent terms). 
We obtain for the coefficients
\bea
c_0&=& \frac{1}{32 p_1p^5}+{\cal O}(1)\cr
(c_5+c_6-c_7-c_8)&=&8p^3+{\cal O}(p_1)\;,\cr
(c_9+c_{10}-c_{11}-c_{12})&=&8p^3+{\cal O}(p_1)\cr
(c_{14}-c_{15})&=& 0+{\cal O}(p_1^2)\;,\cr
(c_{16}-c_{17})&=& -4p_1p^4+{\cal O}(p_1^2)\;,\cr
(c_{18}-c_{19})&=&-4p_1p^4+{\cal O}(p_1^2)\;,
\eea
and the terms of order $p_1^2$ in the first two lines are harder to calculate, but for invariance we must have the difference 
of the first two lines to be $8p_1 p^2+{\cal O}(p_1^2)$, such that 
\bea
\langle j^a_\mu(p_1)j^b_\nu(-p_2)j^c_\rho(-p_3)\rangle&\simeq&  \epsilon^{abc}\frac{N^2}{8p}\left[-\left(\delta_{\mu\rho}-\frac{p_\mu 
p_\rho}{p^2}\right)p_\nu-\left(\delta_{\mu\nu}-\frac{p_\mu p_\nu}{p^2}\right)p_\rho\right]\cr
&\simeq& \frac{1}{2}\epsilon^{abd}\langle j_\mu^d(p_2)j_\nu^c(-p_2)\rangle p_{3\rho}-\frac{1}{2}\epsilon^{adc}\langle 
j_\mu^d(p_3)j_\rho^b(-p_3)\rangle p_{2\nu}.\cr
&&
\eea

This is what replaces the (inflationary) expectation \cite{Maldacena:2002vr} in the case of the {\em scalar} 3-point function
in the same limit,
\be
\langle\zeta_{p_1}\zeta_{-p_2}\zeta_{-p_3}\rangle\sim n_s\langle \zeta_{p_1}\zeta_{-p_1}\rangle\langle \zeta_{p_2}\zeta_{p_3}
\rangle\;,
\ee
for the case of the holographic 3-point function of currents. 

But actually we see that we can put the result in a form that closely mimics 
the one above for scalars. Indeed, for a two-point function 
that, at least in the perturbative regime is of the type 
\be
\langle j_\mu^a (p) j_\n^b(-b)\rangle=K \delta^{ab}p^{n_m}\left(\delta_{\mu\nu}-\frac{p_\mu p_\nu}{p^2}\right)\equiv
\frac{t}{2\pi}\delta^{ab}p \left(\delta_{\mu\nu}-\frac{p_\mu p_\nu}{p^2}\right)\;,
\ee
with $n_m\in \mathbb{R}$ a real power (which in perturbation theory is such that $|n_m-1|\ll 1$) and $K=N^2/4$, and $t\equiv 2
\pi K p^{n_m-1}$,
we can easily see that we have 
\be
\frac{\d}{\d p_\mu}\langle j_\mu^a(p)j_\nu^b(-p) \rangle\simeq-K \delta^{ab} p^{n_m}(d-1)\frac{p_\nu}{p^2}\;,
\ee
so that we can put the 3-point function in the form
\bea
\langle j^a_\mu(p_1)j^b_\nu(-p_2)j^c_\rho(-p_3)\rangle&\simeq&
-\frac{1}{2}\epsilon^{ade}\langle j_\mu^d(p_2)j_\nu^b(-p_2)\rangle \frac{1}{K (d-1)p_3^{n_m-2}}
\frac{\d}{\d p_{3\sigma}}\langle j_\sigma^e (p_3)j_\rho^c(-p_3)\rangle \cr
&&+\frac{1}{2}\epsilon^{ade}\langle j_\mu^e (p_3)j_\rho^c (-p_3)\rangle\frac{1}{K(d-1) p_2^{n_m-2}} \frac{\d}{\d p_{2\sigma}}
\langle j_\sigma^d(p_2)j_\nu^b(-p_2)\rangle.\cr&&\label{3pointspecial}
\eea

In the physical case, $K=N^2/4$, $n_m\simeq 1$, $d=3$ so the overall coefficient is $\simeq 1/(4K)=1/N^2$. 

In order to go to the magnetic currents, we need again to apply the $Sl(2,\mathbb{Z})$ S-duality, which, in the absence of 
a parity violating term, $w=0$ for the 2-point function, amounted to $t\rightarrow 1/t$ and $j_\mu^a\rightarrow \tilde j_\mu^a$, 
so we obtain 
\bea
\langle \tilde j^a_\mu(p_1)\tilde j^b_\nu(-p_2)\tilde j^c_\rho(-p_3)\rangle&\simeq&
-\frac{1}{2}\epsilon^{ade}\langle \tilde j_\mu^d(p_2)\tilde j_\nu^b(-p_2)\rangle \frac{4\pi^2 K p_3^{n_m}}{(d-1)}
\frac{\d}{\d p_{3\sigma}}\langle \tilde j_\sigma^e (p_3)\tilde j_\rho^c(-p_3)\rangle \cr
&&+\frac{1}{2}\epsilon^{ade}\langle \tilde j_\mu^e (p_3)\tilde j_\rho^c (-p_3)\rangle
\frac{4\pi^2K p_2^{n_m}}{(d-1)} \frac{\d}{\d p_{2\sigma}}
\langle \tilde j_\sigma^d(p_2)\tilde j_\nu^b(-p_2)\rangle.\cr&&
\eea

This would be the 3-point function giving, through the holographic map, the cosmological monopole distribution, 
and its non-Gaussianity, and in principle, if monopoles would 
be discovered in the universe, this could be measured. As we said, its IR finiteness corresponds to the fact that it 
does not have any cosmological singularity (since the IR of the quantum field theory corresponds to the early times in 
cosmology), although this fact is better tested at two-loop.

\section{Discussion of the two-loop result}

The 2-loop contribution would give anomalous dimensions, like in the 2-point function case and, more generally, 
would give nontrivial momentum dependence through dimensional transmutation. The 2-loop calculation would, 
among other things, verify if the equation (\ref{3pointspecial}) still holds at 2-loop, when $n_m$ is nontrivial 
(different than one), just like in the scalar case.

In this paper we do not compute the two-loop contribution. However, in this section we will show that all two-loop 
diagrams with the quartic vertex are zero in dimensional regularization, meaning that the final result is completely 
independent of the explicit form of the potential. This is also true for the 2-point function of the current operator, 
as described in \cite{Nastase:2020lvn}. This means that the results obtained in this toy model are actually universal 
in the set of the phenomenological actions, as long as they admit vortex solutions (corresponding to monopoles in 4 
dimensions). 

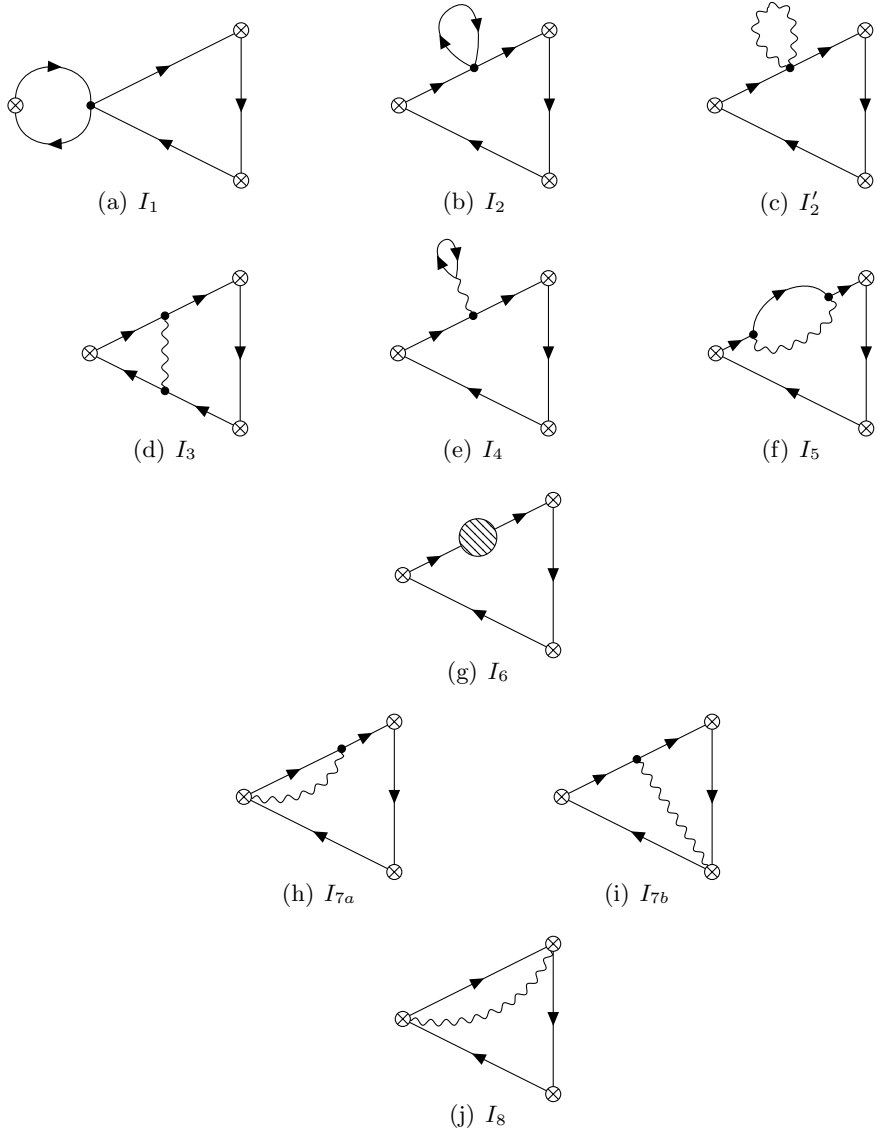
\begin{figure}[h!]
    \centering
    \hspace{-1cm}\subfigure[$I_{1}$]{
    \begin{tikzpicture}[baseline = (m.base),arrowlabel/.style={
        /tikzfeynman/momentum/.cd, % means that the following keys are read from the /tikzfeynman/momentum family
        arrow shorten=#1,arrow distance=2.5mm
      },
      arrowlabel/.default=0.4]
        \def\leglength{1}
        \begin{feynman} [inline =(a.base) ]
            \vertex[crossed dot] (l1) at (-2,0){};
            \vertex[dot] (a) at (-\leglength,0) {};
            \vertex[crossed dot]  (b) at ( \leglength,\leglength) {};
            \vertex[crossed dot]  (c) at (\leglength,- \leglength) {};
    
    \diagram*{
        (l1) -- [fermion,half left] (a)
          -- [fermion,half left] (l1),
        (a) -- [fermion] (b) -- [fermion] (c) -- [fermion] (a),
    };
        \end{feynman}
    \end{tikzpicture}
    }\hspace{1.5cm}
    \subfigure[$I_2$]{
    \begin{tikzpicture}[baseline = (m.base),arrowlabel/.style={
        /tikzfeynman/momentum/.cd, % means that the following keys are read from the /tikzfeynman/momentum family
        arrow shorten=#1,arrow distance=2.5mm
      },
      arrowlabel/.default=0.4]
        \def\leglength{1}
        \begin{feynman} [inline =(a.base) ]
            \vertex[crossed dot] (a) at (-\leglength,0) {};
            \vertex[crossed dot]  (b) at ( \leglength,\leglength) {};
            \vertex[crossed dot]  (c) at (\leglength,- \leglength) {};
            \vertex[dot] (d) at (0,0.5) {};
            \vertex (t) at (-0.3,1.3);
    \diagram*{
        (a) -- [fermion] (d) -- [fermion,out=155, in=210, min distance=0.4cm] (t) --[fermion,out=30, in=75, min distance=0.4cm] (d) -- [fermion] (b) -- [fermion] (c)  -- [fermion] (a),
         %(l1) [momentum ={[arrowlabel]$q+p_2$}] (a),
         };
        \end{feynman}
    \end{tikzpicture}
    }\hspace{1.6cm}
    \subfigure[$I_2^\prime$]{
    \begin{tikzpicture}[baseline = (m.base),arrowlabel/.style={
        /tikzfeynman/momentum/.cd, % means that the following keys are read from the /tikzfeynman/momentum family
        arrow shorten=#1,arrow distance=2.5mm
      },
      arrowlabel/.default=0.4]
        \def\leglength{1}
        \begin{feynman} [inline =(a.base) ]
            \vertex[crossed dot] (a) at (-\leglength,0) {};
            \vertex[crossed dot]  (b) at ( \leglength,\leglength) {};
            \vertex[crossed dot]  (c) at (\leglength,- \leglength) {};
            \vertex[dot] (d) at (0,0.5) {};
            \vertex (t) at (-0.3,1.3);
    \diagram*{
        (a) -- [fermion] (d) -- [photon,out=155, in=210, min distance=0.4cm] (t) --[photon,out=30, in=75, min distance=0.4cm] (d) -- [fermion] (b) -- [fermion] (c)  -- [fermion] (a),
         %(l1) [momentum ={[arrowlabel]$q+p_2$}] (a),
         };
        \end{feynman}
    \end{tikzpicture}
    } \\
    \subfigure[$I_3$]{
    \begin{tikzpicture}[baseline = (m.base),arrowlabel/.style={
        /tikzfeynman/momentum/.cd, % means that the following keys are read from the /tikzfeynman/momentum family
        arrow shorten=#1,arrow distance=2.5mm
      },
      arrowlabel/.default=0.4]
        \def\leglength{1}
        \begin{feynman} [inline =(a.base) ]
            \vertex[crossed dot] (a) at (-\leglength,0) {};
            \vertex[crossed dot]  (b) at ( \leglength,\leglength) {};
            \vertex[crossed dot]  (c) at (\leglength,- \leglength) {};
            \vertex[dot] (d) at (0,0.5) {};
            \vertex[dot] (e) at (0,-0.5) {};
           
    \diagram*{
        (a) -- [fermion] (d) -- [fermion]  (b) -- [fermion] (c)  -- [fermion] (e) -- [fermion] (a),
        (d) -- [photon] (e),
         %(l1) [momentum ={[arrowlabel]$q+p_2$}] (a),
         };
        \end{feynman}
    \end{tikzpicture}
    }\hspace{1.5cm}
    \subfigure[$I_4$]{
    \begin{tikzpicture}[baseline = (m.base),arrowlabel/.style={
        /tikzfeynman/momentum/.cd, % means that the following keys are read from the /tikzfeynman/momentum family
        arrow shorten=#1,arrow distance=2.5mm
      },
      arrowlabel/.default=0.4]
        \def\leglength{1}
        \begin{feynman} [inline =(a.base) ]
            \vertex[crossed dot] (a) at (-\leglength,0) {};
            \vertex[crossed dot]  (b) at ( \leglength,\leglength) {};
            \vertex[crossed dot]  (c) at (\leglength,- \leglength) {};
            \vertex[dot] (d) at (0,0.5) {};
            \vertex (t1) at (-0.23,1);
            \vertex (t2) at (-0.4,1.5);
    \diagram*{
        (a) -- [fermion] (d) -- [photon] (t1),
        (d) -- [fermion] (b) -- [fermion] (c)  -- [fermion] (a),
        (t1) -- [fermion,out=155, in=210, min distance=0.2cm] (t2) --[fermion,out=30, in=75, min distance=0.2cm] (t1)
         %(l1) [momentum ={[arrowlabel]$q+p_2$}] (a),
         };
        \end{feynman}
    \end{tikzpicture}
    } \hspace{1.5cm}
    \subfigure[$I_5$]{
    \begin{tikzpicture}[baseline = (m.base),arrowlabel/.style={
        /tikzfeynman/momentum/.cd, % means that the following keys are read from the /tikzfeynman/momentum family
        arrow shorten=#1,arrow distance=2.5mm
      },
      arrowlabel/.default=0.4]
        \def\leglength{1}
        \begin{feynman} [inline =(a.base) ]
            \vertex[crossed dot] (a) at (-\leglength,0) {};
            \vertex[crossed dot]  (b) at ( \leglength,\leglength) {};
            \vertex[crossed dot]  (c) at (\leglength,- \leglength) {};
            \vertex[dot] (d1) at (-0.5,0.25) {};
            \vertex[dot] (d2) at (0.5,0.75) {};
            
    \diagram*{
        (a) -- [fermion] (d1),
        (d2) -- [fermion] (b) -- [fermion] (c)  -- [fermion] (a),
        (d1) -- [fermion,out=90, in=135, min distance=0.4cm] (d2),
        (d2) -- [photon,out=305, in=270, min distance=0.3cm] (d1)
         };
        \end{feynman}
    \end{tikzpicture}
    }\\
    \subfigure[$I_6$]{
    \begin{tikzpicture}[baseline = (m.base),arrowlabel/.style={
        /tikzfeynman/momentum/.cd, % means that the following keys are read from the /tikzfeynman/momentum family
        arrow shorten=#1,arrow distance=2.5mm
      },
      arrowlabel/.default=0.4]
        \def\leglength{1}
        \begin{feynman} [inline =(a.base) ]
            \vertex[crossed dot] (a) at (-\leglength,0) {};
            \vertex[crossed dot]  (b) at ( \leglength,\leglength) {};
            \vertex[crossed dot]  (c) at (\leglength,- \leglength) {};
            \vertex[blob] (d) at (0,0.5) {};
            \vertex (t1) at (-0.23,1);
            \vertex (t2) at (-0.4,1.5);
    \diagram*{
        (a) -- [fermion] (d), 
        (d) -- [fermion] (b) -- [fermion] (c)  -- [fermion] (a),
        
         %(l1) [momentum ={[arrowlabel]$q+p_2$}] (a),
         };
        \end{feynman}
    \end{tikzpicture}
    }\\
    \subfigure[$I_{7a}$]{
    \begin{tikzpicture}[baseline = (m.base),arrowlabel/.style={
        /tikzfeynman/momentum/.cd, % means that the following keys are read from the /tikzfeynman/momentum family
        arrow shorten=#1,arrow distance=2.5mm
      },
      arrowlabel/.default=0.4]
        \def\leglength{1}
        \begin{feynman} [inline =(a.base) ]
            \vertex[crossed dot] (a) at (-\leglength,0) {};
            \vertex[crossed dot]  (b) at ( \leglength,\leglength) {};
            \vertex[crossed dot]  (c) at (\leglength,- \leglength) {};
            \vertex[dot] (d) at (0.3,0.64) {};
           
    \diagram*{
        (a) -- [fermion] (d) -- [fermion]  (b) -- [fermion] (c)  -- [fermion] (a),
        (a) -- [photon, out=350, in=260, min distance=0.4cm ] (d),
         %(l1) [momentum ={[arrowlabel]$q+p_2$}] (a),
         };
        \end{feynman}
    \end{tikzpicture}
    } \hspace{1.5cm}
   \begin{comment} \subfigure[$I_{7b}$]{
    \begin{tikzpicture}[baseline = (m.base),arrowlabel/.style={
        /tikzfeynman/momentum/.cd, % means that the following keys are read from the /tikzfeynman/momentum family
        arrow shorten=#1,arrow distance=2.5mm
      },
      arrowlabel/.default=0.4]
        \def\leglength{1}
        \begin{feynman} [inline =(a.base) ]
            \vertex[crossed dot] (a) at (-\leglength,0) {};
            \vertex[crossed dot]  (b) at ( \leglength,\leglength) {};
            \vertex[crossed dot]  (c) at (\leglength,- \leglength) {};
            \vertex[dot] (d) at (-0.3,0.36) {};

    \diagram*{
        (a) -- [fermion] (d) -- [fermion]  (b) -- [fermion] (c)  -- [fermion] (a),
        (b) -- [photon, out=250, in=320, min distance=0.4cm ] (d),
         %(l1) [momentum ={[arrowlabel]$q+p_2$}] (a),
         };
        \end{feynman}
    \end{tikzpicture}
    }\hspace{1.5cm}
\end{comment}
\subfigure[$I_{7b}$]{
    \begin{tikzpicture}[baseline = (m.base),arrowlabel/.style={
        /tikzfeynman/momentum/.cd, % means that the following keys are read from the /tikzfeynman/momentum family
        arrow shorten=#1,arrow distance=2.5mm
      },
      arrowlabel/.default=0.4]
        \def\leglength{1}
        \begin{feynman} [inline =(a.base) ]
            \vertex[crossed dot] (a) at (-\leglength,0) {};
            \vertex[crossed dot]  (b) at ( \leglength,\leglength) {};
            \vertex[crossed dot]  (c) at (\leglength,- \leglength) {};
            \vertex[dot] (d) at (0,0.5) {};

    \diagram*{
        (a) -- [fermion] (d) -- [fermion]  (b) -- [fermion] (c)  -- [fermion] (a),
        (c) -- [photon] (d),
         %(l1) [momentum ={[arrowlabel]$q+p_2$}] (a),
         };
        \end{feynman}
    \end{tikzpicture} 
    }\\
    \subfigure[$I_{8}$]{
    \begin{tikzpicture}[baseline = (m.base),arrowlabel/.style={
        /tikzfeynman/momentum/.cd, % means that the following keys are read from the /tikzfeynman/momentum family
        arrow shorten=#1,arrow distance=2.5mm
      },
      arrowlabel/.default=0.4]
        \def\leglength{1}
        \begin{feynman} [inline =(a.base) ]
            \vertex[crossed dot] (a) at (-\leglength,0) {};
            \vertex[crossed dot]  (b) at ( \leglength,\leglength) {};
            \vertex[crossed dot]  (c) at (\leglength,- \leglength) {};
    
    \diagram*{
        (a) -- [fermion] (b) -- [fermion] (c) -- [fermion] (a),
        (a) -- [photon, out=350, in=250, min distance=0.4cm ] (b),
    };
        \end{feynman}
    \end{tikzpicture}
    }

    \caption{Two-loop diagrams contributing to $ \langle j_{\mu}^{a}(p_{1})j_{\nu}^{b}(-p_{2})j_{\rho}^{c}(-p_{3})\rangle$.}
    \label{figure_diagrams_two_loops}
\end{figure}

All possible diagrams contributing to the 3-point function at two loops are presented in \autoref{figure_diagrams_two_loops}.

\begin{comment}\hspace{1.5cm} 
    \subfigure[$I_{8b}$]{
    \begin{tikzpicture}[baseline = (m.base),arrowlabel/.style={
        /tikzfeynman/momentum/.cd, % means that the following keys are read from the /tikzfeynman/momentum family
        arrow shorten=#1,arrow distance=2.5mm
      },
      arrowlabel/.default=0.4]
        \def\leglength{1}
        \begin{feynman} [inline =(a.base) ]
            \vertex[crossed dot] (a) at (-\leglength,0) {};
            \vertex[crossed dot]  (b) at ( \leglength,\leglength) {};
            \vertex[crossed dot]  (c) at (\leglength,- \leglength) {};
    
    \diagram*{
        (a) -- [fermion] (b) -- [fermion] (c) -- [fermion] (a),
        (b) -- [photon, out=230, in=120, min distance=0.4cm ] (c),
    };
        \end{feynman}
    \end{tikzpicture}
    }
    \subfigure[$I_{8c}$]{
    \begin{tikzpicture}[baseline = (m.base),arrowlabel/.style={
        /tikzfeynman/momentum/.cd, % means that the following keys are read from the /tikzfeynman/momentum family
        arrow shorten=#1,arrow distance=2.5mm
      },
      arrowlabel/.default=0.4]
        \def\leglength{1}
        \begin{feynman} [inline =(a.base) ]
            \vertex[crossed dot] (a) at (-\leglength,0) {};
            \vertex[crossed dot]  (b) at ( \leglength,\leglength) {};
            \vertex[crossed dot]  (c) at (\leglength,- \leglength) {};
    
    \diagram*{
        (a) -- [fermion] (b) -- [fermion] (c) -- [fermion] (a),
        (c) -- [photon, out=110, in=0, min distance=0.4cm ] (a),
    };
        \end{feynman}
    \end{tikzpicture}
    } 
\end{comment}
\FloatBarrier

The only diagrams containing the quartic vertex 
\begin{equation}
    \text{Tr} -4\lambda\left|\Phi_{1}\times\Phi_{2}\right|^{2}
\end{equation}
are $I_1$ and $I_2$. For the first, the integral over the momenta running in the first loop factorizes,
\begin{equation}
    I_1 = \begin{tikzpicture}[baseline = (m.base),arrowlabel/.style={
        /tikzfeynman/momentum/.cd, % means that the following keys are read from the /tikzfeynman/momentum family
        arrow shorten=#1,arrow distance=2.5mm
      },
      arrowlabel/.default=0.4]
        \def\leglength{1}
        \begin{feynman} [inline =(a.base) ]
            \vertex[crossed dot] (l1) at (-2,0){};
            \vertex[dot] (a) at (-\leglength,0) {};
            \vertex[crossed dot]  (b) at ( \leglength,\leglength) {};
            \vertex[crossed dot]  (c) at (\leglength,- \leglength) {};
    
    \diagram*{
        (l1) -- [fermion,half left,momentum={[arrowlabel]$p_1+k$}] (a)
          -- [fermion,half left,momentum={[arrowlabel]$k$}] (l1),
        (a) -- [fermion] (b) -- [fermion] (c) -- [fermion] (a),
    };
        \end{feynman}
    \end{tikzpicture} \sim \int \frac{d^3k}{(2\pi)^3} \frac{(2k + p_1)_\mu}{k^2 (k+p_1)^2 } \int  \frac{d^3q}{(2\pi)^3} \frac{(2q+p_2)_\nu (2q-p_3)_\rho }{q^2(q+p_2)^2(q-p_3)^2},
\end{equation}
where by Lorentz invariance, we have 
\begin{align}
    \int\frac{d^{3}k}{(2\pi)^{3}}\frac{(2k+p_{1})_{\mu}}{k^{2}(k+p_{1})^{2}} & = 2\int\frac{d^{3}k}{(2\pi)^{3}}\frac{k_{\mu}}{k^{2}(k+p_{1})^{2}}+p_{1\mu}\int\frac{d^{3}k}{(2\pi)^{3}}\frac{1}{k^{2}(k+p_{1})^{2}} \nonumber \\
    & = 2\left(-\frac{p_{1\mu}}{2}\int\frac{d^{3}k}{(2\pi)^{3}}\frac{1}{k^{2}(k+p_{1})^{2}}\right)+p_{1\mu}\int\frac{d^{3}k}{(2\pi)^{3}}\frac{1}{k^{2}(k+p_{1})^{2}} \nonumber \\
    & = 0.
\end{align}

The integral for $I_2$ do not contribute, since the integral over the momentum running in the loop also factorize and 
vanishes in dimensional regularization,
\begin{equation}
    I_2 \quad = \quad \begin{tikzpicture}[baseline = (m.base),arrowlabel/.style={
        /tikzfeynman/momentum/.cd, % means that the following keys are read from the /tikzfeynman/momentum family
        arrow shorten=#1,arrow distance=2.5mm
      },
      arrowlabel/.default=0.4]
        \def\leglength{1}
        \begin{feynman} [inline =(a.base) ]
            \vertex[crossed dot] (a) at (-\leglength,0) {};
            \vertex[crossed dot]  (b) at ( \leglength,\leglength) {};
            \vertex[crossed dot]  (c) at (\leglength,- \leglength) {};
            \vertex[dot] (d) at (0,0.5) {};
            \vertex (t) at (-0.3,1.3);
    \diagram*{
        (a) -- [fermion] (d) -- [fermion,out=155, in=210, min distance=0.4cm,momentum={[arrowlabel]$k$}] (t) --[fermion,out=30, in=75, min distance=0.4cm,momentum={[arrowlabel]$k$}] (d) -- [fermion] (b) -- [fermion] (c)  -- [fermion] (a),
         %(l1) [momentum ={[arrowlabel]$q+p_2$}] (a),
         };
        \end{feynman}
    \end{tikzpicture}\quad \sim \quad \int d^3k \frac{1}{k^2} = 0.
\end{equation}

Therefore, all diagrams containing the quartic interaction vanish in dimensional regularization, and the two-loop result 
is independent of the explicit form of the potential. This means that the general results obtained using the toy model 
\eqref{toy_model_horatiu} can be generalized to any class of phenomenological models \eqref{phenomenological_models} with 
only bosonic fields. The quartic potential is only necessary to have vortex solutions to solve the monopole problem.

The same argument is valid for $I_{2\prime}$ and $I_4$: both vanish in dimensional regularization. The diagram $I_6$ is 
the counterm-diagram, but since the 3-point function is one-loop finite, it can be set to zero. The non-zero diagrams are 
$I_3$, $I_5$, $I_7$, and $I_8$. The integral expression for these diagrams are given below for future reference.
\begin{itemize}
    \item[-] $I_3$
\end{itemize}
    \begin{equation}
        (I_3)_{\mu \nu \rho}^{abc} =  \begin{tikzpicture}[baseline = (m.base),arrowlabel/.style={
            /tikzfeynman/momentum/.cd, % means that the following keys are read from the /tikzfeynman/momentum family
            arrow shorten=#1,arrow distance=2.5mm
          },
          arrowlabel/.default=0.4]
            \def\leglength{2}
            \begin{feynman} [inline =(a.base) ]
                \vertex[crossed dot,label = left:{$a,\mu$}] (a) at (-\leglength,0) {};
                \vertex[crossed dot,label = right:{$b,\nu$}]  (b) at ( \leglength,\leglength) {};
                \vertex[crossed dot,label = right:{$c,\rho$}]  (c) at (\leglength,- \leglength) {};
                \vertex[dot,label = above:{$\lambda_1$}] (d) at (0,1) {};
                \vertex[dot,label = right:{$\lambda_2$}] (e) at (0,-1) {};
               
        \diagram*{
            (a) -- [fermion,momentum ={[arrowlabel]$q+p_2$}] (d) -- [fermion, momentum ={[arrowlabel]$r+p_2$}]  (b) -- [fermion,momentum ={[arrowlabel]$r$}] (c)  -- [fermion,momentum ={[arrowlabel]$r-p_3$}] (e) -- [fermion,momentum ={[arrowlabel]$q-p_3$}] (a),
            (d) -- [photon,momentum ={[arrowlabel]$q-r$}] (e),
             %(l1) [momentum ={[arrowlabel]$q+p_2$}] (a),
             };
            \end{feynman}
        \end{tikzpicture}\nonumber
    \end{equation}

    \begin{equation}
        = 2g^2\epsilon^{abc}  \int\frac{d^{d}q}{(2\pi)^{d}}\int\frac{d^{d}r}{(2\pi)^{d}}\frac{\left(2q-p_{3}+p_{2}\right)_{\mu}\left(2r+p_{2}\right)_{\nu}\left(2r-p_{3}\right)_{\rho}\left(2p_{2}+q+r\right)_{\lambda_{1}}\delta_{\lambda_{1}\lambda_{2}}\left(-2p_{3}+q+r\right)_{\lambda_{2}}}{\left(q+p_{2}\right)^{2}\left(r+p_{2}\right)^{2}r^{2}\left(r-p_{3}\right)^{2}\left(q-p_{3}\right)^{2}\left(q-r\right)^{2}}.
    \end{equation}

The integral expression for the variations of this diagram (obtained by connecting different propagators with a gauge propagator) 
can be obtained by just rotating this diagram and relabeling the momenta in each propagator. We will follow this approach for the 
next non-zero diagrams, writing just the integral for the first variation, since all the other possibilities can be easily obtained by 
rotations and reflections of the diagrams.

\begin{itemize}
    \item[-] $I_5$
\end{itemize}
    \begin{equation}
       (I_5)_{\mu \nu \rho}^{abc} =  \begin{tikzpicture}[baseline = (m.base),arrowlabel/.style={
            /tikzfeynman/momentum/.cd, % means that the following keys are read from the /tikzfeynman/momentum family
            arrow shorten=#1,arrow distance=2.5mm
          },
          arrowlabel/.default=0.4]
            \def\leglength{2}
            \begin{feynman} [inline =(a.base) ]
                \vertex[crossed dot,label = left:{$a,\mu$}] (a) at (-\leglength,0) {};
                \vertex[crossed dot,label = right:{$b,\nu$}]  (b) at ( \leglength,\leglength) {};
                \vertex[crossed dot,label = right:{$c,\rho$}]  (c) at (\leglength,- \leglength) {};
                \vertex[dot] (d1) at (-1,0.5) {};
                \vertex[dot] (d2) at (1,1.5) {};
                
        \diagram*{
            (a) -- [fermion,momentum ={[arrowlabel]$q+p_2$}] (d1),
            (d2) -- [fermion,momentum ={[arrowlabel]$q+p_2$}] (b) -- [fermion,momentum ={[arrowlabel]$q$}] (c)  -- [fermion,momentum ={[arrowlabel]$q-p_3$}] (a),
            (d1) -- [fermion,out=90, in=135, min distance=0.4cm,momentum ={[arrowlabel]$q+r+p_2$}] (d2),
            (d2) -- [photon,out=305, in=270, min distance=0.3cm,momentum ={[arrowlabel]$r$}] (d1)
             };
            \end{feynman}
        \end{tikzpicture} \nonumber 
    \end{equation}

    \begin{equation}
        = 2\epsilon^{abc}\int\frac{d^{d}q}{(2\pi)^{d}}\int\frac{d^{d}r}{(2\pi)^{d}}\frac{\left(2q+p_{2}-p_{3}\right)_{\mu}\left(2q+p_{2}\right)_{\nu}\left(2q-p_{3}\right)_{\rho}\left(2q+2p_{2}+r\right)_{\lambda_{1}}\left(2q+2p_{2}+r\right)_{\lambda_{2}}}{\left(q+p_{2}\right)^{4}\left(q-p_{3}\right)^{2}\left(p_{2}+q+r\right)^{2}q^{2}}. \nonumber 
    \end{equation}

    \begin{itemize}
        \item[-] $I_{7a}$
    \end{itemize}
        \begin{equation*}
            (I_{7a})_{\mu \nu \rho}^{abc} = \begin{tikzpicture}[baseline = (m.base),arrowlabel/.style={
                /tikzfeynman/momentum/.cd, % means that the following keys are read from the /tikzfeynman/momentum family
                arrow shorten=#1,arrow distance=2.5mm
              },
              arrowlabel/.default=0.4]
                \def\leglength{2}
                \begin{feynman} [inline =(a.base) ]
                    \vertex[crossed dot,label = left:{$a,\mu$}] (a) at (-\leglength,0) {};
                    \vertex[crossed dot,label = right:{$b,\nu$}]  (b) at ( \leglength,\leglength) {};
                    \vertex[crossed dot,label = right:{$c,\rho$}]  (c) at (\leglength,- \leglength) {};
                    \vertex[dot] (d) at (0.6,1.28) {};
                   
            \diagram*{
                (a) -- [fermion, momentum ={[arrowlabel]$q + r + p_2$}] (d) -- [fermion, momentum ={[arrowlabel]$q + p_2$}]  (b) -- [fermion, momentum ={[arrowlabel]$q$}] (c)  -- [fermion, momentum ={[arrowlabel]$q - p_3$}] (a),
                (a) -- [photon, out=350, in=260, min distance=0.4cm, reversed momentum' ={[arrowlabel]$r$} ] (d),
                 
                 };
                \end{feynman}
            \end{tikzpicture}
        \end{equation*}
    
    \begin{equation}
        =-4g^{2}\epsilon^{abc}\int\frac{d^{d}q}{(2\pi)^{d}}\int\frac{d^{d}r}{(2\pi)^{d}}\frac{\eta_{\mu\lambda_{1}}\delta_{\lambda_{1}\lambda_{2}}\left(2q+2p_{2}+r\right)_{\lambda_{2}}\left(2q+p_{2}\right)_{\nu}\left(2q-p_{3}\right)_{\rho}}{\left(q+p_{2}+r\right)^{2}\left(q+p_{2}\right)^{2}\left(q-p_{3}\right)^{2}q^{2}r^{2}}.
    \end{equation}
    
\begin{itemize}
    \item[-] $I_{7b}$
\end{itemize}
    
\begin{equation}
    (I_{7b})_{\mu \nu \rho}^{abc}  =  \begin{tikzpicture}[baseline = (m.base),arrowlabel/.style={
         /tikzfeynman/momentum/.cd, % means that the following keys are read from the /tikzfeynman/momentum family
            arrow shorten=#1,arrow distance=2.5mm
        },
          arrowlabel/.default=0.4]
    \def\leglength{2}
    \begin{feynman} [inline =(a.base) ]
        \vertex[crossed dot,label = left:{$a,\mu$}] (a) at (-\leglength,0) {};
            \vertex[crossed dot,label = right:{$b,\nu$}]  (b) at ( \leglength,\leglength) {};
            \vertex[crossed dot,label = right:{$c,\rho$}]  (c) at (\leglength,- \leglength) {};
        \vertex[dot] (d) at (0,1) {};

        \diagram*{
            (a) -- [fermion, momentum ={[arrowlabel]$q + p_2$}] (d) -- [fermion, momentum ={[arrowlabel]$q + p_2 - r$}]  (b) -- [fermion, momentum ={[arrowlabel]$q -r$}] (c)  -- [fermion,  momentum ={[arrowlabel]$q - p_3$}] (a),
            (c) -- [photon, reversed  momentum ={[arrowlabel]$r$}] (d),
             %(l1) [momentum ={[arrowlabel]$q+p_2$}] (a),
             };
        \end{feynman}
    \end{tikzpicture} 
\end{equation}
\begin{equation}
   =-4g^{2}\epsilon^{abc}\int\frac{d^{d}q}{(2\pi)^{d}}\int\frac{d^{d}r}{(2\pi)^{d}}\frac{\left(2q+p_{2}-p_{3}\right)_{\mu}\left(2q-2r+p_{2}\right)_{\nu}\left(2q+2p_{2}-r\right)_{\rho}}{\left(q+p_{2}\right)^{2}\left(q+p_{2}+r\right)^{2}\left(q-r\right)^{2}\left(q-p_{3}\right)^{2}r^{2}}.
\end{equation}

\begin{itemize}
    \item[-] $I_{8}$
\end{itemize}

\begin{equation}
    (I_8)_{\mu \nu \rho}^{abc}\begin{tikzpicture}[baseline = (m.base),arrowlabel/.style={
        /tikzfeynman/momentum/.cd, % means that the following keys are read from the /tikzfeynman/momentum family
            arrow shorten=#1,arrow distance=2.5mm}, arrowlabel/.default=0.4]
    \def\leglength{2}
        \begin{feynman} [inline =(a.base) ]
            \vertex[crossed dot,label = left:{$a,\mu$}] (a) at (-\leglength,0) {};
            \vertex[crossed dot,label = right:{$b,\nu$}]  (b) at ( \leglength,\leglength) {};
            \vertex[crossed dot,label = right:{$c,\rho$}]  (c) at (\leglength,- \leglength) {};
        
            \diagram*{
            (a) -- [fermion,momentum ={[arrowlabel]$q+r+p_2$}] (b) -- [fermion,momentum ={[arrowlabel]$q$}] (c) -- [fermion,momentum ={[arrowlabel]$q-p_3$}] (a),
            (a) -- [photon, out=350, in=250, min distance=0.4cm,reversed momentum ={[arrowlabel]$r$} ] (b),
            };
        \end{feynman}
    \end{tikzpicture}
\end{equation}
\begin{equation}
    =8g^{2}2\epsilon^{abc}\eta_{\mu\nu}\int\frac{d^{d}q}{(2\pi)^{d}}\int\frac{d^{d}r}{(2\pi)^{d}}\frac{\left(2q-p_{3}\right)_{\rho}}{\left(q+p_{2}+r\right)^{2}\left(q-p_{3}\right)^{2}q^{2}r^{2}}.
\end{equation}

\section{Discussion and conclusion}

In this paper we have computed the one-loop 3-point function for the current operator in a toy model with global $SO(3)$ symmetry
for the phenomenological approach to holographic cosmology. This toy model was used  previously in order 
to solve the monopole problem in holographic cosmology. 

Since the result can be used for quantum field theory reasons as well, and as a test for our understanding, we started with 
two particular cases, of two lightlike momenta ($p_1^2=p_3^2=0$), and one lightlike momentum ($p_1^2=0$). 
In these cases, we have found IR divergences, which were treated using mass regularization. The results were checked to 
satisfy the transverse Ward identities. Finally, the calculation at generic external momenta was done and found to be (UV and) IR 
finite, and also respect the transverse Ward identities. 

When applying the result to holographic cosmology, applying an S-duality on the 3-point function of $SO(3)$ (electric) currents 
gives the 3-point function of magnetic currents, giving, through the holographic map, 
the monopole distribution 3-point function, specifically 
its non-Gaussianity. We took the special limit, 
previously used in inflation for scalar perturbations, $p_{2\mu}\simeq -p_{3\mu}$, $p_2^2\simeq p_3^2\gg p_1^2$, and 
found that the 3-point function reduces to a combination of the two-point functions and the anomalous dimension $\delta(\tilde j)
=n_m-1$, where $n_m$ is the monopole analog of $n_s$ for scalar fluctuations in inflation. The IR finiteness of the 
quantum field theory 3-point function is mapped to the absence of cosmological singularity for the monopole 3-point function. 

%After reviewing the phenomenological holographic approach to cosmology, we present the results for $\langle j_{\mu}^{a}(p_{2}+p_{3})j_{\nu}^{b}(-p_{2})j_{\rho}^{c}(-p_{3})\rangle$ at one-loop for two particular cases first: with $p_1^2=p_3^2 = 0$ and only $p_1^2=0$. In both cases, the result is IR divergent, which is a reflex of the choice of external light-like momenta. We have treated the divergences using mass regularization. We highlight that the 3-point function for the second case satisfies the transverse Ward identity when contracted with $p_1^\mu$ since all divergences are proportional to $p_{1\mu}$ and $p_1^2=0$ in this case.

%After dealing with these particular cases, we performed the full calculation. The final result is UV and IR finite and also satisfies the transverse Ward identity.

The two-loop contribution would be needed both to find a nontrivial $n_m$ ($\neq 1$) and to better test the IR finiteness (absence
of cosmological singularity). We left that for future work, though we set up the problem. Moreover,  
by analyzing the two-loop diagrams, we conclude that the two-loop result for the 3-point function is independent of the form 
of the scalar potential since all diagrams with insertions of the quartic scalar interaction vertex vanish in dimensional regularization. 
This was already true for the 2-point function, where it led to the conclusion that the fact that the 
magnetic current operator is relevant 
was generic, and not restricted to the toy model. We can now extend this finding to the 3-point function and say that the 
3-point monopole distribution profile, and the resulting monopole non-Gaussianity, 
is independent of the quartic scalar interaction, thus is generic, and not restricted to the toy model.

For the future, it would be good to do the full calculation of the two-loop 3-point function, as well as introduce fermions in the model, 
to make the result more general. The methods used to calculate the Feynman diagram integrals 
in these 3-dimensional field theories are interesting in their own right, and could be used in the future in other cases.

It will also be interesting to calculate the $\langle T J J\rangle $\footnote{See also the previous calculation 
of a $\langle T J J \rangle $ correlator, though not within this particular context, in \cite{Bzowski:2013sza,Bzowski:2017poo,Armillis:2009pq}.} 
and the $\langle T T J\rangle$ correlators. Since $T_{\mu\nu}$ couples both to scalar and tensor perturbations in cosmology, in practice this 
should give a cross-correlation between primordial perturbations of the metric and monopoles, which can be considerably larger than the 
$\langle JJJ\rangle$ correlator computed here. 

%Therefore, this work extends the conclusion that the results obtained using toy models in holographic cosmology are very general: they are valid for any purely bosonic phenomenological model with $SO(3)$ symmetry and potential allowing vortex solutions, dual to monopole configurations in cosmology. It would be nice to also study how the 2 and 3-point function change under the addition of fermions in the toy model, to check if the solution to the monopole problem can be extended in this case. The problem of connecting this result to the 3-point function of the monopole field in cosmology is still open, and it will be addressed in the future.

\section*{Acknowledgements}

We would like to thank Kostas Skenderis for discussions.
The work of HN is supported in part by  CNPq grant 301491/2019-4 and FAPESP grant 2019/21281-4.
 HN would also like to thank the ICTP-SAIFR for their support through FAPESP grant 2021/14335-0.
The work of MC is supported by FAPESP grant 2022/02791-4.

\appendix

\section{Calculation of the 1-loop integrals in momentum space} 
\label{appendix_integrals_in_momentum_space}

In this appendix we present the details of the calculation of the integrals \eqref{set_of_integrals_1} and 
\eqref{set_of_integrals_2} in dimensional regularization using the Feynman parametrization.
\begin{itemize}
    \item[-] Integral $I_0$
\end{itemize}

\begin{equation}
    I_{0} = 2\int_{0}^{1}dx\int_{0}^{1-x}dy\int\frac{d^{d}q}{(2\pi)^{d}}\frac{1}{\mathcal{D}^{3}},
\end{equation}
where
\begin{eqnarray}
    \mathcal{D} &=& xq^{2}+y(q+p_{2})^{2}+(1-x-y)(q-p_{3})^{2} \nonumber \\
    &=& q^{2}+2q\cdot\left[yp_{2}-(1-x-y)p_{3}\right]+yp_{2}^{2}+(1-x-y)p_{3}^{2}.
\end{eqnarray}

Changing variables:
\begin{equation} 
    \bar{q}=q+yp_{2}-(1-x-y)p_{3},
\end{equation} 
we eliminate terms with $\bar q \cdot p_i$,
\begin{equation}
    \mathcal{D} = \bar{q}^{2}+(1-x)xp_{3}^{2}+(p_{2}+p_{3})\cdot(p_{2}+p_{3}-2xp_{3})y-(p_{2}+p_{3})^{2}y^{2},
\end{equation}
and using the conservation of momenta, we can simplify the denominator,
\begin{eqnarray}
    \mathcal{D}	&=& \bar{q}^{2}+(1-x)xp_{3}^{2}+p_{1}\cdot(p_{1}-2xp_{3})y-p_{1}^{2}y^{2} \nonumber \\
	&=& \bar{q}^{2}+(1-x)xp_{3}^{2}+p_{1}^{2}y-p_{1}^{2}y^{2}-2xyp_{1}\cdot p_{3} \nonumber \\
	&=& \bar{q}^{2}+xyp_2^2 + (1-x-y)(xp_3^2 + yp_1^2) \nonumber \\
	&=& \bar{q}^{2}+\Delta^{2},
\end{eqnarray}
where 
\begin{equation}
    \Delta^2 \equiv xyp_2^2 + (1-x-y)(xp_3^2 + yp_1^2). 
\end{equation}

Now we can use the general result in dimensional regularization
\begin{equation}
\int\frac{d^{d}q}{(2\pi)^{d}}\frac{1}{(q^{2}+\Delta^{2})^{n}}=\frac{\Gamma(n-d/2)}{(4\pi)^{d/2}\Gamma(n)}(\Delta^{2})^{d/2-n}
\end{equation}
to write
\begin{equation}
    I_{0} = \frac{\Gamma(3-d/2)}{(4\pi)^{d/2}\Gamma(3)}2\int_{0}^{1}dx\int_{0}^{1-x}dy\left[xyp_2^2 
    + (1-x-y)(xp_3^2 + yp_1^2)\right]^{d/2-3} ,
\end{equation}
and since there are no poles in the gamma function for $d=3$, we can set $d=3$ everywhere,
\begin{eqnarray}
    I_{0} &=& 2\frac{\Gamma(3-3/2)}{(4\pi)^{3/2}\Gamma(3)}\int_{0}^{1}dx\int_{0}^{1-x}dy\left[xyp_2^2 
    + (1-x-y)(xp_3^2 + yp_1^2)\right]^{3/2-3}\nonumber \\
    &=& \frac{1}{(4\pi)^{3/2}}\frac{\sqrt{\pi}}{2}\int_{0}^{1}dx\int_{0}^{1-x}dy\frac{1}{\left[xyp_2^2 
    + (1-x-y)(xp_3^2 + yp_1^2)\right]^{3/2}}\nonumber \\
    &=& \frac{1}{16\pi}I^{(0,0,3/2)}(p_{1,}p_{2},p_{3}),
\end{eqnarray}
where
\begin{equation}
    I^{(0,0,3/2)}(p_{1,}p_{2},p_{3}) \equiv \int_{0}^{1}dx\int_{0}^{1-x}dy\frac{1}{\left[xyp_2^2 + (1-x-y)(xp_3^2 + yp_1^2)\right]^{3/2}}.
\end{equation}

\begin{itemize}
    \item[-] Integral $I_\mu$ 
\end{itemize}

\begin{eqnarray}
    I_{\mu} &=& \int\frac{d^{d}q}{(2\pi)^{d}}\frac{q_{\mu}}{q^{2}(q+p_{2})^{2}(q-p_{3})^{2}} \nonumber \\
    &=& 2\int_{0}^{1}dx\int_{0}^{1-x}dy\int\frac{d^{d}q}{(2\pi)^{d}}\frac{q_{\mu}}{\left[xq^{2}
    +y(q+p_{2})^{2}+(1-x-y)(q-p_{3})^{2}\right]^{3}} \nonumber \\
    &=& 2\int_{0}^{1}dx\int_{0}^{1-x}dy\int\frac{d^{d}\bar{q}}{(2\pi)^{d}}\frac{\bar{q}_{\mu}
    +(1-x)p_{3\mu}-yp_{1\mu}}{\left[\bar{q}^{2}+\Delta^{2}\right]^{3}} \nonumber \\
    &=& 2\int_{0}^{1}dx\int_{0}^{1-x}dy\left[\int\frac{d^{d}\bar{q}}{(2\pi)^{d}}\frac{\bar{q}_{\mu}}
    {\left[\bar{q}^{2}+\Delta^{2}\right]^{3}}+\left[p_{3\mu}(1-x)-yp_{1\mu}\right]\int\frac{d^{d}\bar{q}}
    {(2\pi)^{d}}\frac{1}{\left[\bar{q}^{2}+\Delta^{2}\right]^{3}}\right] \nonumber \\
    &=& 2\frac{\Gamma(3-d/2)}{(4\pi)^{d/2}\Gamma(3)}\int_{0}^{1}dx\int_{0}^{1-x}dy
    \frac{p_{3\mu}(1-x)-yp_{1\mu}}{\left[xyp_2^2 + (1-x-y)(xp_3^2 + yp_1^2)\right]^{3-d/2}} \nonumber \\
    &=& \frac{1}{16\pi}f_{\mu}(p_{1},p_{2},p_{3}),
\end{eqnarray}
where 
\begin{equation}
    \int \frac{d^d \bar q}{(2pi)^d} \frac{\bar q_\mu}{(\bar q^2 + \Delta^2)^3} = 0
\end{equation}
by Lorentz invariance, and 
\begin{equation}
    f_{\mu}(p_{1},p_{2},p_{3}) \equiv \int_{0}^{1}dx\int_{0}^{1-x}dy\frac{p_{3\mu}(1-x)-yp_{1\mu}}{\left[xyp_2^2 
    + (1-x-y)(xp_3^2 + yp_1^2)\right]^{3/2}}.
\end{equation}

\begin{itemize}
    \item[-] Integral $I_{\mu \nu}$ 
\end{itemize}

\begin{eqnarray}
    I_{\mu\nu} &=& \int\frac{d^{d}q}{(2\pi)^{d}}\frac{q_{\mu}q_{\nu}}{q^{2}(q+p_{2})^{2}(q-p_{3})^{2}}\nonumber \\
    &=& 2\int_{0}^{1}dx\int_{0}^{1-x}dy\int\frac{d^{d}q}{(2\pi)^{d}}\frac{q_{\mu}q_{\nu}}{\left[xq^{2}+y(q+p_{2})^{2}
    +(1-x-y)(q-p_{3})^{2}\right]^{3}}\nonumber \\
    &=& 2\int_{0}^{1}dx\int_{0}^{1-x}dy\int\frac{d^{d}\bar{q}}{(2\pi)^{d}}\frac{\left[\bar{q}-yp_{2}
    +(1-x-y)p_{3}\right]_{\mu}\left[\bar{q}-yp_{2}+(1-x-y)p_{3}\right]_{\nu}}{\left(\bar{q}^{2}+\Delta^{2}\right)^{3}}. \nonumber \\
\end{eqnarray}

Expanding the numerator,
\begin{eqnarray}
    && \left[\bar{q}_{\mu}-yp_{2\mu}+(1-x-y)p_{3\mu}\right]\left[\bar{q}_{\nu}-yp_{2\nu}+(1-x-y)p_{3\nu}\right] =  \nonumber \\
	&=& \bar{q}_{\mu}\bar{q}_{\nu}+\bar{q}_{\mu}\left[(1-x-y)p_{3\nu}-yp_{2\nu}\right]
	+\bar{q}_{\nu}\left[(1-x-y)p_{3\mu}-yp_{2\mu}\right] \nonumber \\
	&& +\left[(1-x-y)p_{3\mu}-yp_{2\mu}\right]\left[(1-x-y)p_{3\nu}-yp_{2\nu}\right].
\end{eqnarray}

But since terms with only one $\bar q$ vanish, we can already drop them from the calculation,
\begin{eqnarray}
    I_{\mu\nu}&=& 2\int_{0}^{1}dx\int_{0}^{1-x}dy\left\{ \left[(1-x-y)^{2}p_{3\mu}p_{3\nu}
    +y^{2}p_{2\mu}p_{2\nu}-y(1-x-y)\left(p_{3\mu}p_{2\nu}+p_{2\mu}p_{3\nu}\right)\right]\right. \nonumber \\
    && \left.\times \int\frac{d^{d}\bar{q}}{(2\pi)^{d}}\frac{1}{\left(\bar{q}^{2}+\Delta^{2}\right)^{3}} 
    +\int\frac{d^{d}\bar{q}}{(2\pi)^{d}}\frac{\bar{q}_{\mu}\bar{q}_{\nu}}{\left(\bar{q}^{2}+\Delta^{2} \right)^{3}}\right\} \nonumber \\
    &=& 2\int_{0}^{1}dx\int_{0}^{1-x}dy \nonumber \\ 
    && \left\{ \frac{\Gamma(3-d/2)}{(4\pi)^{d/2}\Gamma(3)}\frac{(1-x-y)^{2}p_{3\mu}p_{3\nu}
    +y^{2}p_{2\mu}p_{2\nu}-y(1-x-y)\left(p_{3\mu}p_{2\nu}+p_{2\mu}p_{3\nu}\right)}{\left[xyp_2^2 
    + (1-x-y)(xp_3^2 + yp_1^2)\right]^{3-d/2}}\right. \nonumber \\
    && \left.+\int\frac{d^{d}\bar{q}}{(2\pi)^{d}}\frac{\bar{q}_{\mu}\bar{q}_{\nu}}{\left(\bar{q}^{2}+\Delta^{2}\right)^{3}}\right\}.
\end{eqnarray}

By Lorentz invariance, the last integral is
\begin{equation}
    \int\frac{d^{d}\bar{q}}{(2\pi)^{d}}\frac{\bar{q}_{\mu}\bar{q}_{\nu}}{\left(\bar{q}^{2}+\Delta^{2}\right)^{3}} 
    = \frac{\delta_{\mu\nu}}{2(4-d)}\frac{\Gamma(3-d/2)}{(4\pi)^{d/2}}\left(\Delta^{2}\right)^{d/2-2}, 
    \label{result_qmu_qnu_integral_lorenzt_invariance}
\end{equation}
and we have 
\begin{eqnarray}
    I_{\mu\nu}  &=& 2\int_{0}^{1}dx\int_{0}^{1-x}dy \nonumber \\
    && \times \left\{ \frac{\Gamma(3-d/2)}{(4\pi)^{d/2}\Gamma(3)}\frac{(1-x-y)^{2}p_{3\mu}p_{3\nu}+y^{2}p_{2\mu}p_{2\nu}-y(1-x-y)
    \left(p_{3\mu}p_{2\nu}+p_{2\mu}p_{3\nu}\right)}{\left[xyp_2^2 + (1-x-y)(xp_3^2 + yp_1^2)\right]^{3-d/2}}\right. \nonumber \\
    && \left.+\frac{\Gamma(3-d/2)}{(4\pi)^{d/2}\Gamma(3)}\frac{\Gamma(3)}{2(4-d)}\frac{\delta_{\mu\nu}}{\left[xyp_2^2 + (1-x-y)
    (xp_3^2 + yp_1^2)\right]^{2-d/2}}\right\}  \nonumber \\ 
    &=& \frac{1}{16\pi}\left[f_{\mu\nu}(p_{1},p_{2},p_{3})+\delta_{\mu\nu}f_{0}^{\prime}(p_{1},p_{2},p_{3})\right],
\end{eqnarray}
where 
\begin{equation}
    f_{\mu\nu}(p_{1},p_{2},p_{3}) \equiv \int_{0}^{1}dx\int_{0}^{1-x}dy\frac{\left[(1-x)p_{3\mu}-yp_{1\mu}\right]\left[(1-x)
    p_{3\nu}-yp_{1\nu}\right]}{\left[xyp_2^2 + (1-x-y)(xp_3^2 + yp_1^2)\right]^{3/2}},
\end{equation}
\begin{equation}
    f_{0}^{\prime}(p_{1},p_{2},p_{3})\equiv\int_{0}^{1}dx\int_{0}^{1-x}dy\frac{1}{\left[xyp_2^2 + (1-x-y)(xp_3^2 + yp_1^2)\right]^{1/2}}.
\end{equation}

\begin{itemize}
    \item[-] Integral $I_{\mu \nu \rho}$
\end{itemize}

\begin{eqnarray}
    I_{\mu\nu \rho} &=& \int\frac{d^{d}q}{(2\pi)^{d}}\frac{q_{\mu}q_{\nu}q_{\rho}}{q^{2}(q+p_{2})^{2}(q-p_{3})^{2}} \nonumber\\
    &=& 2\int_{0}^{1}dx\int_{0}^{1-x}dy\int\frac{d^{d}q}{(2\pi)^{d}}\frac{q_{\mu}q_{\nu}q_{\rho}}{\left[xq^{2}+y(q+p_{2})^{2}
    +(1-x-y)(q-p_{3})^{2}\right]^{3}}\nonumber \\
    &=& 2\int_{0}^{1}dx\int_{0}^{1-x}dy\int\frac{d^{d}\bar{q}}{(2\pi)^{d}}\frac{1}{\left(\bar{q}^{2}+\Delta^{2}\right)^{3}} \nonumber \\
    && \times \left[\bar{q}-yp_{2}+(1-x-y)p_{3}\right]_{\mu}\left[\bar{q}-yp_{2}+(1-x-y)p_{3}\right]_{\nu}\left[\bar{q}-yp_{2}
    +(1-x-y)p_{3}\right]_{\rho}. \nonumber \\
\end{eqnarray}

Expanding the numerator 
\begin{eqnarray}
    && \left[\bar{q}-yp_{2}+(1-x-y)p_{3}\right]_{\mu}\left[\bar{q}-yp_{2}+(1-x-y)p_{3}\right]_{\nu}\left[\bar{q}-yp_{2}
    +(1-x-y)p_{3}\right]_{\rho}\nonumber \\
    &=&\bar{q}_{\mu}\bar{q}_{\nu}\bar{q}_{\rho}+\bar{q}_{\mu}\bar{q}_{\nu}\left[(1-x-y)p_{3\rho}-yp_{2\rho}\right]
    +\left[(1-x-y)p_{3\nu}-yp_{2\nu}\right]\bar{q}_{\mu}\bar{q}_{\rho} \nonumber \\
    &&+\bar{q}_{\mu}\left[(1-x-y)p_{3\nu}-yp_{2\nu}\right]\left[(1-x-y)p_{3\rho}-yp_{2\rho}\right] \nonumber \\
    &&+\left[(1-x-y)p_{3\mu}-yp_{2\mu}\right]\bar{q}_{\nu}\bar{q}_{\rho}+\bar{q}_{\nu}\left[(1-x-y)p_{3\mu}
    -yp_{2\mu}\right]\left[(1-x-y)p_{3\rho}-yp_{2\rho}\right] \nonumber \\
    &&+\left[(1-x-y)p_{3\mu}-yp_{2\mu}\right]\left[(1-x-y)p_{3\nu}-yp_{2\nu}\right]\bar{q}_{\rho} \nonumber \\
    &&+\left[(1-x-y)p_{3\mu}-yp_{2\mu}\right]\left[(1-x-y)p_{3\nu}-yp_{2\nu}\right]\left[(1-x-y)p_{3\rho}-yp_{2\rho}\right].
\end{eqnarray}

Again, terms with an odd number of $\bar q$s will vanish by Lorentz invariance, so we just drop them. Therefore
\begin{eqnarray}
    I_{\mu\nu\rho} &=& 2\int_{0}^{1}dx\int_{0}^{1-x}dy\left\{ \left[(1-x-y)p_{3\rho}-yp_{2\rho}\right]\int\frac{d^{d}\bar{q}}{(2\pi)^{d}}
    \frac{\bar{q}_{\mu}\bar{q}_{\nu}}{\left(\bar{q}^{2}+\Delta^{2}\right)^{3}}\right. \nonumber \\
    && \hspace{-1.5cm} +\left[(1-x-y)p_{3\nu}-yp_{2\nu}\right]\int\frac{d^{d}\bar{q}}{(2\pi)^{d}}\frac{\bar{q}_{\mu}\bar{q}_{\rho}}
    {\left(\bar{q}^{2}+\Delta^{2}\right)^{3}}+\left[(1-x-y)p_{3\mu}-yp_{2\mu}\right]\int\frac{d^{d}\bar{q}}{(2\pi)^{d}}\frac{\bar{q}_{\nu}
    \bar{q}_{\rho}}{\left(\bar{q}^{2}+\Delta^{2}\right)^{3}} \nonumber \\
    &&\hspace{-1.5cm}  \left.  \left[(1-x-y)p_{3\mu}-yp_{2\mu}\right]\left[(1-x-y)p_{3\nu}-yp_{2\nu}\right]\left[(1-x-y)p_{3\rho}-yp_{2\rho}
    \right]\int\frac{d^{d}\bar{q}}{(2\pi)^{d}}\frac{1}{\left(\bar{q}^{2}+\Delta^{2}\right)^{3}} \right\}, \nonumber \\
\end{eqnarray}
and using again the result \eqref{result_qmu_qnu_integral_lorenzt_invariance}, we find
\begin{eqnarray}
    I_{\mu\nu\rho} &=& 2\int_{0}^{1}dx\int_{0}^{1-x}dy\left\{ \left[(1-x-y)p_{3\rho}-yp_{2\rho}\right]\frac{\delta_{\mu\nu}
    \Gamma(3)}{2(4-d)}\frac{\Gamma(3-d/2)}{(4\pi)^{d/2}\Gamma(3)}\left(\Delta^{2}\right)^{d/2-2}\right.\nonumber \\
    && +\left[(1-x-y)p_{3\nu}-yp_{2\nu}\right]\frac{\delta_{\mu\rho}\Gamma(3)}{2(4-d)}\frac{\Gamma(3-d/2)}{(4\pi)^{d/2}
    \Gamma(3)}\left(\Delta^{2}\right)^{d/2-2}\nonumber \\
    && +\left[(1-x-y)p_{3\mu}-yp_{2\mu}\right]\frac{\delta_{\nu\rho}\Gamma(3)}{2(4-d)}
    \frac{\Gamma(3-d/2)}{(4\pi)^{d/2}\Gamma(3)}\left(\Delta^{2}\right)^{d/2-2}\nonumber \\
    && +\left[(1-x-y)p_{3\mu}-yp_{2\mu}\right]\left[(1-x-y)p_{3\nu}-yp_{2\nu}\right]\left[(1-x-y)p_{3\rho}-yp_{2\rho}\right] \nonumber \\
    && \times \frac{\Gamma(3-d/2)}{(4\pi)^{3/2}\Gamma(3)}(\Delta^{2})^{d/2-3}\nonumber \\
    &=& \frac{1}{16\pi}\left\{ \delta_{\mu\nu}\int_{0}^{1}dx\int_{0}^{1-x}dy\frac{\left[(1-x-y)p_{3\rho}-yp_{2\rho}\right]}
    {\left[xyp_2^2 + (1-x-y)(xp_3^2 + yp_1^2)\right]^{1/2}}\right. \nonumber \\
    && +\delta_{\mu\rho}\int_{0}^{1}dx\int_{0}^{1-x}dy\frac{\left[(1-x-y)p_{3\nu}-yp_{2\nu}\right]}{\left[xyp_2^2 
    + (1-x-y)(xp_3^2 + yp_1^2)\right]^{1/2}} \nonumber \\
    && +\delta_{\nu\rho}\int_{0}^{1}dx\int_{0}^{1-x}dy\frac{\left[(1-x-y)p_{3\mu}-yp_{2\mu}\right]}{\left[xyp_2^2 
    + (1-x-y)(xp_3^2 + yp_1^2)\right]^{1/2}} \nonumber \\
    && \left.+\int_{0}^{1}dx\int_{0}^{1-x}dy\frac{\left[(1-x-y)p_{3\mu}-yp_{2\mu}\right]\left[(1-x-y)p_{3\nu}
    -yp_{2\nu}\right]\left[(1-x-y)p_{3\rho}-yp_{2\rho}\right]}{\left[xyp_2^2 + (1-x-y)(xp_3^2 + yp_1^2)\right]^{3/2}}\right\} \nonumber \\
    &=& \frac{1}{16\pi}\left[\delta_{\mu\nu}f_{\rho}^{\prime}(p_{1},p_{2},p_{3})+\delta_{\mu\rho}
    f_{\nu}^{\prime}(p_{1},p_{2},p_{3})+\delta_{\nu\rho}f_{\mu}^{\prime}(p_{1},p_{2},p_{3})+f_{\mu\nu\rho}(p_{1},p_{2},p_{3})\right],
\end{eqnarray}
where 
\begin{eqnarray}
    f_{\mu}^{\prime}(p_{1,}p_{2},p_{3}) &=&\int_{0}^{1}dx\int_{0}^{1-x}dy\frac{(1-x-y)p_{3\mu}-yp_{2\mu}}
    {\left[xyp_2^2 + (1-x-y)(xp_3^2 + yp_1^2)\right]^{1/2}} \\
    f_{\mu\nu\rho}(p_{1,}p_{2},p_{3})&=& \int_{0}^{1}dx\int_{0}^{1-x}dy\frac{1}{\left[xyp_2^2 + (1-x-y)(xp_3^2 
    + yp_1^2)\right]^{3/2}}, \nonumber \\
    && \times\left[(1-x-y)p_{3\mu}-yp_{2\mu}\right]\left[(1-x-y)p_{3\nu}-yp_{2\nu}\right]\left[(1-x-y)p_{3\rho}
    -yp_{2\rho}\right]. \nonumber \\
\end{eqnarray}

All in all, using the notation \eqref{notation_integrals_feynman_parameters}, we find
\begin{eqnarray*}
    f_{\mu}(p_{1},p_{2},p_{3}) &=& p_{3\mu}\left( I^{(0,0,3/2)} - I^{(1,0,3/2)}  \right)-p_{1\mu} I^{(0,1,3/2)}, \\
    f_{\mu\nu}(p_{1},p_{2},p_{3}) &=& \left(f_{0} - 2 I^{(1,0,3/2)} +  I^{(2,0,3/2)} \right)p_{3\mu}p_{3\nu} \nonumber \\
    && \quad +\left(I^{(1,1,3/2)}-I^{(0,1,3/2)}\right)\left(p_{3\mu}p_{1\nu}+p_{1\mu}p_{3\nu}\right) +I^{(0,2,3/2)}p_{1\mu}p_{1\nu}, \\ 
    f_{\mu}^{\prime}(p_{1},p_{2},p_{3}) &=&  p_{3\mu}\left( I^{(0,0,1/2)} - I^{(1,0,1/2)}  \right) - p_{1\mu} I^{(0,1,1/2)}, \\
    f_{\mu\nu\rho} (p_{1},p_{2},p_{3}) &=& \left(f_{0}-3I^{(1,0,3/2)}+3I^{(2,0,3/2)}-I^{(3,0,3/2)}\right)
    p_{3\mu}p_{3\nu}p_{3\rho} \nonumber \\
    && -\left(I^{(0,1,3/2)}+I^{(2,1,3/2)}-2I^{(1,1,3/2)}\right)(p_{3\mu}p_{3\nu}p_{1\rho}+p_{3\mu}
    p_{1\nu}p_{3\rho}+p_{1\mu}p_{3\nu}p_{3\rho}) \nonumber \\
    && \quad +\left(I^{(0,2,3/2)}-I^{(1,2,3/2)}\right)(p_{3\mu}p_{1\nu}p_{1\rho}+p_{1\mu}p_{3\nu}
    p_{1\rho}+p_{1\mu}p_{1\nu}p_{3\rho}) \nonumber  \\
    && \quad \quad - I^{(0,3,3/2)}p_{1\mu}p_{1\nu}p_{1\rho} .
\end{eqnarray*}

\section{Integrals in Feynman Parameters}

\label{appendix_integral_Feynman_parameters}

Now we present the details of the calculation of the integrals \eqref{all_integrals_feynman_parameters}.

\subsection{Useful Formulas}

All integrals we want to solve are of the form
\begin{equation}
 I^{(a,b,c)}(p_1,p_2,p_3,m)=\int_{0}^{1}dx\int_{0}^{1-x}dy\frac{x^{a}y^{b}}{\left(xyp_2^2 + (1-x-y)(xp_3^2 
 + yp_1^2)+m^2\right)^{c}}, \label{general_form_integrals_feynman_parameters}
\end{equation}

First, we note that
\begin{equation}
    I^{(a,b,c)}(p_1,p_2,p_3,m) = I^{(b,a,c)}(p_3,p_2,p_1,m).
\end{equation}

To prove this, we first exchange the order of integration on the left hand side. By doing this, the domain of integration changes from 
\begin{equation}
    0\leq x \leq 1, \quad \text{and} \quad 0\leq y \leq 1-x,
\end{equation}
to
\begin{equation}
    0\leq y \leq 1, \quad \text{and} \quad 0\leq x \leq 1-y,
\end{equation}
meaning that  
\begin{align}
    I^{(a,b,c)}(p_1,p_2,p_3,m) &= \int_{0}^{1}dx\int_{0}^{1-x}dy\frac{x^{a}y^{b}}{\left(xyp_2^2 + (1-x-y)(xp_3^2 + yp_1^2)+m^2\right)^{c}} \nonumber \\
    & = \int_{0}^{1}dy\int_{0}^{1-y}dx\frac{x^{a}y^{b}}{\left(xyp_2^2 + (1-x-y)(xp_3^2 + yp_1^2)+m^2\right)^{c}}.
\end{align}

Now, we just rename the variables of integration $x\leftrightarrow y$ to obtain 
\begin{align}
    I^{(a,b,c)}(p_1,p_2,p_3,m) &= \int_{0}^{1}dx\int_{0}^{1-x}dy\frac{y^{a}x^{b}}{\left(yxp_2^2 
    + (1-y-x)(yp_3^2 + xp_1^2)+m^2\right)^{c}} \nonumber \\
    &= I^{(b,a,c)}(p_3,p_2,p_1,m), \label{relation_integrals}
\end{align}
as we wanted to show. We will use this relation to solve less integrals in some cases we considered.

It will be useful to define  
\begin{equation}
I^{(a,b,c)}(p_1,p_2,p_3,m) = \int_0^1 dx x^a I^{(b,c)}(p_1,p_2,p_3,m)(x),
\end{equation}
where
\begin{equation}
I^{(b,c)}(p_1,p_2,p_3,m)(x) = \int_0^{1-x} \frac{y^b}{(\alpha y^2 + \beta(x) y 
+ \gamma(x))^{c}},\label{integral_in_y_feynman_paramters}
\end{equation}
\begin{equation}
\alpha = -p_1^2, \hspace{1cm} \beta(x) = p_1^2 + x(p_2^2-p_1^2-p_3^2), \hspace{1cm} \gamma(x) 
= m^2 + (1-x)xp_3^2. \label{expressions_for_alpha_beta_gamma}
\end{equation}

In order to solve the integrals \eqref{all_integrals_feynman_parameters}, we will just need the 
solution of \eqref{integral_in_y_feynman_paramters} for $b=\{0,1\}$, and for  $c=\{3/2,1/2\}$. 

\begin{itemize}
    \item $b=0$ and $c=3/2$
\end{itemize}

\begin{align}
     \int_0^{1-x} \frac{1}{(\alpha y^2 + \beta y + \gamma)^{3/2}} &= \left.- \frac{2(\beta +2\alpha y)}
     {(\beta^2 - 4 \alpha \gamma)\sqrt{\alpha y^2 + \beta y + \gamma}}\right|_0^{1-x} \nonumber \\
    &= \frac{2}{\beta^2 - 4 \alpha \gamma} \left[ \frac{\beta}{\sqrt{\gamma }} - \frac{\beta +2(1-x)\alpha}
    {\sqrt{(1-x)^2\alpha +(1-x)\beta + \gamma}} \right],
\end{align}
and after substituting the expressions \eqref{expressions_for_alpha_beta_gamma}, we find 
\begin{align}
    I^{(0,3/2)}(p_1,p_2,p_3,m)(x) &= 2\left[ x(p_{2}^{2}-p_{3}^{2}) \left( \sqrt{(1-x)xp_{2}^{2}+m^{2}} 
    - \sqrt{(1-x)xp_{3}^{2}+m^{2}}  \right)  \right. \nonumber \\
    & \left.  +  (1-x)p_{1}^{2} \left( \sqrt{(1-x)xp_{2}^{2}+m^{2}}+\sqrt{(1-x)xp_{3}^{2}+m^{2}} \right) \right]\nonumber \\
    &  \times  \frac{1}{\sqrt{ \left( (1-x)xp_{2}^{2}+m^{2} \right) \left( (1-x)xp_{3}^{2}+m^{2} \right) }} \nonumber \\
    & \times \frac{1}{4m^{2}p_{1}^{2}  +  \left( (1-x)p_{1}^{2}+x(p_{2}^{2}-p_{3}^{2}) \right) 
     \left((1-x)p_{1}^{2}+x(p_{2}^{2}+p_{3}^{2}) \right)}. \label{I032x}
\end{align}

\begin{itemize}
    \item $b=1$ and  $c=3/2$
\end{itemize}

\begin{align}
    \int_0^{1-x} \frac{y}{(\alpha y^2 + \beta y + \gamma)^{3/2}} & = \left. \frac{ 2 (2 \gamma + \beta y)}
    {(\beta^2 - 4 \alpha \gamma)\sqrt{\alpha y^2 + \beta y + \gamma}}\right|_0^{1-x} \nonumber \\
    & = - \frac{2}{\beta^2 - 4 \alpha \gamma} \left[ 2\sqrt{\gamma}  - \frac{(1-x)\beta  + 2 \gamma }
    {\sqrt{(1-x)^2\alpha +(1-x)\beta + \gamma}} \right],
\end{align}
and after substituting \eqref{expressions_for_alpha_beta_gamma}, we find 
\begin{align}
    I^{(1,3/2)}(p_1,p_2,p_3,m)(x) &= \left[ \frac{4m^2 + 2(1-x) \left( (1-x)p_1^2 + x (p_2^2+p_3^2) \right)}
    { \sqrt{(1-x)xp_2^2 + m^2}} - 4 \sqrt{(1-x)xp_3^2 + m^2}\right] \nonumber \\
    & \times \frac{1}{4m^{2}p_{1}^{2}  +  \left( (1-x)p_{1}^{2}+x(p_{2}^{2}-p_{3}^{2}) \right)  
    \left((1-x)p_{1}^{2}+x(p_{2}^{2}+p_{3}^{2}) \right)}. \label{I132x}
\end{align}

\begin{itemize}
    \item $b=0$ and  $c=1/2$
\end{itemize}

\begin{align}
    \int^{1-x}_0 dy \frac{1}{(\alpha y^2 + \beta y + \gamma)^{1/2}} &= \left. \frac{1}{\sqrt{\gamma}} 
    \log \left( 2\alpha y + 2 \sqrt{\alpha ( \alpha y^2 + \beta y + \gamma ) } + \beta \right) \right|_0^{1-x} \nonumber \\
    &= \frac{1}{\sqrt{\alpha}} \left[ \log \left( 2(1-x)\alpha + 2\sqrt{\alpha ) \alpha y^2 + \beta y+ \gamma)} 
    + \beta \right)  \right. \nonumber \\
    & \left.- \log \left( 2\sqrt{\alpha \gamma} + \beta \right)\frac{}{} \right],
\end{align}
which gives
\begin{align}
    I^{(0,1/2)}(p_1,p_2,p_3,m)(x) &= \frac{i}{p_1} \left[ \log \left( (1-x)p_1^2 + (p_2^2 - p_3^2)x + 
    2p_1 \sqrt{-m^2 -(1-x)x p_3^2 } \right) \right. \nonumber \\
    & -  \log \left( -(1-x)p_1^2 + (p_2^2 - p_3^2)x + 2p_1 \sqrt{-m^2 -(1-x)xp_2^2}  \right). \label{I012x}
\end{align}

Note the presence of an overall complex number. We also have a complex number appearing inside the argument 
of the logarithm as a consequence of the terms $\sqrt{-m^2 -(1-x)xp_2^2}$ and $\sqrt{-m^2 -(1-x)xp_2^2}$. 
After integration over x, these complex numbers will generate inverse trigonometric functions, such that the final result is real.

\begin{itemize}
    \item $b=1$ and  $c=1/2$
\end{itemize}

\begin{align}
    \int_{0}^{1-x}dy\frac{y}{(\alpha y^{2}+\beta y+\gamma)^{1/2}} &= \left. \frac{1}{\alpha} \sqrt{\alpha y^{2}
    +\beta y+\gamma} - \frac{\beta}{2\alpha^{3/2}}\tanh^{-1} \left( \frac{2\alpha y+\beta}{2\sqrt{\alpha(\alpha 
    y^{2}+\beta y+\gamma)}}\right) \right|_{0}^{1-x} \nonumber \\
    & = \frac{1}{\alpha} \left( \sqrt{\alpha(1-x)^{2}+\beta(1-x)+\gamma} - \sqrt{\gamma} \right)  \nonumber \\
    & -\frac{\beta}{2\alpha^{3/2}} \left[ \tanh^{-1} \left( \frac{2\alpha(1-x)+\beta}{2 \sqrt{\alpha(\alpha(1-x)^{2}
    +\beta(1-x)+\gamma)}} \right) - \tanh^{-1} \left( \frac{\beta}{2 \sqrt{\alpha\gamma}} \right)  \right], \nonumber \\ \label{I112x}
\end{align}
which in terms of the external momenta gives 
\begin{align}
    I^{(1,1/2)}(p_{1},p_{2},p_{3},m)(x) &= -\frac{1}{2p_{2}^{3}} \left\{  2p_{1} \left( \sqrt{(1-x)xp_{2}^{2}+m^{2}} 
    - \sqrt{(1-x)xp_{3}^{2}+m^{2}} \right) \right. \nonumber \\
    & + \left( (1-x)p_{1}^{2}+(p_{2}^{2}-p_{3}^{2})x \right) \left[ \cot^{-1}\left(-\frac{2p_{1}\sqrt{(1-x)xp_{2}^{2}
    +m^{2}}}{(1-x)p_{1}^{2}-(p_{2}^{2}-p_{3}^{2})x}\right) \right. \nonumber \\
    & \left. \left. - \cot^{-1}\left(\frac{2p_{1}\sqrt{(1-x)xp_{3}^{2}+m^{2}}}{(1-x)p_{1}^{2}+(p_{2}^{2}+p_{3}^{2})x} \right)\right]\right\}. 
\end{align}

These are all the $y-$integrals we will need in order to compute the eight independent scalar integrals in the Feynman parameters. 
The last step before perform the $x-$integration is to specialize the results above for the three cases we are interested in.

\subsection{$p_1^2 = p_3^2 = 0$}

In this case, the integrals \eqref{I032x},\eqref{I132x}, \eqref{I012x}, and \eqref{I112x} are
\begin{align}
    I^{(0,3/2)}(0,p_2,0,m)(x) & = \frac{2}{p_2^2 x} \left(  \frac{1}{m} - \frac{1}{\sqrt{(1-x)x p_2^2 + m^2}}\right),  \\
    I^{(1,3/2)}(0,p_2,0,m)(x) &= \frac{2}{p_2^4 x^2} \left( \frac{(1-x)xp_2^2 + 2m^2}{\sqrt{(1-x)xp_2^2 + m^2 }} - 2m \right), \\
    I^{(0,1/2)}(0,p_2,0,m)(x) &= \frac{2}{p_2^2 x} \left( \sqrt{(1-x)xp_2^2 + m^2} - m \right),  \\
    I^{(1,1/2)}(0,p_2,0,m)(x) &= \frac{2}{3p_2^4 x^2} \left[ 2m^3 + \sqrt{(1-x)xp_2^2 + m^2} \left( (1-x)x p_2^2 - 2m^2 \right) \right].
\end{align}

Now we are able to solve the $x-$integrals.

%%%%%%%%%%%%%%%%%%%%%%%%%%  I(0,0,3/2) %%%%%%%%%%%%%%%%%%%%%%%%%%
\begin{itemize} 
\item $I^{(0,0,3/2)}$
\end{itemize}

\begin{align}
    I^{(0,0,3/2)}(p_2,m) & = \int_0^1 dx I^{(0,3/2)}(0,p_2,0,m)(x) \nonumber \\
    & = \frac{2}{p_{2}^{2}}\left(\frac{1}{m}\int_{0}^{1}dx\frac{1}{x}-\int_{0}^{1}dx\frac{1}{x\sqrt{(1-x)xp_{2}^{2}+m^{2}}}\right) 
    \nonumber \\
    &= \frac{2}{mp_{2}^{2}}\left[\log x+\frac{1}{2}\log\left(1+\frac{2m^{2}+p_{2}^{2}x}{2m\sqrt{(1-x)xp_{2}^{2}+m^{2}}}\right)\right. 
    \nonumber \\
    & \quad\quad\quad\left.-\frac{1}{2}\log\left(1-\frac{2m^{2}+p_{2}^{2}x}{2m\sqrt{(1-x)xp_{2}^{2}+m^{2}}}\right)\right]_{0}^{1}. 
    \nonumber \\ 
\end{align}

%
\begin{comment}
\begin{eqnarray}
I^{(0,0,3/2)} &=& \int_{0}^{1}dx\int_{0}^{1-x}dy\frac{1}{\left(xyp_{2}^{2}+m^{2}\right)^{3/2}} \nonumber \\
&=& -\frac{2}{p_{2}^{2}}\int_{0}^{1}dx\left.\frac{1}{x\sqrt{xyp_{2}^{2}+m^{2}}}\right|_{0}^{1-x} \nonumber \\
&=& -\frac{2}{p_{2}^{2}}\int_{0}^{1}dx\frac{1}{x}\left(\frac{1}{\sqrt{(1-x)xp_{2}^{2}+m^{2}}}-\frac{1}{m}\right) \nonumber \\
&=& \frac{2}{p_{2}^{2}}\left(\frac{1}{m}\int_{0}^{1}dx\frac{1}{x}-\int_{0}^{1}dx\frac{1}{x\sqrt{(1-x)xp_{2}^{2}+m^{2}}}\right) \nonumber \\
&=& \frac{2}{mp_{2}^{2}}\left[\log x+\frac{1}{2}\log\left(1+\frac{2m^{2}+p_{2}^{2}x}{2m\sqrt{(1-x)xp_{2}^{2}+m^{2}}}\right)\right. \nonumber \\
&& \quad\quad\quad\left.-\frac{1}{2}\log\left(1-\frac{2m^{2}+p_{2}^{2}x}{2m\sqrt{(1-x)xp_{2}^{2}+m^{2}}}\right)\right]_{0}^{1}. \nonumber \\ 
\end{eqnarray}
\end{comment}

The the lower limit of integration is not trivial since $\log(0) = -\infty$. For $x=1$ we have
\begin{align}
    & \left.\log x+\frac{1}{2}\log\left(1+\frac{2m^{2}+p_{2}^{2}x}{2m\sqrt{(1-x)xp_{2}^{2}+m^{2}}}\right)-\frac{1}{2}\log\left(1-\frac{2m^{2}+p_{2}^{2}x}{2m\sqrt{(1-x)xp_{2}^{2}+m^{2}}}\right)\right|^{1}\nonumber \\ 
    &  \quad \quad =  \frac{1}{2}\log\left(1+\frac{p_{2}^{2}}{4m^{2}}\right)-\frac{1}{2}\log\left(\frac{p_{2}^{2}}{4m^{2}}\right)-\frac{i\pi}{2},
\end{align}
while for $x=0$, we expand around small $x$ and obtain, after taking the $x\rightarrow 0$ limit,
\begin{eqnarray}
    &&\left.\log x+\frac{1}{2}\log\left(1+\frac{2m^{2}+p_{2}^{2}x}{2m\sqrt{(1-x)xp_{2}^{2}+m^{2}}}\right)
    -\frac{1}{2}\log\left(1-\frac{2m^{2}+p_{2}^{2}x}{2m\sqrt{(1-x)xp_{2}^{2}+m^{2}}}\right)\right|_{0} \nonumber \\
    && \hspace{3cm} =  -\frac{1}{2}\log\frac{p_{2}^{2}}{4m^{2}}-\frac{1}{2}\log\left(1+\frac{p_{2}^{2}}{4m^{2}}\right)-\frac{1}{2}i\pi.
\end{eqnarray}

Hence, the integral is
\begin{equation}
    I^{(0,0,3/2)}(p_2,m) = \frac{2}{mp_{2}^{2}}\log\left(1+\frac{p_{2}^{2}}{4m^{2}}\right).
\end{equation}

Expanding around small $m$ and taking the limit $m\rightarrow 0$, we find

\begin{equation}
    I^{(0,0,3/2)}(p_2,m) = 2\lim_{m\rightarrow0}\frac{\log\left(p_{2}^{2}/4m^{2}\right)}{mp_{2}^{2}} .\label{I(0,0,3/2)}
\end{equation}

This result is divergent as $m\rightarrow 0$.

%%%%%%%%%%%%%%%%%%%%%%%%%%  I(1,0,3/2) %%%%%%%%%%%%%%%%%%%%%%%%%%

\begin{itemize}
    \item $I^{(1,0,3/2)}$:
\end{itemize}
\begin{align}
    I^{(1,0,3/2)}(p_2,m) & = \int_0^1 dx x I^{(0,3/2)}(0,p_2,0,m)(x) \nonumber \\
    &= -\frac{2}{p_{2}^{2}}\left(\int_{0}^{1}dx\frac{1}{\sqrt{(1-x)xp_{2}^{2}+m^{2}}}-\frac{1}{m}\right) \nonumber \\
    & = -\frac{2}{p_{2}^{2}}\left[\frac{2}{p_{2}}\cot^{-1}\left(\frac{2 m}{p_{2}}\right)-\frac{1}{m}\right] \nonumber \\
    & = -\frac{4}{p_{2}^{3}}\cot^{-1}\left(\frac{2m}{p_{2}}\right)+\frac{2}{mp_{2}^{2}} \nonumber \\
    &\overset{m\rightarrow 0}{=}  \frac{2}{p_2 ^2} \lim_{m\rightarrow 0} \frac{1}{m} -\frac{2\pi}{p_2^3}. \label{I(1,0,3/2}
\end{align}

\begin{itemize}
    \item $I^{(2,0,3/2)}$:
\end{itemize}

\begin{align}
    I^{(2,0,3/2)}(p_2,m) & = \int_0^1 dx x^2  I^{(0,3/2)}(0,p_2,0,m)(x) \nonumber \\
    & =  -\frac{2}{p_{2}^{2}}\int_{0}^{1}dxx\left(\frac{1}{\sqrt{(1-x)xp_{2}^{2}+m^{2}}}-\frac{1}{m}\right) \nonumber \\
    & =  -\frac{2}{p_{2}^{2}}\left( -\frac{1}{2m} + \int_{0}^{1}dx\frac{x}{\sqrt{(1-x)xp_{2}^{2}+m^{2}}}\right)  \nonumber \\
    & =  -\frac{2}{p_{2}^{3}}\cot^{-1}\left(\frac{2m}{p}\right)+\frac{1}{m p_{2}^{2}} \nonumber \\
    &\overset{m\rightarrow 0}{=}  \frac{1}{p_2 ^2} \lim_{m\rightarrow 0} \frac{1}{m} - \frac{\pi}{p_2^3}.
\end{align}

\begin{itemize}
    \item $I^{(3,0,3/2)}$:
\end{itemize}

\begin{align}
    I^{(3,0,3/2)}(p_2,m) & = \int_0^1 dx x^3  I^{(0,3/2)}(0,p_2,0,m)(x) \nonumber \\
    &= -\frac{2}{p_{2}^{2}}\left\{ \int_{0}^{1}dx\frac{x^{2}}{\sqrt{(1-x)xp_{2}^{2}+m^{2}}}-\frac{1}{3m}\right\} \nonumber \\
    &= \frac{m}{p_{2}^{4}}-\frac{\left(4m^{2}+3p_{2}^{2}\right)}{2p_{2}^{5}} \cot^{-1}\left(\frac{2m}{p_{2}}\right)
    +\frac{2}{3mp_{2}^{2}} \nonumber \\
    &\overset{m \rightarrow 0}{=} \frac{2}{3p_{2}^{2}}\lim_{m\rightarrow0}\frac{1}{m} -\frac{3\pi}{4p_{2}^{3}}.
\end{align}

\begin{itemize}
    \item $I^{(1,1,3/2)}$:
\end{itemize}

\begin{align}
    I^{(1,1,3/2)}(p_2,m) & = \int_0^1 dx x  I^{(1,3/2)}(0,p_2,0,m)(x) \nonumber \\
    &= \frac{2}{p_2^{4}}\left(\int_{0}^{1}dx\frac{(1-x)xp_{2}^{2}+2m^{2}}{x\sqrt{(1-x)xp_{2}^{2}+m^{2}}}
    +2m\lim_{x\rightarrow0}\log x\right) \nonumber \\
    &= \frac{2}{p_2 ^3} \cot^{-1} \left( \frac{2m}{p_2} \right) -\frac{4m}{p_2^4}\log \left( 1+\frac{p_2^2}{4m^2} \right) \nonumber \\
    &\overset{m \rightarrow 0}{=} \frac{\pi}{p_2^3}.
\end{align}

\begin{itemize}
    \item $I^{(2,1,3/2)}$:
\end{itemize}

\begin{align}
    I^{(2,1,3/2)}(p_2,m) & = \int_0^1 dx x^2  I^{(1,3/2)}(0,p_2,0,m)(x) \nonumber \\
    &= \frac{2}{p_{2}^{4}}\left( \int_{0}^{1}dx\frac{(1-x)xp_{2}^{2}+2m^{2}}{\sqrt{(1-x)xp_{2}^{2}+m^{2}}}-2m\right) \nonumber \\
    &= \frac{2}{p_{2}^{4}}\left( \frac{m}{2}+\frac{\left(12m^{2}+p^{2}\right)\cot^{-1}(2m/p)}{4p_2 } -2m \right)  \nonumber \\
    &= \frac{\left(12m^{2}+p_{2}^{2}\right)}{2p_{2}^{5}}\cot^{-1}\left(\frac{2m}{p_{2}}\right)-\frac{3m}{p_{2}^{4}} \nonumber \\
    &\overset{m \rightarrow 0}{=}  \frac{\pi}{4p_2^3 }.
\end{align}
\begin{comment}
\begin{eqnarray}
    I^{(2,1,3/2)} &=& \int_{0}^{1}dx\int_{0}^{1-x}dy\frac{x^{2}y}{\left( xyp_{2}^{2}+m^{2}\right)^{3/2}} \nonumber \\
    &=& \frac{2}{p_{2}^{4}}\int_{0}^{1}dx\left( \frac{xyp_{2}^{2}+2m^{2}}{\sqrt{xyp_{2}^{2}+m^{2}}}\right)_{0}^{1-x} \nonumber \\
    &=& \frac{2}{p_{2}^{4}}\left( \int_{0}^{1}dx\frac{(1-x)xp_{2}^{2}+2m^{2}}{\sqrt{(1-x)xp_{2}^{2}+m^{2}}}-2m\right) \nonumber \\
    &=& \frac{2}{p_{2}^{4}}\left( \frac{m}{2}+\frac{\left(12m^{2}+p^{2}\right)\cot^{-1}(2m/p)}{4p_2 } -2m \right)  \nonumber \\
    &=& \frac{\left(12m^{2}+p_{2}^{2}\right)}{2p_{2}^{5}}\cot^{-1}\left(\frac{2m}{p_{2}}\right)-\frac{3m}{p_{2}^{4}} \nonumber \\
    &\overset{m \rightarrow 0}{=}&  \frac{\pi}{4p_2^3 }.
\end{eqnarray}
\end{comment}

%%%%%%%%%%%%%%%%%%%%%%%%%%  I(0,0,1/2) %%%%%%%%%%%%%%%%%%%%%%%%%%

\begin{itemize}
    \item $I^{(0,0,1/2)}$, $I^{(1,0,1/2)}$:
\end{itemize}

These integrals are finite for  $m=0$ and are simpler to calculate. However, 
for completeness, we will give the solution for $m\neq 0$,
\begin{eqnarray}
    I^{(0,0,1/2)} &=& \frac{2}{p_{2}}\cot^{-1}\left(\frac{2m}{p_{2}}\right)+\frac{m}{p_{2}^{2}}
    \left[\log16+2\log\left(\frac{m^{2}}{4m^{2}+p^{2}}\right)\right], \\ 
    I^{(1,0,1/2)} &=& -\frac{m}{p_{2}^{2}}+\frac{2m^{2}}{p_{2}^{3}}\cot^{-1}\left(\frac{2m}{p_{2}}\right)
    +\frac{1}{2p_{2}}\cot^{-1}\left(\frac{2m}{p_{2}}\right).
\end{eqnarray}

Taking the massless limit,
\begin{equation}
    I^{(0,0,1/2)} = \frac{\pi}{p_2}, \quad I^{(1,0,1/2)} =  \frac{\pi}{4p_2}.
\end{equation}

Finally, with these results in hand we note the following relations between these integrals,
\begin{align}
   I^{(1,0,3/2)} &= 2I^{(1,0,3/2)}, \\
   I^{(3,0,3/2)} &= \frac{1}{3} \left( I^{(1,0,3/2)} - I^{(2,1,3/2)} \right), \\
   I^{(1,1,3/2)} &= 4I^{(2,1,3/2)} = \frac{4}{p_2^2} I^{(1,0,1/2)}, \\
   I^{(0,0,1/2)} &= 4 I^{(1,0,1/2)}
\end{align}

Using these relations together with \eqref{relation_integrals} derived above, we find
\begin{align}
    f_{\mu}(p_{1},p_{2},p_{3}) &= I^{(0,0,3/2)}p_{3\mu}-I^{(1,0,3/2)}\left(p_{1\mu}+p_{3\mu}\right), \\
    f_{\mu\nu}(p_{1},p_{2},p_{3}) &= \left(I^{(0,0,3/2)}-3I^{(2,0,3/2)}\right)p_{3\mu}p_{3\nu} \nonumber \\
    & +\left(I^{(1,1,3/2)}-I^{(1,0,3/2)}\right)\left(p_{3\mu}p_{1\nu}+p_{1\mu}p_{3\nu}\right)+I^{(2,0,3/2)}p_{1\mu}p_{1\nu}, \\
    f_{\mu}^{\prime}(p_{1},p_{2},p_{3}) &= I^{(0,0,1/2)}p_{3\mu}-I^{(1,0,1/2)}\left(p_{1\mu}+p_{3\mu}\right), \\
    f_{\mu\nu\rho}(p_{1},p_{2},p_{3}) &= \left(I^{(0,0,3/2)}-3I^{(2,0,3/2)}-I^{(3,0,3/2)}\right)
    p_{3\mu}p_{3\nu}p_{3\rho}-I^{(3,0,3/2)}p_{1\mu}p_{1\nu}p_{1\rho} \nonumber \\
    &-\left(I^{(1,0,3/2)}-7I^{(2,1,3/2)}\right)(p_{3\mu}p_{3\nu}p_{1\rho}+p_{3\mu}p_{1\nu}
    p_{3\rho}+p_{1\mu}p_{3\nu}p_{3\rho}) \nonumber \\
    &+\left(I^{(2,0,3/2)}-I^{(2,1,3/2)}\right)(p_{3\mu}p_{1\nu}p_{1\rho}+p_{1\mu}p_{3\nu}p_{1\rho}+p_{1\mu}p_{1\nu}p_{3\rho}).
\end{align}

\subsection{$p_1^2 = 0$}

For $p_1^2=0$ but $p_3^2 \neq 0$, the integrals \eqref{I032x},\eqref{I132x}, \eqref{I012x}, and \eqref{I112x} are
\begin{align}
    I^{(0,3/2)}(0,p_2,p_3,m)(x) &= \frac{2}{(p_{2}^{2}-p_{3}^{2})x}\left(\frac{1}{\sqrt{(1-x)xp_{3}^{2}+m^{2}}}-
    \frac{1}{\sqrt{(1-x)xp_{2}^{2}+m^{2}}}\right) \\
    I^{(1,3/2)}(0,p_2,p_3,m)(x) &= \frac{2}{\left(p_{2}^{2}-p_{3}^{2}\right)} \frac{1}{x^{2}}\left(\frac{(1-x)
    x\left(p_{2}^{2}+p_{3}^{2}\right)+2m^{2}}{\sqrt{(1-x)xp_{2}^{2}+m^{2}}}-2\sqrt{(1-x)xp_{3}^{2}+m^{2}}\right)   \\
    I^{(0,1/2)}(0,p_2,p_3,m)(x) &=  \frac{2}{(p_{2}^{2}-p_{3}^{2})} \frac{1}{x}\left(\sqrt{(1-x)xp_{2}^{2}
    +m^{2}}-\sqrt{(1-x)xp_{3}^{2}+m^{2}}\right)   \\
    I^{(1,1/2)}(0,p_2,p_3,m)(x) &= -\frac{2}{3\left(p_{2}^{2}-p_{3}^{2}\right)^{2}}\int_{0}^{1}
    dx\frac{\left[\left(3p_{3}^{2}-p_{2}^{2}\right)(1-x)x+2m^{2}\right]\sqrt{(1-x)xp_{2}^{2}+m^{2}}}{x^{2}}
\end{align}

Now we use these results to solve the remaing integrals on $x$.
%%%%%%%%%%%%%%%%%%%%%%%%%%  I(0,0,3/2) %%%%%%%%%%%%%%%%%%%%%%%%%%
\begin{itemize}
    \item$I^{(0,0,3/2)}$
\end{itemize}
\begin{align}
    I^{(0,0,3/2)}(0,p_2,p_3,m) & = \int_0^1 dx I^{(0,3/2)}(0,p_2,p_3,m)(x) \nonumber \\
     &=-\frac{2}{(p_{2}^{2}-p_{3}^{2})}\int_{0}^{1}dx\frac{1}{x}\left(\frac{1}{\sqrt{(1-x)xp_{2}^{2}
     +m^{2}}}-\frac{1}{\sqrt{(1-x)xp_{3}^{2}+m^{2}}}\right) \nonumber \\
    &= \frac{2}{m(p_{2}^{2}-p_{3}^{2})}\left[\cot^{-1}\left(\frac{2m\sqrt{(1-x)xp_{2}^{2}+m^{2}}}
    {p_{2}^{2}x+2m^{2}}\right)\right. \nonumber \\
    & \quad\quad\quad\quad\quad\quad\quad\quad\quad\quad\quad \left.\cot^{-1}\left(\frac{2m
    \sqrt{(1-x)xp_{3}^{2}+m^{2}}}{p_{3}^{2}x+2m^{2}}\right)\right]_{0}^{1} \nonumber \\
    &= \frac{2}{m(p_{2}^{2}-p_{3}^{2})}\log\left(\frac{1-p_{2}^{2}/4m^{2}}{1-p_{3}^{2}/4m^{2}}\right) \nonumber \\
    &\overset{m\rightarrow 0}{=} \frac{2}{(p_2^2 - p_3^2)}\log \left(\frac{p_2 ^2}{p_3 ^2}  \right) \lim_{m\rightarrow 0} \frac{1}{m}.
\end{align}

\begin{comment}
\begin{eqnarray}
    I^{(0,0,3/2)}(0,p_2,p_3,m)  &=&  \int_{0}^{1}dx\int_{0}^{1-x}dy\frac{1}{\left[x(1-x-y)p_{3}^{2}+xyp_{2}^{2}+m^{2}\right]^{3/2}}\nonumber \\
    &=& -\frac{2}{(p_{2}^{2}-p_{3}^{2})}\int_{0}^{1}dx\left.\frac{1}{x\sqrt{x(1-x-y)p_{3}^{2}+xyp_{2}^{2}+m^{2}}}\right|_{0}^{1-x}  \nonumber \\
    &=& -\frac{2}{(p_{2}^{2}-p_{3}^{2})}\int_{0}^{1}dx\frac{1}{x}\left(\frac{1}{\sqrt{(1-x)xp_{2}^{2}+m^{2}}}-\frac{1}{\sqrt{(1-x)xp_{3}^{2}+m^{2}}}\right) \nonumber \\
    &=& \frac{2}{m(p_{2}^{2}-p_{3}^{2})}\left[\cot^{-1}\left(\frac{2m\sqrt{(1-x)xp_{2}^{2}+m^{2}}}{p_{2}^{2}x+2m^{2}}\right)\right. \nonumber \\&&
   \quad\quad\quad\quad\quad\quad\quad\quad\quad\quad\quad \left.\cot^{-1}\left(\frac{2m\sqrt{(1-x)xp_{3}^{2}+m^{2}}}{p_{3}^{2}x+2m^{2}}\right)\right]_{0}^{1} \nonumber \\
    &=& \frac{2}{m(p_{2}^{2}-p_{3}^{2})}\log\left(\frac{1-p_{2}^{2}/4m^{2}}{1-p_{3}^{2}/4m^{2}}\right) \nonumber \\
    &\overset{m\rightarrow 0}{=}&\frac{2}{(p_2^2 + p_3^2)}\log \left(\frac{p_2 ^2}{p_3 ^2}  \right) \lim_{m\rightarrow 0} \frac{1}{m}.
\end{eqnarray}
\end{comment}

%%%%%%%%%%%%%%%%%%%%%%%%%%  I(1,0,3/2) %%%%%%%%%%%%%%%%%%%%%%%%%%
\begin{itemize}
    \item $I^{(1,0,3/2)}$
\end{itemize}

\begin{align}
    I^{(1,0,3/2)}(0,p_2,p_3,m) & = \int_0^1 dx x I^{(0,3/2)}(0,p_2,p_3,m)(x) \nonumber \\
    &= \frac{2}{(p_{2}^{2}-p_{3}^{2})}\int_{0}^{1}dx\left(\frac{1}{\sqrt{(1-x)xp_{3}^{2}+m^{2}}}
    -\frac{1}{\sqrt{(1-x)xp_{2}^{2}+m^{2}}}\right) \nonumber \\
    &= \frac{4}{(p_{2}^{3}p_{3}-p_{2}p_{3}^{3})}\left[\cot^{-1}\left(\frac{2m}{p_{3}}\right)
    -\cot^{-1}\left(\frac{2m}{p_{2}}\right)\right] \nonumber \\
    &\overset{m\rightarrow 0}{=} \frac{2\pi}{p_{2}^{2}p_{3} + p_{2}p_{3}^{2}}.
\end{align}

\begin{comment}
\begin{eqnarray}
    I^{(1,0,3/2)} &=& \int_{0}^{1}dx\int_{0}^{1-x}dy\frac{x}{\left[x(1-x-y)p_{3}^{2}+xyp_{2}^{2}+m^{2}\right]^{3/2}} \nonumber \\
    &=& -\frac{2}{(p_{2}^{2}-p_{3}^{2})}\int_{0}^{1}dx\left.\frac{1}{\sqrt{x(1-x-y)p_{3}^{2}+xyp_{2}^{2}+m^{2}}}\right|_{0}^{1-x} \nonumber \\
    &=& \frac{2}{(p_{2}^{2}-p_{3}^{2})}\int_{0}^{1}dx\left(\frac{1}{\sqrt{(1-x)xp_{3}^{2}+m^{2}}}-\frac{1}{\sqrt{(1-x)xp_{2}^{2}+m^{2}}}\right) \nonumber \\
    &=& \frac{4}{(p_{2}^{3}p_{3}-p_{2}p_{3}^{3})}\left[\cot^{-1}\left(\frac{2m}{p_{3}}\right)-\cot^{-1}\left(\frac{2m}{p_{2}}\right)\right] \nonumber \\
    &\overset{m\rightarrow 0}{=}& \frac{2\pi}{p_{2}^{2}p_{3} + p_{2}p_{3}^{2}}.
\end{eqnarray}
\end{comment}

%%%%%%%%%%%%%%%%%%%%%%%%%%%%%%%%%%%%%%%%%%%%%%%%%%%%%%%%%%%%%%%%%
%%%%%%%%%%%%%%%%%%%%%%%%%%  I(2,0,3/2) %%%%%%%%%%%%%%%%%%%%%%%%%%
%%%%%%%%%%%%%%%%%%%%%%%%%%%%%%%%%%%%%%%%%%%%%%%%%%%%%%%%%%%%%%%%%

\begin{itemize}
    \item$I^{(2,0,3/2)}$
\end{itemize}
\begin{align}
    I^{(2,0,3/2)}(0,p_2,p_3,m) & = \int_0^1 dx x^2 I^{(0,3/2)}(0,p_2,p_3,m)(x) \nonumber \\
    &= \frac{2}{(p_{2}^{2}-p_{3}^{2})}\int_{0}^{1}dx\left(\frac{x}{\sqrt{(1-x)xp_{3}^{2}+m^{2}}}
    -\frac{x}{\sqrt{(1-x)xp_{2}^{2}+m^{2}}}\right) \nonumber \\
    &= \frac{1}{p_{2}^{2}p_{3}^{2}(p_{2}^{2}-p_{3}^{2})}\left[2\left(p_{3}^{2}\sqrt{(1-x)xp_{2}^{2}
    +m^{2}}-p_{2}^{2}\sqrt{(1-x)xp_{3}^{2}+m^{2}}\right)\right. \nonumber \\
    & \left.+p_{2}p_{3}^{2}\tan^{-1}\left(\frac{p_{2}\left(1-2x\right)}{2\sqrt{(1-x)xp_{2}^{2}+m^{2}}}
    \right)-p_{3}p_{2}^{2}\tan^{-1}\left(\frac{p_{3}\left(1-2x\right)}{2\sqrt{(1-x)xp_{3}^{2}+m^{2}}}\right)\right]_{0}^{1} \nonumber \\
    &= \frac{2}{(p_{2}^{3}p_{3}-p_{2}p_{3}^{3})}\left[\cot^{-1}\left(\frac{2m}{p_{3}}\right)
    -\cot^{-1}\left(\frac{2m}{p_{2}}\right)\right] \nonumber \\
    &\overset{m\rightarrow 0}{=} \frac{\pi}{p_{2}^{2}p_{3}+p_{2}p_{3}^{2}}.
\end{align}
\begin{comment}
\begin{eqnarray}
    I^{(2,0,3/2)}(0,p_2,p_3,m) &=&\int_{0}^{1}dx\int_{0}^{1-x}dy\frac{x^{2}}{\left[x(1-x-y)p_{3}^{2}+xyp_{2}^{2}+m^{2}\right]^{3/2}} \nonumber \\
    &=&-\frac{2}{(p_{2}^{2}-p_{3}^{2})}\int_{0}^{1}dx\left.\frac{x}{\sqrt{x(1-x-y)p_{3}^{2}+xyp_{2}^{2}+m^{2}}}\right|_{0}^{1-x} \nonumber \\
    &=&\frac{2}{(p_{2}^{2}-p_{3}^{2})}\int_{0}^{1}dx\left(\frac{x}{\sqrt{(1-x)xp_{3}^{2}+m^{2}}}-\frac{x}{\sqrt{(1-x)xp_{2}^{2}+m^{2}}}\right) \nonumber \\
    &=&\frac{1}{p_{2}^{2}p_{3}^{2}(p_{2}^{2}-p_{3}^{2})}\left[2\left(p_{3}^{2}\sqrt{(1-x)xp_{2}^{2}+m^{2}}-p_{2}^{2}\sqrt{(1-x)xp_{3}^{2}+m^{2}}\right)\right. \nonumber \\
    && \left.+p_{2}p_{3}^{2}\tan^{-1}\left(\frac{p_{2}\left(1-2x\right)}{2\sqrt{(1-x)xp_{2}^{2}+m^{2}}}\right)-p_{3}p_{2}^{2}\tan^{-1}\left(\frac{p_{3}\left(1-2x\right)}{2\sqrt{(1-x)xp_{3}^{2}+m^{2}}}\right)\right]_{0}^{1} \nonumber \\
    &=&\frac{2}{(p_{2}^{3}p_{3}-p_{2}p_{3}^{3})}\left[\cot^{-1}\left(\frac{2m}{p_{3}}\right)-\cot^{-1}\left(\frac{2m}{p_{2}}\right)\right] \nonumber \\
    &\overset{m\rightarrow 0}{=}& \frac{\pi}{p_{2}^{2}p_{3}+p_{2}p_{3}^{2}}.
\end{eqnarray}
\end{comment}

%%%%%%%%%%%%%%%%%%%%%%%%%%%%%%%%%%%%%%%%%%%%%%%%%%%%%%%%%%%%%%%%%
%%%%%%%%%%%%%%%%%%%%%%%%%%  I(3,0,3/2) %%%%%%%%%%%%%%%%%%%%%%%%%%
%%%%%%%%%%%%%%%%%%%%%%%%%%%%%%%%%%%%%%%%%%%%%%%%%%%%%%%%%%%%%%%%%

\begin{itemize}
    \item$I^{(3,0,3/2)}$
\end{itemize}
\begin{align}
I^{(3,0,3/2)}(0,p_2,p_3,m) &= \int_0^1 dx x^3 I^{(0,3/2)}(0,p_2,p_3,m)(x) \nonumber \\
&= \frac{2}{(p_{2}^{2}-p_{3}^{2})}\int_{0}^{1}dx\left(\frac{x^{2}}{\sqrt{(1-x)xp_{3}^{2}+m^{2}}}
-\frac{x^{2}}{\sqrt{(1-x)xp_{2}^{2}+m^{2}}}\right) \nonumber \\
&= \frac{1}{4p_{2}^{3}p_{3}^{3}\left(p_{2}^{2}-p_{3}^{2}\right)}\left[2\left(3+2x\right)p_{2}p_{3}
\left(p_{3}^{2}\sqrt{(1-x)xp_{2}^{2}+m^{2}}-p_{2}^{2}\sqrt{(1-x)xp_{3}^{2}+m^{2}}\right)\right. \nonumber \\
& \hspace*{1cm}+\left(4m^{2}+3p_{2}^{2}\right)p_{3}^{3}\tan^{-1}\left(\frac{p_{2}\left(1-2x\right)}
{2\sqrt{(1-x)xp_{2}^{2}+m^{2}}}\right) \nonumber \\
& \hspace*{2cm} \left.-\left(4m^{2}+3p_{3}^{2}\right)p_{2}^{3}\tan^{-1}\left(\frac{p_{3}
\left(1-2x\right)}{2\sqrt{(1-x)xp_{3}^{2}+m^{2}}}\right)\right]^1_0 \nonumber \\
&=  -\frac{1}{2p_{2}^{3}p_{3}^{3}\left(p_{2}^{2}-p_{3}^{2}\right)}\left[2mp_{2}p_{3}
\left(p_{2}^{2}-p_{3}^{2}\right)+\left(4m^{2}+3p_{2}^{2}\right)p_{3}^{3}\cot^{-1}\left(\frac{2m}{p_{2}}\right)\right. \nonumber \\
& \hspace{2cm} \left.-\left(4m^{2}+3p_{3}^{2}\right)p_{2}^{3}\cot^{-1}\left(\frac{2m}{p_{3}}\right)\right] \nonumber \\
&\overset{m\rightarrow 0}{=} \frac{3\pi}{4(p_{2}^{2}p_{3}+p_{2}p_{3}^{2})}.
\end{align}

\begin{itemize}
    \item $I^{(0,1,3/2)}$
\end{itemize}

\begin{align}
    I^{(0,1,3/2)}(0,p_2,p_3,m) &= \int_0^1 dx I^{(1,3/2)}(0,p_2,p_3,m)(x) \nonumber \\
    &= \frac{2}{\left(p_{2}^{2}-p_{3}^{2}\right)}\int_{0}^{1}dx\frac{1}{x^{2}}\left(\frac{(1-x)x\left(p_{2}^{2}
    +p_{3}^{2}\right)+2m^{2}}{\sqrt{(1-x)xp_{2}^{2}+m^{2}}}-2\sqrt{(1-x)xp_{3}^{2}+m^{2}}\right) \nonumber \\
    &= \frac{1}{mp_{2}\left(p_{2}^{2}-p_{3}^{2}\right)x}\left[2p_{2}p_{3}^{2}x\left(\coth^{-1}
    \left(\frac{2m\sqrt{(1-x)xp_{3}^{2}+m^{2}}}{p_{3}^{2}x+2m^{2}}\right)\right.\right. \nonumber \\
    & \hspace{-3cm}\left. -\coth^{-1}\left(\frac{2m\sqrt{(1-x)xp_{2}^{2}+m^{2}}}{p_{2}^{2}x
    +2m^{2}}\right)\right)-4mp_{2}\left(\sqrt{(1-x)xp_{2}^{2}+m^{2}}-\sqrt{(1-x)xp_{3}^{2}+m^{2}}\right) \nonumber \\
    & \hspace{-3cm} \left.+p_{3}x\tan^{-1}\left(\frac{p_{3}\left(1-2x\right)}{2\sqrt{(1-x)xp_{3}^{2}+m^{2}}}\right) +2m\left(p_{2}
    ^{2}+p_{3}^{2}\right)x\tan^{-1}\left(\frac{p_{2}\left(1-2x\right)}{2\sqrt{(1-x)xp_{2}^{2}+m^{2}}}\right)\right]_{0}^{1} \nonumber \\
    &= -\frac{2}{mp_{2}\left(p_{2}^{2}-p_{3}^{2}\right)}\left[2m\left(p_{2}^{2}+p_{3}^{2}\right)\cot^{-1}\left(\frac{2m}{p_{2}}\right) 
    \right. \nonumber \\
    & \hspace*{2cm} \left.-p_{2}\left(p_{2}^{2}-p_{3}^{2}+4mp_{3}\cot^{-1}\left(\frac{2m}{p_{3}}\right)+p_{3}^{2}\log 
    \left(\frac{1+p_{3}^{2}/4m^{2}}{1+p_{2}^{2}/4m^{2}} \right)\right)\right] \nonumber \\
    &\overset{m\rightarrow 0}{=} -\frac{2\pi}{p_{2}\left(p_{2}+p_{3}\right)^{2}}+\frac{2}{\left(p_{2}^{2}-p_{3}^{2}\right)}
    \left[1-\frac{p_{3}^{2}}{\left(p_{2}^{2}-p_{3}^{2}\right)}\log\left(\frac{p_{2}^{2}}{p_{3}^{2}}\right)\right]\lim_{m\rightarrow0}\frac{1}{m}.
\end{align}

%%%%%%%%%%%%%%%%%%%%%%%%%%%%%%%%%%%%%%%%%%%%%%%%%%%%%%%%%%%%%%%%%
%%%%%%%%%%%%%%%%%%%%%%%%%%  I(0,2,3/2) %%%%%%%%%%%%%%%%%%%%%%%%%%
%%%%%%%%%%%%%%%%%%%%%%%%%%%%%%%%%%%%%%%%%%%%%%%%%%%%%%%%%%%%%%%%%

\begin{itemize}
    \item$I^{(0,2,3/2)}$
\end{itemize}
\begin{align}
I^{(0,2,3/2)}(0,p_2,p_3,m) &= \int_{0}^{1}dx\int_{0}^{1-x}dy\frac{y^{2}}{\left[x(1-x-y)p_{3}^{2}+xyp_{2}^{2}
+m^{2}\right]^{3/2}} \nonumber \\
&= -\frac{2}{3\left(p_{2}^{2}-p_{3}^{2}\right)}\int_{0}^{1}dx\frac{1}{x^{3}\sqrt{(1-x)xp_{2}^{2}+m^{2}}}\left[8m^{4}
+4m^{2}\left(p_{2}^{2}+3p_{3}^{2}\right)(1-x)x\right. \nonumber \\
& \left.-\left(p_{2}^{4}-4p_{2}^{2}p_{3}^{2}-3p_{3}^{4}\right)\left(1-x\right)^{2}x^{2}-8\sqrt{(1-x)xp_{2}^{2}
+m^{2}}\left((1-x)xp_{2}^{2}+m^{2}\right)^{3/2}\right] \nonumber \\
&\overset{m\rightarrow0}{=} -\frac{p_{2}+3p_{3}}{p_{2}\left(p_{2}+p_{3}\right)^{3}}\pi+\frac{1}{\left(p_{2}^{2}
-p_{3}^{2}\right)^{3}}\left[p_{2}^{4}-4p_{2}^{2}p_{3}^{2}+3p_{3}^{4}+2p_{3}^{4}\log\left(\frac{p_{2}^{2}}
{p_{2}^{3}}\right)\right]\lim_{m\rightarrow0}\frac{1}{m}. \nonumber \\
\end{align}

%%%%%%%%%%%%%%%%%%%%%%%%%%  I(0,3,3/2) %%%%%%%%%%%%%%%%%%%%%%%%%%
\begin{itemize}
    \item $I^{(0,3,3/2)}$
\end{itemize}

\begin{align}
    I^{(0,3,3/2)}(0,p_2,p_3,m) &= \int_{0}^{1}dx\int_{0}^{1-x}dy\frac{y^{3}}{\left[x(1-x-y)p_{3}^{2}
    +xyp_{2}^{2}+m^{2}\right]^{3/2}} \nonumber \\
    &\overset{m\rightarrow0}{=}-\frac{3\pi\left(p_{2}^{2}+4p_{2}p_{3}+5p_{3}^{2}\right)}{4p_{2}\left(p_{2}+p_{3}\right)^{4}}\nonumber \\
    & \quad\quad\quad\quad\quad+\frac{2p_{2}^{6}-9p_{2}^{4}p_{3}^{2}+18p_{2}^{2}p_{3}^{4}-11p_{3}^{6}-6p_{3}^{2}
    \log\left(p_{2}^{2}/p_{3}^{2}\right)}{3\left(p_{2}^{2}-p_{3}^{2}\right)^{4}}\lim_{m\rightarrow0}\frac{1}{m} .   \nonumber \\
\end{align}

%%%%%%%%%%%%%%%%%%%%%%%%%%  I(1,1,3/2) %%%%%%%%%%%%%%%%%%%%%%%%%%
\begin{itemize}
    \item $I^{(1,1,3/2)}$
\end{itemize}

\begin{align}
I^{(1,1,3/2)}(0,p_2,p_3,m) &= \int_0^1 dx x I^{(1,3/2)}(0,p_2,p_3,m)(x) \nonumber \\
&= \int_{0}^{1}dx\int_{0}^{1-x}dy\frac{xy}{\left[x(1-x-y)p_{3}^{2}+xyp_{2}^{2}+m^{2}\right]^{3/2}} \nonumber \\
&= -\frac{2}{\left(p_{2}^{2}-p_{3}^{2}\right)}\int_{0}^{1}dx\left(\frac{x(2x+y-2)p_{3}^{2}-xyp_{2}^{2}
-2m^{2}}{x\sqrt{x(1-x-y)p_{3}^{2}+xyp_{2}^{2}+m^{2}}}\right)_{0}^{1-x} \nonumber \\
&= \frac{2}{\left(p_{2}^{2}-p_{3}^{2}\right)}\int_{0}^{1}dx\frac{1}{x}\left(\frac{(1-x)x\left(p_{2}^{2}
+p_{3}^{2}\right)+2m^{2}}{\sqrt{(1-x)xp_{2}^{2}+m^{2}}}-2\sqrt{(1-x)xp_{3}^{2}+m^{2}}\right) \nonumber \\
&= \frac{2}{p_{2}\left(p_{2}^{2}-p_{3}^{2}\right)}\left[\left(p_{2}^{2}+p_{3}^{2}\right)
\cot^{-1}\left(\frac{2m}{p_{2}}\right)-2p_{2}p_{3}\cot^{-1}\left(\frac{2m}{p_{3}}\right)\right. \nonumber \\
& \hspace{7cm} \left.-2mp_{2}\log\left(\frac{1+p_{2}/4m^{2}}{1+p_{3}^{2}/4m^{2}}\right)\right] \nonumber \\
&\overset{m\rightarrow0}{=} \frac{\pi}{p_{2}\left(p_{2}+p_{3}\right)^{2}}.
\end{align}

%%%%%%%%%%%%%%%%%%%%%%%%%%  I(1,2,3/2) %%%%%%%%%%%%%%%%%%%%%%%%%%
\begin{itemize}
    \item $I^{(1,2,3/2)}$
\end{itemize}
\begin{align}
I^{(1,2,3/2)}(0,p_2,p_3,m) &= \int_{0}^{1}dx\int_{0}^{1-x}dy\frac{xy^{2}}{\left[x(1-x-y)p_{3}^{2}
+xyp_{2}^{2}+m^{2}\right]^{3/2}} \nonumber \\
&= -\frac{2}{3\left(p_{2}^{2}-p_{3}^{2}\right)}\int_{0}^{1}dx\frac{1}{x^{2}\sqrt{(1-x)xp_{2}^{2}+m^{2}}}
\left[8m^{4}+4m^{2}\left(p_{2}^{2}+3p_{3}^{2}\right)(1-x)x\right. \nonumber \\
&  \hspace{-1cm}\left.-\left(p_{2}^{4}-6p_{2}^{2}p_{3}^{2}-3p_{3}^{4}\right)\left(1-x\right)^{2}x^{2}
-8\sqrt{(1-x)xp_{2}^{2}+m^{2}}\left((1-x)xp_{2}^{2}+m^{2}\right)^{3/2} \right] \nonumber \\
&\overset{m\rightarrow0}{=} \frac{p_{2}+3p_{3}}{4p_{2}\left(p_{2}+p_{3}\right)^{3}}\pi.
\end{align}

%%%%%%%%%%%%%%%%%%%%%%%%%%  I(2,1,3/2) %%%%%%%%%%%%%%%%%%%%%%%%%%
\begin{itemize}
    \item $I^{(2,1,3/2)}$
\end{itemize}
\begin{align}
    I^{(2,1,3/2)}(0,p_2,p_3,m) &= \int_0^1  dx x^2 I^{(1,3/2)}(0,p_2,p_3,m)(x) \nonumber \\
    &= \frac{2}{\left(p_{2}^{2}-p_{3}^{2}\right)^{2}}\int_{0}^{1}dx\left(\frac{(1-x)x\left(p_{2}^{2}
    +p_{3}^{2}\right)+2m^{2}}{\sqrt{(1-x)xp_{2}^{2}+m^{2}}}-2\sqrt{(1-x)xp_{3}^{2}+m^{2}}\right) \nonumber \\
    &= -\frac{1}{\left(p_{2}^{2}-p_{3}^{2}\right)^{2}}\left[\left(1-\frac{p_{3}^{2}}{p_{2}^{2}}\right)m-\frac{4m^{2}\left(3p_{2}^{2}-p_{3}
    ^{2}\right)+p_{2}^{2}\left(p_{2}^{2}+p_{3}^{2}\right)}{2p_{2}^{2}}\cot^{-1}\left(\frac{2m}{p_{2}}\right)\right. \nonumber \\
    & \hspace*{2cm }\left.+\frac{4m^{2}+p_{3}^{2}}{p_{3}}\cot^{-1}\left(\frac{2m}{p_{3}}\right)\right] \nonumber \\
    &\overset{m\rightarrow0}{=} \frac{\pi}{4p_{2}\left(p_{2}+p_{3}\right)^{2}}.
\end{align}

%%%%%%%%%%%%%%%%%%%%%%%%%%  I(0,0,1/2) %%%%%%%%%%%%%%%%%%%%%%%%%%
\begin{itemize}
    \item $I^{(0,0,1/2)}$
\end{itemize}
\begin{align}
I^{(0,0,1/2)}(0,p_2,p_3,m) &= \int_0^1  dx I^{(0,1/2)}(0,p_2,p_3,m)(x) \nonumber \\
&= \frac{2}{(p_{2}^{2}-p_{3}^{2})}\int_{0}^{1}dx\frac{1}{x}\left(\sqrt{(1-x)xp_{2}^{2}+m^{2}}
-\sqrt{(1-x)xp_{3}^{2}+m^{2}}\right) \nonumber \\
&= \frac{2}{m(p_{2}^{2}-p_{3}^{2})}\left[\cot^{-1}\left(\frac{2m\sqrt{(1-x)xp_{2}^{2}+m^{2}}}
{p_{2}^{2}x+2m^{2}}\right)\right.\nonumber \\
& \hspace{5cm}\left.-\cot^{-1}\left(\frac{2m\sqrt{(1-x)xp_{3}^{2}+m^{2}}}{p_{3}^{2}x+2m^{2}}\right)\right]_{0}^{1} \nonumber \\
&= \frac{2}{p_{2}^{2}-p_{3}^{2}}\left[p_{2}\cot^{-1}\left(\frac{2m}{p_{2}}\right)-p_{3}\cot^{-1}\left(\frac{2m}{p_{3}}\right)\right]
-\frac{2m}{p_{2}^{2}-p_{3}^{2}}\log\left(\frac{1-p_{2}^{2}/4m^{2}}{1-p_{3}^{2}/4m^{2}}\right) \nonumber \\
&\overset{m\rightarrow0}{=} \frac{\pi}{p_{2}+p_{3}}.
\end{align}

%%%%%%%%%%%%%%%%%%%%%%%%%%  I(1,0,1/2) %%%%%%%%%%%%%%%%%%%%%%%%%%
\begin{itemize}
    \item $I^{(1,0,1/2)}$
\end{itemize}
\begin{align}
I^{(1,0,1/2)}(0,p_2,p_3,m)  &= \int_0^1  dx x I^{(0,1/2)}(0,p_2,p_3,m)(x) \nonumber \\
&=  \frac{2}{(p_{2}^{2}-p_{3}^{2})}\int_{0}^{1}dx\left(\sqrt{(1-x)xp_{2}^{2}+m^{2}}-\sqrt{(1-x)xp_{3}^{2}+m^{2}}\right)\nonumber \\
&= \frac{1}{2\left(p_{2}^{3}p_{3}-p_{2}p_{3}^{3}\right)}\left[\left(p_{2}^{2}+4m^{2}\right)p_{3}\cot^{-1}\left(\frac{2m}{p_{2}}
\right)-\left(p_{3}^{2}+4m^{2}\right)p_{2}\cot^{-1}\left(\frac{2m}{p_{3}}\right)\right]\nonumber \\
&\overset{m\rightarrow0}{=} \frac{\pi}{4\left(p_{2}+p_{3}\right)}.
\end{align}

%%%%%%%%%%%%%%%%%%%%%%%%%%  I(0,1,1/2) %%%%%%%%%%%%%%%%%%%%%%%%%%
\begin{itemize}
    \item $I^{(0,1,1/2)}$
\end{itemize}
\begin{align}
    I^{(0,1,1/2)}(0,p_2,p_3,m) &= \int_0^1  dx I^{(1,1/2)}(0,p_2,p_3,m)(x) \nonumber \\
    &= -\frac{2}{3\left(p_{2}^{2}-p_{3}^{2}\right)^{2}}\int_{0}^{1}dx\frac{\left[\left(3p_{3}^{2}-p_{2}^{2}\right)(1-x)x
    +2m^{2}\right]\sqrt{(1-x)xp_{2}^{2}+m^{2}}}{x^{2}}\nonumber \\
    & = \frac{1}{2p_{2}\left(p_{2}^{2}-p_{3}^{2}\right)^{2}}\left[\left(p_{2}^{4}-3p_{2}^{2}p_{3}^{2}+4m^{2}
    \left(p_{2}^{2}+p_{3}^{2}\right)\right)\cot^{-1}\left(\frac{2m}{p_{2}}\right)\right.\nonumber \\
    &\left.-2p_{2}\left(4m^{2}p_{3}-p_{3}^{3}\right)\cot^{-1}\left(\frac{2m}{p_{3}}\right)+m\left(p_{3}^{2}
    -p_{2}^{2}+2p_{3}^{2}\log\left(\frac{1-p_{2}^{2}/4m^{2}}{1-p_{3}^{2}/4m^{2}}\right)\right)\right]\nonumber \\
    & \overset{m\rightarrow0}{=}\frac{p_{2}+2p_{3}}{4\left(p_{2}+p_{3}\right)^{2}} \pi.
\end{align}

\subsection{Third and final case}

%%%%%%%%%%%%%%%%%%%%% I(0,0,3/2) %%%%%%%%%%%%%%%%%%%%%

\begin{itemize}
    \item $I^{(0,0,3/2)}$
\end{itemize}
\begin{align}
    I^{(0,0,3/2)}(p_1,p_2,p_3) &= \int_0^1 dx I^{(0,3/2)}(p_1,p_2,p_3)(x)  \nonumber \\
    &= \frac{2\left(p_{2}+p_{3}\right)}{p_{2}p_{3}}\int_{0}^{1}dx\frac{1}{\sqrt{(1-x)x}\left[p_{1}^{2}(1-x)
    +\left(p_{2}+p_{3}\right){}^{2}x\right]} \nonumber \\
    & = \left. \frac{4}{p_1 p_2 p_3 \sqrt{1-x}} \sqrt{x-1} \tanh ^{-1}\left(\frac{p_2+p_3}{p_1 } 
    \sqrt{\frac{x}{x-1}}\right) \right|_0^1.
\end{align}

Evaluate the lower limit is easy since $\tanh^{-1}(0) = 0$. For the upper limit of integration, however, the limit $x\rightarrow 1$ gives
\begin{equation}
    \lim_{x\rightarrow 1}  \frac{4}{p_1 p_2 p_3 \sqrt{1-x}} \sqrt{x-1} \tanh ^{-1}\left(\frac{p_2+p_3}{p_1 } \sqrt{\frac{x}{x-1}}\right) 
    = \frac{2 \pi }{p_1 p_2 p_3}.
\end{equation}

Hence, the integral is 
\begin{equation}
    I^{(0,0,3/2)}(p_1,p_2,p_3) = \frac{2 \pi }{p_1 p_2 p_3}.
\end{equation}

%%%%%%%%%%%%%%%%%%%%% I(1,0,3/2) %%%%%%%%%%%%%%%%%%%%%
\begin{itemize}
    \item $I^{(1,0,3/2)}$
\end{itemize}
\begin{align}
    I^{(1,0,3/2)}(p_1,p_2,p_3) &= \int_0^1 dx x I^{(0,3/2)}(p_1,p_2,p_3)(x)  \nonumber \\
    &= \frac{2\left(p_{2}+p_{3}\right)}{p_{2}p_{3}}\int_{0}^{1}dx\frac{x}{\sqrt{(1-x)x}\left[p_{1}^{2}(1-x)+\left(p_{2}
    +p_{3}\right){}^{2}x\right]}\nonumber \\
    & = \frac{4\sqrt{x-1}}{p_{2}p_{3}\left[\left(p_{2}+p_{3}\right){}^{2}-p_{1}^{2}\right]\sqrt{1-x}}\left[\left(p_{2}
    +p_{3}\right)\sinh^{-1}\left(\sqrt{x-1}\right)\right. \nonumber \\
    &\quad\quad\quad\quad\quad\quad\quad\quad\quad\quad\quad\quad\quad\quad\quad \left.\left.-p_{1}\tanh^{-1}\left(\frac{p_{2}+p_{3}}{p_{1}}\sqrt{\frac{x}{x-1}}\right)\right]\right|_{0}^{1}.
\end{align}

For $x=0$, this gives 
\begin{equation}
    -\frac{2 \pi  \left(p_2+p_3\right)}{p_2 p_3 \left[\left(p_2+p_3\right)^2-p_1^2\right]},
\end{equation}
by direct substitution. For $x=1$ we take the limit, obtaining
\begin{equation}
    -\frac{2 \pi  p_1}{p_2 p_3 \left[\left(p_2+p_3\right)^2-p_1^2\right]},
\end{equation}
such that  
\begin{align}
    I^{(1,0,3/2)}(p_1,p_2,p_3) &=-\frac{2 \pi  p_1}{p_2 p_3 \left[\left(p_2+p_3\right)^2-p_1^2\right]} + \frac{2 \pi  \left(p_2+p_3\right)}{p_2 p_3 \left[\left(p_2+p_3\right)^2-p_1^2\right]} \nonumber \\
    &= \frac{2 \pi }{p_2 p_3 \left(p_1+p_2+p_3\right)}.
\end{align}

%%%%%%%%%%%%%%%%%%%%% I(2,0,3/2) %%%%%%%%%%%%%%%%%%%%%
\begin{itemize}
    \item $I^{(2,0,3/2)}$
\end{itemize}

\begin{align}
    I^{(2,0,3/2)}(p_1,p_2,p_3) &= \int_0^1 dx x^2 I^{(0,3/2)}(p_1,p_2,p_3)(x)  \nonumber \\
    &= \frac{2\left(p_{2}+p_{3}\right)}{p_{2}p_{3}}\int_{0}^{1}dx\frac{x^2}{\sqrt{(1-x)x}\left[p_{1}^{2}(1-x)+\left(p_{2}
    +p_{3}\right){}^{2}x\right]}\nonumber \\
    & = \frac{2}{p_{2}p_{3}\left[\left(p_{2}+p_{3}\right)^{2}-p_{1}^{2}\right]^{2}}\left\{ \left(p_{2}+p_{3}\right)
    p_{1}^{2}\left[\sqrt{(1-x)x}-3\sin^{-1}\left(\sqrt{x}\right)\right]\right.\nonumber \\
    & \left.\left.-\left(p_{2}+p_{3}\right){}^{3}\left[\sqrt{(1-x)x}-\sin^{-1}\left(\sqrt{x}\right)\right]+2p_{1}^{3}\tan^{-1}
    \left(\frac{p_{2}+p_{3}}{p_{1}}\sqrt{\frac{x}{1-x}}\right)\right\} \right|_{0}^{1}.
\end{align}

The lower limit is zero, whereas for the upper limit we take $x\rightarrow 1$, obtaining
\begin{equation}
    I^{(2,0,3/2)}(p_1,p_2,p_3) = \frac{2 p_1+p_2+p_3}{p_2 p_3 \left(p_1+p_2+p_3\right)^2} \pi.
\end{equation}

%%%%%%%%%%%%%%%%%%%%% I(3,0,3/2) %%%%%%%%%%%%%%%%%%%%%
\begin{itemize}
    \item $I^{(3,0,3/2)}$
\end{itemize}

\begin{align}
    I^{(3,0,3/2)}(p_1,p_2,p_3) &= \int_0^1 dx x^3 I^{(0,3/2)}(p_1,p_2,p_3)(x)  \nonumber \\
    &= \frac{2\left(p_{2}+p_{3}\right)}{p_{2}p_{3}}\int_{0}^{1}dx\frac{x^3}{\sqrt{(1-x)x}\left[p_{1}^{2}(1-x)+\left(p_{2}
    +p_{3}\right)^{2}x\right]} \nonumber \\
    &= \frac{p_{2}+p_{3}}{2p_{2}p_{3}\left[-p_{1}^{2}+\left(p_{2}+p_{3}\right)^{2}\right]^{3}}\left\{ -\frac{8p_{1}^{5}}
    {\left(p_{2}+p_{3}\right)}\tan^{-1}\left(\frac{p_{2}+p_{3}}{p_{1}}\sqrt{\frac{x}{1-x}}\right)\right.\nonumber \\
    & +\sqrt{(1-x)x}\left[p_{1}^{2}(2x+7)-\left(p_{2}+p_{3}\right)^{2}(2x+3)\right]\left[-p_{1}^{2}+\left(p_{2}+p_{3}\right)^{2}\right]
    \nonumber \\
    & \left.\left.+\left(15p_{1}^{4}-10\left(p_{2}+p_{3}\right)^{2}p_{1}^{2}+3\left(p_{2}+p_{3}\right)^{4}\right)\sin^{-1}\left(\sqrt{x}\right)
    \right\}\frac{}{} \right|_{0}^{1}.
\end{align}

Again the lower limit is zero, and the upper limit gives 
\begin{equation}
    I^{(3,0,3/2)}(p_1,p_2,p_3) = \frac{ 8 p_1^2+9 \left(p_2+p_3\right) p_1+3 \left(p_2+p_3\right)^2}{4 p_2 p_3 
    \left(p_1+p_2+p_3\right)^3}\pi .
\end{equation}

%%%%%%%%%%%%%%%%%%%%% I(1,1,3/2) %%%%%%%%%%%%%%%%%%%%%
\begin{itemize}
    \item $I^{(1,1,3/2)}$
\end{itemize}
\begin{align}
    I^{(1,1,3/2)}(p_1,p_2,p_3) &= \int_0^1 dx x I^{(1,3/2)}(p_1,p_2,p_3)(x)  \nonumber \\
    &=\frac{2}{p_{2}}\int_{0}^{1}dx\frac{\sqrt{(1-x)x}}{p_{1}^{2}(1-x)+(p_{2}+p_{3})^{2}x} \nonumber \\
    &=\frac{2}{p_{2}\left[-p_{1}^{2}+\left(p_{2}+p_{3}\right)^{2}\right]^{2}}\left\{ p_{1}^{2}
    \left[\sqrt{x\left(x-1\right)}+\sinh^{-1}\left(\sqrt{x-1}\right)\right]\right.\nonumber \\
    &-2p_{1}\left(p_{2}+p_{3}\right)\tanh^{-1}\left(\frac{p_{2}+p_{3}}{p_{1}}\sqrt{\frac{x}{x-1}}\right)\nonumber \\
    &\left.\left.+\left(p_{2}+p_{3}\right)^{2}\left[\sinh^{-1}\left(\sqrt{x-1}\right)-\sqrt{x(x-1)}\right]\right\} \frac{}{}\right|_{0}^{1}.
\end{align}

The lower limit gives  
\begin{equation}
    -\frac{p_{1}^{2}+\left(p_{2}+p_{3}\right)^{2}}{p_{2}\left(-p_{1}+p_{2}+p_{3}\right)^{2}\left(p_{1}+p_{2}+p_{3}\right)^{2}}\pi,
\end{equation}
whereas for the upper limit we find, after taking the limit $x\rightarrow 1$,
\begin{equation}
    -\frac{2 \pi  p_1 \left(p_2+p_3\right)}{p_2 \left(-p_1+p_2+p_3\right)^2 \left(p_1+p_2+p_3\right)^2},
\end{equation}
such that 
\begin{equation}
    I^{(1,1,3/2)}(p_1,p_2,p_3) = \frac{\pi }{p_2 \left(p_1+p_2+p_3\right)^2}.
\end{equation}

%%%%%%%%%%%%%%%%%%%%% I(2,1,3/2) %%%%%%%%%%%%%%%%%%%%%
\begin{itemize}
    \item $I^{(2,1,3/2)}$
\end{itemize}
\begin{align}
    I^{(2,1,3/2)}(p_1,p_2,p_3) &= \int_0^1 dx x^2 I^{(1,3/2)}(p_1,p_2,p_3)(x)  \nonumber \\
    & \hspace*{-2cm} = \frac{2}{p_{2}}\int_{0}^{1}dx\frac{x \sqrt{(1-x)x}}{p_{1}^{2}(1-x)+(p_{2}+p_{3})^{2}x} \nonumber \\
    & \hspace*{-2cm} = -\frac{\sqrt{x-1}}{2p_{2}\left[-p_{1}^{2}+(p_{2}+p_{3})^{2}\right]^{3}\sqrt{1-x}}\left\{ p_{1}
    ^{4}\left[(2x+3)\sqrt{x\left(x-1\right)}+3\sinh^{-1}\left(\sqrt{x-1}\right)\right]\right. \nonumber \\
    & \hspace*{-2cm} -8p_{1}^{3}\left(p_{2}+p_{3}\right)\tanh^{-1}\left(\frac{p_{2}+p_{3}}{p_{1}}\sqrt{\frac{x}{x-1}}\right) \nonumber \\
    & \hspace*{-2cm}-2p_{1}^{2}\left(p_{2}+p_{3}\right){}^{2}\left[\sqrt{x\left(x-1\right)}(2x+1)-3\sinh^{-1}\left(\sqrt{x-1}\right)\right] 
    \nonumber \\
    & \hspace*{-2cm} \left.\left.+\left(p_{2}+p_{3}\right){}^{4}+\left[\sqrt{x\left(x-1\right)}(2x-1)-\sinh^{-1}\left(\sqrt{x-1}\right)\right]\right\} 
    \right|_{0}^{1}.
\end{align}

For the lower limit we find 
\begin{equation}
    \frac{\pi  \left(3 p_1^4+6 \left(p_2+p_3\right)^2 p_1^2-\left(p_2+p_3\right)^4\right)}{4 p_2 \left(-p_1+p_2
    +p_3\right)^3 \left(p_1+p_2+p_3\right)^3},
\end{equation}
whereas for the upper limit, again we have to take the limit $x\rightarrow 1$ to obtain 
\begin{equation}
    \frac{2 \pi  p_1^3 \left(p_2+p_3\right)}{p_2 \left(-p_1+p_2+p_3\right)^3 \left(p_1+p_2+p_3\right)^3},
\end{equation}
such that  
\begin{equation}
    I^{(2,1,3/2)}(p_1,p_2,p_3) = \frac{  3 p_1+p_2+p_3}{4 p_2 \left(p_1+p_2+p_3\right)^3}\pi.
\end{equation}

%%%%%%%%%%%%%%%%%%%%% I(0,0,1/2) %%%%%%%%%%%%%%%%%%%%%
\begin{itemize}
    \item $I^{(0,0,1/2)}$
\end{itemize}
\begin{align}
    I^{(0,0,1/2)}(p_1,p_2,p_3) &= \int_0^1 dx  I^{(0,1/2)}(p_1,p_2,p_3)(x)  \nonumber \\
    &= \frac{i}{p_{1}}\int_{0}^{1}dx\log\left(\frac{p_{1}^{2}(1-x)+2ip_{2}p_{1}\sqrt{(1-x)x}+\left(p_{3}^{2}-p_{2}^{2}\right)x}
    {-p_{1}^{2}(1-x)+2ip_{3}p_{1}\sqrt{(1-x)x}+\left(p_{3}^{2}-p_{2}^{2}\right)x}\right) \nonumber \\
    & =  \frac{\pi }{p_1+p_2+p_3}.
\end{align}

%%%%%%%%%%%%%%%%%%%%% I(1,0,1/2) %%%%%%%%%%%%%%%%%%%%%
\begin{itemize}
    \item $I^{(1,0,1/2)}$
\end{itemize}
\begin{align}
    I^{(1,0,1/2)}(p_1,p_2,p_3) &= \int_0^1 dx x I^{(0,1/2)}(p_1,p_2,p_3)(x)  \nonumber \\
    &= \frac{i}{p_{1}}\int_{0}^{1}dx x \log\left(\frac{p_{1}^{2}(1-x)+2ip_{2}p_{1}\sqrt{(1-x)x}+\left(p_{3}^{2}
    -p_{2}^{2}\right)x}{-p_{1}^{2}(1-x)+2ip_{3}p_{1}\sqrt{(1-x)x}+\left(p_{3}^{2}-p_{2}^{2}\right)x}\right) \nonumber \\
    &= \frac{\pi  \left(2 p_1+p_2+p_3\right)}{4 \left(p_1+p_2+p_3\right)^2}.
\end{align}

\section{Match of our results to previous constraints on the 3-point function}\label{previous}

Here we check the consistency of our results for the general one-loop 3-point function with \cite{Bzowski:2013sza,Bzowski:2017poo}.

In \cite{Bzowski:2017poo}, it was shown that the full correlator can be written in terms of a longitudinal part 
(containing the 2-point function $\langle JJ\rangle$) and a transverse-traceless part, which in turn depends only 
on two form factors. The decomposition, using the notation in that paper, is  
\begin{align}
    & \left \langle \left \langle J^{\mu_1 a_1} ( \boldsymbol{p}_1 )  J^{\mu_2 a_2}(\boldsymbol{p}_2)  J^{\mu_3 a_3}(\boldsymbol{p}_3) \right \rangle \right \rangle \nonumber \\
    & =\left \langle \left \langle j^{\mu_1 a_1}(\boldsymbol{p}_1)  j^{\mu_2 a_2}(\boldsymbol{p}_2)  j^{\mu_3 a_3}(\boldsymbol{p}_3) \right \rangle \right \rangle \nonumber \\
    & \quad + \left( \left[ \frac{p_1^{\mu_1}}{p_1^2}\left(g f^{a_1 b a_3}\left\langle\left\langle J^{\mu_3 b} ( \boldsymbol{p}_2 )  J^{\mu_2 a_2} ( -\boldsymbol{p}_2 ) \right\rangle\right\rangle-g f^{a_1 a_2 b}\left\langle\left\langle J^{\mu_2 b} ( \boldsymbol{p}_3 )  J^{\mu_3 a_3} ( -\boldsymbol{p}_3 ) \right\rangle\right\rangle\right)\right]\right. \nonumber \\
    & \quad \quad \left.+\left[\left(\mu_1, a_1, \boldsymbol{p}_1\right) \leftrightarrow\left(\mu_2, b_2, \boldsymbol{p}_2\right)\right]+\left[\left(\mu_1, a_1, \boldsymbol{p}_1\right) \leftrightarrow\left(\mu_3, a_3, \boldsymbol{p}_3\right)\right]\right) \nonumber \\
    & \quad  +\left(\left[\frac{p_1^{\mu_1} p_2^{\mu_2}}{p_1^2 p_2^2} g f^{a_1 a_2 b} p_{2 \alpha}\left\langle\left\langle J^{\alpha b} (\boldsymbol{p}_3 ) J^{\mu_3 a_3} (-\boldsymbol{p}_3 )\right\rangle\right\rangle\right]\right. \nonumber \\
    &  \quad \quad \left.+\left[\left(\mu_1, a_1, \boldsymbol{p}_1\right) \leftrightarrow\left(\mu_3, a_3, \boldsymbol{p}_3\right)\right]+\left[\left(\mu_2, a_2, \boldsymbol{p}_2\right) \leftrightarrow\left(\mu_3, a_3, \boldsymbol{p}_3\right)\right]\right) \;,
    & \label{full_correlator_decomposition}
\end{align}
where the transverse-traceless part is given by
\begin{align}
    & \left \langle \left \langle j^{\mu_1 a_1} (\boldsymbol{p}_1)  j^{\mu_2 a_2} (\boldsymbol{p}_2)  j^{\mu_3 a_3} (\boldsymbol{p}_3) \right \rangle \right \rangle  \nonumber \\
    & =\pi_{\alpha_1}^{\mu_1}\left(\boldsymbol{p}_1\right) \pi_{\alpha_2}^{\mu_2}\left(\boldsymbol{p}_2\right) \pi_{\alpha_3}^{\mu_3}\left(\boldsymbol{p}_3\right)\left[A_1^{a_1 a_2 a_3} p_2^{\alpha_1} p_3^{\alpha_2} p_1^{\alpha_3}\right. \nonumber \\
    & \quad \quad \left.+A_2^{a_1 a_2 a_3} \delta^{\alpha_1 \alpha_2} p_1^{\alpha_3}+A_2^{a_3 a_1 a_2}\left(p_3, p_1, p_2\right) \delta^{\alpha_1 \alpha_3} p_3^{\alpha_2}+A_2^{a_2 a_3 a_1}\left(p_2, p_3, p_1\right) \delta^{\alpha_2 \alpha_3} p_2^{\alpha_1}\right] . \nonumber \\
    &
\end{align}

Considering the different sign convention for momenta in \cite{Bzowski:2017poo}, with $p_1=-p_2-p_3$, but again with 
cyclic $ \boldsymbol{p_1} \leftrightarrow (\mu, \rho)$, $ \boldsymbol{p_2} \leftrightarrow (\nu, \mu)$, 
$ \boldsymbol{p_3} \leftrightarrow (\rho,\nu) $, we obtain the coefficient functions as 
\begin{align}
    & A_1^{a_1 a_2 a_3}=\text { coefficient of } p_2^{\mu_1} p_3^{\mu_2} p_1^{\mu_3} \nonumber \\
    & A_2^{a_1 a_2 a_3}=\text { coefficient of } \delta^{\mu_1 \mu_2} p_1^{\mu_3}\;,
\end{align}
or in our convention
\begin{align}
    & A_1^{abc}=\text { coefficient of } p_{2 \mu} p_{3\nu} p_{1 \rho} \nonumber \\
    & A_2^{abc} = \text { coefficient of } \delta_{\mu \nu} p_{1 \rho }\;,
\end{align}
leading to 
\begin{align}
    A_1^{abc}(p_1, p_2, P_3) & = N^2 \epsilon^{abc} c_0(c_7-c_{11}) = - N^2\epsilon^{abc} \frac{1}{2(p_1+p_2+p_3)^3} \\
    A_2^{abc}(p_1, p_2, P_3) & =  N^2 \epsilon^{abc} c_0c_8 = - N^2\epsilon^{abc} \frac{p_1 + p_2 + 2p_3}{4 (p_1+p_2+p_3)^2} \label{form_factors}.
\end{align}

Finally, it remains to show consistency of our result with \eqref{full_correlator_decomposition}. We first rewrite 
\eqref{full_correlator_decomposition} in our notation of indices and momenta:
\begin{align}
    &  \langle j_{\mu}^a ( p_1 )  j_{\nu}^b (-p_2)  j_{\rho}^c (-p_3)  \rangle  =   \langle j_{\mu}^a (p_1)  j_{\nu}^b (p_2)  j_{\rho}^c (p_3) \rangle_{\text{transverse-traceless}} \nonumber \\
    & + \frac{p_{1\mu}}{p_1^2} \left[\frac{1}{2} \epsilon^{adc} \langle j_\rho^d(p_2) j_\nu^b (-p_2) \rangle  - \frac{1}{2} \epsilon^{abd}  \langle j_\nu^d(p_3) j_\rho^c (-p_3) \rangle\right]  \nonumber \\ 
    & - \frac{p_{2\nu}}{p_2^2}  \left[\frac{1}{2} \epsilon^{bdc} \langle j_\rho^d(p_1) j_\mu^a (-p_1) \rangle  - \frac{1}{2} \epsilon^{bad}  \langle j_\mu^d(p_3) j_\rho^c (-p_3) \rangle\right]  \nonumber \\
    & - \frac{p_{3\rho}}{p_3^2}  \left[\frac{1}{2} \epsilon^{cda} \langle j_\mu^d(p_2) j_\nu^b (-p_2) \rangle  -  \frac{1}{2} \epsilon^{cbd}  \langle j_\nu^d(p_1) j_\mu^a (-p_1) \rangle\right]  \nonumber \\
    & + \frac{p_{1\mu} p_{2\mu}}{p_1^2 p_2^2} \left( \frac{1}{2} \epsilon^{abd} p_{2\alpha} \langle j_\alpha^d(p_3) j_\rho^c (-p_3) \rangle \right) \nonumber \\
    & - \frac{p_{3\rho} p_{2\nu}}{p_3^2 p_2^2} \left( \frac{1}{2} \epsilon^{cbd} p_{2\alpha} \langle j_\alpha^d(p_1) j_\mu^a (-p_1) \rangle \right) \nonumber \\
    & + \frac{p_{1\mu } p_{2\nu}}{p_1^2 p_3^2} \left( \frac{1}{2} \epsilon^{acd} p_{3\alpha} \langle j_\alpha^d(p_2) j_\nu^b (-p_2) \rangle \right) 
\end{align}
where 
\begin{align}
    & \langle j_{\mu}^a (p_1)  j_{\nu}^b (p_2)  j_{\rho}^c (p_3) \rangle_{\text{transverse-traceless}} = \pi_{\mu}^\alpha(p_1) \pi_{\nu}^\beta(p_2) \pi_{\rho}^\gamma(p_3) \left[ A^{abc}_1(p_1, p_2, p_3) p_{2\alpha} p_{3\nu} p_{1\rho} \right. \nonumber \\
    & \left. + A_2^{abc}(p_1,p_2,p_3) \delta_{\alpha \beta} p_{1\gamma} - A_2^{cab}(p_3,p_1,p_2) \delta_{\alpha \gamma} p_{3\beta } - A_2^{bca}(p_2,p_3,p_1) \delta_{\beta \gamma} p_{2\alpha } \right]\;,
\end{align}
and then use the software MATHEMATICA together with the package "xAct" for tensorial calculations, and find indeed 
consistency with \cite{Bzowski:2017poo}, with the form factors \eqref{form_factors}.

\bibliography{3pntfctref}
\bibliographystyle{utphys}

\end{document}